# Contagion Source Detection in Epidemic and Infodemic Outbreaks: Mathematical Analysis and Network Algorithms

**Chee Wei Tan**
Nanyang Technological University
cheewei.tan@ntu.edu.sg

**Pei-Duo Yu**
Chung Yuan Christian University
peiduoyu@cycu.edu.tw

**now**

the essence of knowledge

Boston — Delft

# Contents

# Contagion Source Detection in Epidemic and Infodemic Outbreaks: Mathematical Analysis and Network Algorithms

Chee Wei Tan[1] and Pei-Duo Yu[2]

[1] *Nanyang Technological University, Singapore; cheewei.tan@ntu.edu.sg*
[2] *Chung Yuan Christian University, Taiwan; peiduoyu@cycu.edu.tw*

---

## ABSTRACT

The rapid spread of infectious diseases and online rumors share similarities in terms of their speed, scale, and patterns of contagion. Although these two phenomena have historically been studied separately, the COVID-19 pandemic has highlighted the devastating consequences that simultaneous crises of epidemics and misinformation can have on the world. Soon after the outbreak of COVID-19, the World Health Organization launched a campaign against the COVID-19 Infodemic, which refers to the dissemination of pandemic-related false information online that causes widespread panic and hinders recovery efforts. Undoubtedly, *nothing spreads faster than fear.*

Networks serve as a crucial platform for viral spreading, as the actions of highly influential users can quickly render others susceptible to the same. The potential for contagion in epidemics and rumors hinges on the initial source, underscoring the need for rapid and efficient digital contact

---

tracing algorithms to identify superspreaders or Patient Zero. Similarly, detecting and removing rumor mongers is essential for preventing the proliferation of harmful information in online social networks. Identifying the source of large-scale contagions requires solving complex optimization problems on expansive graphs. Accurate source identification and understanding the dynamic spreading process requires a comprehensive understanding of surveillance in massive networks, including topological structures and spreading veracity. Ultimately, the efficacy of algorithms for digital contact tracing and rumor source detection relies on this understanding.

This monograph provides an overview of the mathematical theories and computational algorithm design for contagion source detection in large networks. By leveraging network centrality as a tool for statistical inference, we can accurately identify the source of contagions, trace their spread, and predict future trajectories. This approach provides fundamental insights into surveillance capability and asymptotic behavior of contagion spreading in networks. Mathematical theory and computational algorithms are vital to understanding contagion dynamics, improving surveillance capabilities, and developing effective strategies to prevent the spread of infectious diseases and misinformation.

# 1

## Introduction

### 1.1 Epidemics and Rumors

The spreading of epidemics and rumors on networks share many important features [28], [29], [36], [58]. The underlying network interaction cannot be directly observed and often has to be implicitly inferred from macroscopic phenomenons. Driven by the same human collective crowd behavior, these network dynamics can lead to common network effects like the small-world phenomenon and percolation thresholds. The study of epidemics and rumor spreading is thus an important part of applied probability theory and graph theory related to the analysis of the evolution of large systems arising in networks. Even though the process of spreading information in online social networks differs from that of disease epidemics, the proliferation of fake news and recent disinformation campaigns in online social networks has emerged in recent years as a formidable cybersecurity threat that can have catastrophic real-world consequences like a pandemic [45], [92], [135], [149].

Though epidemics and rumor spreading have been separately studied in the past with a longer history for the stochastic theory of epidemic spreading, the COVID-19 pandemic has been the first of simultaneous global crises in which both the epidemic and overabundance of mis-information devastatingly wreak havoc on the world. The COVID-19 pandemic is the first pandemic in history in which humans rely heavily on the Internet and online social networks to stay connected amidst the prolonged lockdown and social distancing measures in place. It has also spawned an epidemic of online misinformation, undermining the efficacy of online social networks that humans crucially rely on and disrupting public health risk communications. Shortly after the COVID-19 pandemic started, the World Health Organization (WHO) declared war against the COVID-19 Infodemic, which is the viral spreading of pandemic-related misinformation or disinformation in social media [54].

Spreading processes are dynamic cascading phenomena where the action of some users increases the susceptibility of other users to the same; this results in the successive spread of a disease virus or rumor from an initial few users to a much larger set of users [28], [29], [36], [58]. When a new infectious virus spreads, public healthcare authorities want to identify persons who may have come into contact with an infected person and to trace close social contacts in order to stop ongoing transmission or reduce the spread of infection. When rumors like false treatment for the COVID-19 disease spread in online social networks, this can prevent humans from adopting the right behaviors to reduce the COVID-19 pandemic risk. Once misinformation morphed into disinformation attacks, it can be disruptive and deadly. These simultaneous crises require both public healthcare and cyber security experts to work together to fight infodemics by identifying sources of misinformation.

An objective of interests is to unravel the dynamical spreading process to root out the malicious source quickly, accurately, and reliably with only limited observation data of infected nodes in the network. Just like epidemic countermeasures like digital contact tracing and policies to identify *Patient Zero* in an outbreak,[1] building resilience to catastrophic viral misinformation is of huge importance to a safe and functioning cyberspace because of the highly-connected online social networks.

---

[1] Contact tracing apps based on the Bluetooth wireless radio standard are arguably one of the defining technologies for surveillance during the COVID-19 pandemic [65], [91].

To accurately detect or predict the causation of contagion in large networks, it is crucial to identify the superspreaders and the origin of disease viruses. Similarly, it is essential to determine who is spreading rumors and disinformation to cause division and influence decisions among users. This raises questions about the provenance of such information [92], [100], [150]. Additionally, the implications of network surveillance and the response to contain a contagion must be considered. With the emergence of new communication platforms, new avenues for spreading misinformation and disinformation arise. Identifying the source of contagion can have far-reaching consequences, such as timely responses to the next pandemic or promoting a safe cyberspace.

Numerous fundamental questions remain unanswered in the statistical inference of infection sources in networks. The theory of stochastic processes over large networks is still evolving, and the computational aspects of estimation and detection in networks have not yet been systematically examined, with source identification understood only in the simplest graph topology cases. It is remarkable that even though human social interaction or online social networks are not designed with the intention of spreading a payload (such as an infectious disease virus or rumor) as rapidly as possible, the process of viral spreading over large networks is not fully comprehended [92], [100], [150]. Are there specific network structures, quantifiable measures of user influence that promote viral spreading? If so, what particular features could aid in the development of better digital contact tracing strategies or interventions to counter the spread of malicious rumors?

As we strive to comprehend the spread of contagions across large networks, it is crucial to recognize the potential for cross-pollination of ideas between different types of networks, each with distinct interaction graph structures, initial nodes, and nature of user interactions [28], [29], [36], [58]. For instance, in [6], researchers proposed intervention strategies based on a generative model of viral misinformation spread using infectious disease spreading dynamics. Moreover, when network topology abstraction is sufficiently random, it may provide insights into network phenomena based on percolation theory, as noted in [38].

## 1.2 Propagated Epidemics and Contact Tracing

Tracing the origins of propagated epidemics can be traced back to the investigation of the 1854 London cholera epidemics by John Snow (1813–1858), who is widely recognized as a pioneer of modern epidemiology [5], [49], [50]. His work in tracking the source of the cholera outbreak was a significant breakthrough in epidemiological research. By creating detailed dot distribution maps of household deaths due to cholera, Snow was able to identify the source of the epidemic - a water pump located in Broad Street, Golden Square. Snow's methodical tracing effort was one of the earliest applications of inferential statistics to the study of epidemics [5], [49], [50]. Additionally, his heroic intervention in persuading the parish's vestrymen to remove the water pump symbolizes one of the earliest examples of public health action. It is important to note that Snow's contribution to epidemiology was not only a significant scientific achievement but also a landmark event in the history of public health. The removal of the water pump resulted in the rapid cessation of the cholera epidemic, saving countless lives and laying the foundation for modern epidemiological research.

Nowadays, epidemiologists agree that it is necessary to employ contact tracing to stop an infectious disease from spreading: Once a person has been diagnosed as infected, public health authorities fan out to trace the recent contacts of this person for the purpose of monitoring or quarantine. This process repeats if one of those contacts exhibits symptoms until all the contacts who have been exposed are out of circulation. Contact tracing can be effective in the early stage of an epidemic. However, the COVID-19 pandemic had revealed severe deficiencies in public health protection due to asymptomatic infections. Prior study [22] shows that asymptomatic infections need to be considered in analyzing the spread of the disease. The COVID-19 disease is highly contagious, wideranging with long incubation periods and transmissible within 6 feet. Its speed and scale of infection had overwhelmed most contact tracing capabilities which are labor-intensive, cost-inefficient and very slow [45], [86]. A new public health innovation, *digital contact tracing*, then came to the scene. Digital contact tracing leverages a plethora of mobile apps to contact trace people and to provide exposure notifications [8], [17], [46], [65], [82], [91], [94], [95].

Current contact tracing practices focus primarily on finding recent contacts of index cases, while overlooking the source of origin. In fact, source inference is an important factor that explains the initial success of backward contact tracing adopted by countries like Japan and Australia in the early days of the COVID-19 pandemic [12], [17], [86], showing that, whenever there is a sudden outbreak, tracing transmission events rather than infectious individuals can efficiently and effectively prevent infection waves.

There are several challenging unsolved problems in digital contact tracing [17], [81], [91], [139]. First, what is the fundamental relationship between infectiousness and the agility of contact tracing? Can contact tracing be faster than the spreading of an infectious disease? Second, how to quadruple the speed of contact tracing? Can backward contact tracing complement forward contact tracing to find Patient Zero or the superspreaders accurately? Third, can we design disease surveillance networks so as to provide timely prediction and early warning capability to automate digital contact tracing upon the arrival of future epidemics?

## 1.3 Disinformation and Rumor Source Detection

Online social networks like Twitter, Facebook, and YouTube are critical online platforms for spreading news and the diffusion of all kinds of information. They can however cause misinformation and disinformation to spread faster and more rampantly than the traditional "word-of-mouth" mechanism [3], [15], [52], [62], [92], [96], [98], [110], [135], [149], [159]. In fact, false news spreads faster than the truth in a Twitter network [150]. Misinformation is inaccurate or unreliable information that is spread regardless of an intent to mislead. On the other hand, disinformation is intentionally-fabricated misinformation (e.g., hoax news) that is spread with the intent to influence people to make certain decisions or to further an agenda. A malicious rumor monger can now "infect" people across geographical regions on a massive scale faster than ever before. Online rumors, misinformation, and disinformation can thus disrupt livelihood and have serious real-world repercussions.

Recent examples are political mobilization messages spreading in social media that sparked off waves of demonstrations and protests in the Middle East (dubbed the "Arab Spring" or the Twitter revolution) in 2010-2012. In 2013, a bogus Tweet that the White House was attacked went viral after it was sent out by the Associated Press Twitter account that was hacked [11]. This incident momentarily crashed the stock market, demonstrating how online disinformation can cause flash crash and allowing computer hackers to profiteer in the process. A similarly severe incident happened in 2020 when computer hackers seized control of dozens of Twitter accounts belonging to high-profile users like Barack Obama and Elon Musk to tweet out a "double your bitcoin" scam, which went viral quickly. Eventually, this cryptocurrency scam led to a theft of bitcoins worth more than US $110,000 before all the scam messages were removed. Such Internet frauds and cybersecurity threats will be more widespread, especially when bots are recruited to sow discord to amplify the spread of disinformation.

Nations worldwide now recognize that the spread of misinformation and disinformation is an imminent cybersecurity threat that should be seriously addressed by law enforcement agencies [130]. However, the distinction between harmless misinformation and disinformation is often blurred. Moreover, rapid advances in deepfake technologies can make hoax news look legitimate and further exacerbate the situation. Rooting out rumor mongers and dispelling disinformation of increased scale and impact will be part of a timely and practical defense strategy that can offer intellectually deep insight to the science of networks.

What can cause the viral spreading of rumors or disinformation? One factor is semantics [52], [101]. For example, hoaxes and prank threats such as bomb threats are considered more serious but are likely short-lived as they can be quickly debunked. On the other hand, some rumors might swirl longer in social media (e.g., workplace rumors like layoffs or the inefficacy of certain pandemic measures) [52], [101], [126], [147]. Another factor is the principle of homophily in which humans have a tendency to associate with similar others, leading to cognitive bias typically known as the "echo chamber" effect [56], [62], [159]. The element of surprise can also affect rumor viscosity as people will tend to spread the information.

## 1.4   Overview of the Monograph

This monograph provides an overview of the surveillance of contagion sources in networks that find applications in digital contact tracing and rumor source detection to combat epidemics and infodemics, respectively. Given data that embeds both network topological structure (e.g., knowing who is connected to whom) and relational patterns on how a disease virus or rumor propagates, the objective is to answer the fundamental question: *how to unravel stochastic spreading processes in the network to find the initial outbreak source quickly, accurately and reliably with high confidence by exploiting the topological and statistical properties of networks.*

The contagion source detection problem was first studied in the seminal work [131]–[133]. Mathematically, the problem is: Given a snapshot observation of the contagion graph (showing how "infected" users are connected), who is the contagion source of the spreading? This problem is formulated as a maximum likelihood estimation problem over graphs and then solved exactly for special cases of degree-regular trees with infinite underlying graphs using a new form of network centrality called rumor centrality. Since then, it has spawned a huge literature on contagion source detection with various extensions such as random trees in [37], [53], to multiple sources in [69], [70], [105], [106], [108], [109], [112], [144], [167], to probabilistic sampling in [77], [120] and detection with multiple observations in [35], [153], belief propagation [40], general graphs with irregularity [154], [155], [165] and the implication of probabilistic spreading models and different graph topological features on solving the contagion source detection problem [4], [40], [84], [103], [114], [127], [137], [155], [168].

Different types of network centrality defined on vertices can resolve different types of network problems. Rumor centrality [132] is designed to solve the contagion source detection problem on infinite-size regular tree networks optimally (cf. Section 3.2). The vertex with the maximum rumor centrality is called the rumor center of a tree graph, and the rumor center was proved to be the same as the distance center [131], and the graph centroid of the tree [142], [163], [164]. Furthermore, it was shown in [72], [73], [85] that the graph centroid is almost surely central in the limit of the random growth process of infection on an underlying infinite regular graph. Aside from the distance centrality, another distance-based centrality, the Jordan center, was proposed to solve the contagion source detection in different scenarios [111], [112]. Dynamic influence due to stochastic spreading and opinion dynamics in online social networks can be characterized by the harmonic influence centrality in [1], [148] and the Shapley centrality in [21]. The protection centrality in [2] and relative centrality in [18] measure how important a set of vertices in a network is with respect to other vertices at the gatekeeper level and community level respectively. Querying this contagion source in a large graph with cost constraints and query complexity has been analyzed in [25], [93], [127]. Centrality measures related to the eigenvectors of the network topology are also important in the study of stochastic processes over large graphs [31], [57], [76], [124].

The bibliography included in this monograph seeks to encompass as many contributions as possible, aiming to provide a balanced overview of the key results and methodologies. Although the monograph may not be a perfect summary of the state-of-the-art (see related surveys in [71], [160] before 2018), it aims to serve as an imperfect yet informative summary, providing a rough illustration of the existing literature in the last 15 years and with relevance to the COVID-19 pandemic and infodemic. We survey the various work in this field with a particular focus on the intricate interplay between contagion source detection and mathematical tools like graph theory, probability theory, combinatorics, and algorithm design for statistical inference in the context of large networks.

This monograph provides a comprehensive overview of contagion source detection problem along with a problem-solving approach called "*network centrality as statistical inference*" that expounds a systematic approach to analyze inferential statistical problems in networks with applications to digital contact tracing and rumor source detection. The framework presented in this work establishes a connection between network centrality and the solution of challenging optimization problems that involve complex combinatorial constraints arising from the interaction of a stochastic process with the underlying network. By leveraging an appropriate network centrality, which induces a metric on each graph node, it is possible to obtain compact measures that quantify the importance of nodes and accurately capture the optimality of stochastic optimization. This framework also enables the utilization of graph algorithm techniques to address these problems effectively [59], [145].

We will discuss how the "*network centrality as statistical inference*" approach can be useful to the graph algorithm design that comes with performance guarantees, computational complexity, detection accuracy, and to address the "big data" regime in which the contagion graph can be very large (as is the case in the COVID-19 pandemic and infodemic). Designing scalable algorithms that uncover the contagion source accurately by leveraging network science and mathematical tools will be important to prevent future pandemics (e.g., 'Disease X' pandemic and infodemic) given the enhanced human connectivity on a global scale. We will conclude with open issues and several promising research directions to address the challenges of surveillance of spreading in networks.

# 2

## Preliminaries and Network Centrality

In this section, we first describe the contagion spreading model along with basic preliminaries on graph theory, combinatorics, probability theory and statistical inference that will be used in subsequent sections. We then introduce some measures of network centrality that are classical in the literature (e.g., [59], [76], [145]) and also new ones pertinent to the contagion source detection problem.

### 2.1 Contagion Spreading Model

In this section, we briefly describe the discrete spreading model of contagion considered in [131]–[133], [161], [162], [165], [166]. The model of interest is called the SI (Susceptible-Infectious) model as it is a simplified version of the classical SIR (Susceptible-Infectious-Recovered) model in epidemiology [83]. In the SI model, each node is either infected by the virus or susceptible to the virus. We assume that an epidemic spreads on a networked structure with a node set $V$. Initially, a single source node $v^{\star} \in V$ initiates the spreading of malicious content (e.g., disease, virus, or rumor) on a networked structure. Once a node is infected, it stays infected and can, in turn, infect its susceptible neighbors. That is, the malicious content can be spread from node $i$ to node $j$ if and only if there is a connection between them. Therefore, we can model the connection between two nodes in a network using a link or a graph edge.

Let $\mathbf{S}$ denote the set of susceptible vertices that have at least one infected neighbor, i.e., those vertices in $\mathbf{S}$ might be infected in the near future. In the real world, there are some people that are more likely to be infected by the virus, and some are more likely to spread the virus to others. We can assume that each person has two parameters say $R_{\mathsf{i}}$ and $R_{\mathsf{s}}$, which correspond to the rate of being infected and the rate of spreading the virus to others respectively. Let $\tau_{ij}$ be the spreading time from node $i$ to node $j$, which are random variables that are independently and exponentially distributed with parameter $\lambda$ (without loss of generality, let $\lambda = 1$). Because of the memoryless property of the exponential distribution, we assume that each newly infected vertex $v$ is randomly chosen from $\mathbf{S}$ with the probability that

$$P(v \text{ is infected}) \propto R_{\mathsf{i}}^v \cdot \sum_u R_{\mathsf{s}}^u,$$

where each $u$ is an infected neighbor of $v$, $R_{\mathsf{i}}^v$ denote the infected rate of $v$ and $R_{\mathsf{s}}^u$ denote the spreading rate of $u$. Hence, the probability of a vertex $v_a$ being infected in the next time period can be computed by

$$P(v_a \text{ is infected}) = \frac{R_{\mathsf{i}}^{v_a} \cdot \sum\limits_{u_a} R_{\mathsf{s}}^{u_a}}{\sum\limits_{v \in \mathbf{S}} [R_{\mathsf{i}}^v \cdot \sum\limits_u R_{\mathsf{s}}^u]}, \tag{2.1}$$

where $u_a$ and $u$ represents each infected neighbor of $v_a$ and $v$ respectively.

Let $G_n \subset G$ denote the infected subgraph with $n$ nodes, which models snapshot observation of the spreading over $G$ when there are $n$ infected nodes. Two parameters $R_{\mathsf{i}}$ and $R_{\mathsf{s}}$ are unknown for all nodes in $G_n$ since we assume that the graph topology of $G_n$ is the only given information. In this monograph, we assume that $R_{\mathsf{i}} = R_{\mathsf{s}} = 1$ for all nodes, and we can simplify the infection probability (2.1) for vertex $v_a$ as

$$P(v_a \text{ is infected}) = \frac{\sum\limits_{u_a} 1}{\sum\limits_{v \in \mathbf{S}} [\sum\limits_u 1]}, \tag{2.2}$$

which implies that $v_a$ is more likely to be infected (or told the rumor) if it has more "infected" neighborhoods. Note that when $G$ is a tree network, this spreading model is equivalent to the one considered in [131]–[133]. We can analyze more general cases, such as graphs with cycles, using the definition in (2.1).

## 2.2 Mathematical Preliminaries and Statistical Inference

In this section, we introduce some mathematical preliminaries and tools in graph theory and algorithms, probability theory, analytic combinatorics and message passing algorithms for statistical inference.

### 2.2.1 Preliminaries on Graph Theory

A graph $G$ consists of a vertex set $V(G) = \{v_1, v_2, \ldots, v_n\}$, and an edge set $E(G) = \{(v_i, v_j) | v_i, v_j \in V(G)\}$. A pair $(v_i, v_j)$ is called an edge in $G$; note that $v_i$ and $v_j$ are not necessarily to be distinct. For a directed graph, each $(v_i, v_j)$ is an ordered pair, and for an undirected graph, we have $(v_i, v_j) = (v_j, v_i)$. When $u$ and $v$ are the endpoints of an edge, they are *adjacent* and are *neighbors*. We define $N(v) = \{u \in V(G) | (u, v) \in E(G)\}$ to be the set of neighbors of $v$. A *subgraph* of a graph $G$ is a graph $H$ with $V(H) \subseteq V(G)$ and $E(H) \subseteq E(G)$. We say $G' = (V(G'), E(G'))$ is an *induced subgraph* of $G$ if $V(G') \in V(G)$ and for all $u, v \in V(G')$, we have $(u, v) \in E(G')$ whenever $(u, v) \in E(G)$.

A *loop* in a graph is a single edge whose endpoints are equal. *Multiple edges* are edges having the same pair of endpoints. A graph without a loop or multiple edges is called *simple graph*. A *path* is a simple graph whose vertices can be ordered so that two vertices are adjacent if and only if they are consecutive in the list. A $u - v$ *path* is a path that starts with $u$ and ends with $v$. The *distance* from $u$ to $v$, denoted as $d(u, v)$, is the number of edges of the shortest $u - v$ path. A *cycle* is a graph with an equal number of vertices and edges whose vertices can be placed around a circle so that two vertices are adjacent if and only if they appear consecutively along the circle.

A graph $G$ is *connected* if for any two vertices $u$, $v$ there is a $u - v$ path in $G$. A *tree* is a connected simple graph without a cycle. A *subtree* of a tree $T$ is a tree $T'$ with $V(T) \subseteq V(T')$ and $E(T) \subseteq E(T')$. The *neighborhood* of a vertex $v$ is a set of all neighbors of v denoted as $N(v)$.

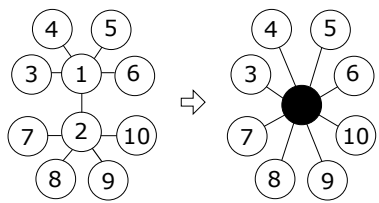

**Figure 2.1:** The graph on the left is the graph before the vertex contraction, and the graph on the right is the result of contracting $v_1$ and $v_2$.

The *degree* of a vertex $v$ is the number of its neighbors denoted as $d(v)$. A graph $G$ is said to be $d$-regular if all vertices in $G$ are of degree $d$. All vertices with degree 1 in a tree are called *leaves*.

In this monograph, a $d$-regular tree is a tree where all its vertices are of degree $d$. Since every vertex in a $d$-regular tree has $d$ children, there is no leaf vertex in a $d$-regular tree, and the size of this tree is infinite. A *rooted tree $T$* is a tree with one vertex $v_r$ chosen as a *root*. For any vertex $v$ in a rooted tree with root $r$, the *parent* of $v$ is its neighbor on $v_r - v$ path; the *children* are its other neighbors and $child(v)$ denote the set of all children of $v$. An *ancestor* of a vertex is any other node on the path from the node to the root. A vertex $u$ is a *descendant* of a vertex $v$ if and only if $v$ is an ancestor of $u$. Given a tree $T$ with root $v_r$, a *branch $T_v^r$* is the subtree with $v$ as a root containing all its descendants in $T_r$. For convenience, let $t_v^r$ denote the size of $T_v^r$. So for a $d$-regular rooted tree with root $v_r$ and size $n$, we have $n - 1 = \sum_{i=1}^{d} t_{v_i}^{v_r}$, where $v_i$ are children of $v_r$ for $i = 1, 2, \dots d$.

The *vertex contraction* of a pair of vertices $v_i$ and $v_j$ is a graph operation whereby $v_i$ and $v_j$ are replaced by a single vertex $v$ that becomes adjacent to the union of the vertices originally adjacent to $v_i$ and $v_j$ [158]. Conceptually, $v_i$ and $v_j$ are lumped together into a single vertex, shown in Figure 2.1.

### 2.2.2   Maximum Likelihood Estimation

In statistical inference, we often have a collection of observed data $X = x$ and the probability density function $f(X = x|\theta)$ of the data. However, the parameter $\theta$ is an unknown value. Hence, how to estimate the value of $\theta$ is the main concern in this section. There is more than one approach to estimating parameter(s) in a density function (e.g. method of moments, method of maximum likelihood, Bayesian approach). We will mainly focus on the method of maximum likelihood [49].

Given the observed data $X = x'$, if we see $f(X = x'|\theta)$ as a function of $\theta$, then we can define a new function $L(\theta|X = x)$ which is called the *likelihood function* [49]. Our goal is to find the most "suitable" parameter $\theta = \theta^\star$ that leads to the outcome $X = x'$, i.e.,

$$L(\theta^\star|X = x') = \max_\theta \{L(\theta|X = x')\}.$$

We call $\theta^\star$ the *maximum likelihood estimator* of $\theta$.

In the following, we give two examples of estimating unknown using the method of maximum likelihood. The first example is coin tossing. We toss a coin 10 times, with each outcome being either "H" or "T". Let the probability of getting "H" be $p$ and $X$ denote the total times that the outcome is "H", then $X$ is a binomial random variable with parameters $(n, p)$ where $n$ is the number of trials. We have the probability mass function of $X$ is $P(X = k) = \binom{n}{k} p^k (1-p)^{n-k}$, where $k = 1, 2, \ldots n$. The problem of interest is to estimate the unknown parameter $p$ based on the observed outcome. Suppose we get six H's and four T's. We can express the probability of getting six H's and four T's as

$$P(X = 6|p) = \binom{10}{6} p^6 (1-p)^4.$$

To best fit the experimental results, we shall maximize the likelihood $P(X = 6|p)$ over $p$. Hence, the function $\binom{10}{6} p^6 (1-p)^4$ reaches its maximum when $p = 0.6$. That is, 0.6 is the best estimation for $p$ based on the observed outcome.

Consider the general case that the coin has been tossed $n$ times, and there are $k$ H's, and then we have

$$P(X = k|p) = \binom{n}{k} p^k (1-p)^{n-k}.$$

The likelihood function is $L(X = k|p) = \binom{n}{k} p^k (1-p)^{n-k}$. Since each trial outcome in the random sequence of tosses is independent, we have $p^\star = k/n$, which coincides with the intuition, i.e., $p^\star$ is the ratio between the number of H and $n$. Hence, the ratio $k/n$ is the maximum likelihood estimator of $p$. In this example, we estimate an unknown parameter of a given distribution.

Let us consider another example of drawing two balls from a bin. The bin has three balls in it: a red one, a white one, and the third one whose color is unknown. We randomly draw two balls from the bin and let a random variable $X$ denote the number of red balls among the drawn balls. Our goal is to infer the color $c \in \{red, white\}$ of the third ball based on the observed data. What is the best estimate of $c$ when $X = 0$? If $c = red$, then we have $P(X = 0) = 0$. If $c = white$, then we have $P(X = 0) = 1/3$. We can conclude that $c = white$ is the best estimation from our observation since it maximizes the likelihood of $X = 0$. To find the most probable color of $c$, we compute the likelihood of the observed sample for all possible $c$. In this example, $c$ is not a parameter of a probability distribution. However, we can still compute its best estimate using the method of maximum likelihood [49].

In Section 3, we consider the contagion source detection problem, which was first formulated as a maximum likelihood estimation problem in [132]. The observed data of this problem is a connected graph $G_n$ which consists of all infected individuals, and the unknown parameter is any vertex $v \in G_n$ that could be the contagion source. In particular, the contagion source detection problem considers the likelihood of observing $G_n$ by assuming that every single vertex in $G_n$ is a probable source. The same formulation was also considered in [88] for a Bayesian inference problem where the observed data can be modeled as a bipartite graph [88]. Unlike the above examples, there is no closed-form formula to describe the likelihood of every trial outcome (i.e., infection event) in the contagion source detection problem. Both the problem formulation and its solution using the topological structure of graphs and message-passing algorithms in [50], [87], [113] present the contagion source detection as a challenging yet interesting maximum likelihood estimation problem over massive graphs.

## 2.3 Simple Line Network Illustration

We consider an example where $G_n$ is a line graph with $n$ labeled vertices as shown in Figure 2.2. To construct $G_n$, we can start from any single vertex $i$ $(1 \leq i \leq n)$ and connect a new vertex to the existing vertices iteratively. How many ways are there to construct $G_n$ by the above-mentioned method? For example, if we start with the vertex labeled $n$, then there is only one way to construct the line graph, i.e., we add the remaining vertices to the graph in the following order $n - 1$, $n - 2$, ..., 2, 1. Let $f_i$ denote the number of ways to construct the line graph starting with vertex $i$, then we have $f_i = \binom{n-1}{n-i}$ for $i = 1, 2, \ldots, n$. Note that $f_i$ is the binomial coefficient. Let $f(x)$ be the ordinary generating function of $f_i$, then we have

$$f(x) = \binom{n-1}{n-1}x^1 + \binom{n-1}{n-2}x^2 + \binom{n-1}{n-3}x^3 + \ldots + \binom{n-1}{0}x^{n-1},$$

which implies $f(x) = x(1 + x)^{n-1}$. Hence, the total number of ways to construct the line graph with $n$ vertices is

$$\binom{n-1}{n-1} + \binom{n-1}{n-2} + \binom{n-1}{n-3} + \ldots + \binom{n-1}{0},$$

which is equivalent to $f(1) = 1 \cdot (1+1)^{n-1} = 2^{n-1}$. Note that $f_i$ reaches its maximum at $i = \lceil \frac{n-1}{2} \rceil$, i.e., when we start from the *most central vertex*, the number of ways to construct the line reaches its maximum value. This most central vertex in the line network can be intuitively interpreted as the most probable spreading source. In the next section, we will provide a more rigorous analysis to further substantiate this observation.

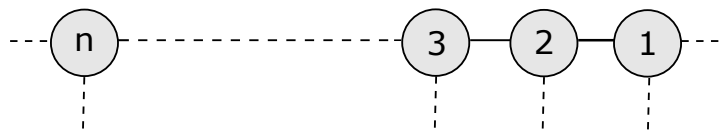

**Figure 2.2:** An example of $G_n$ being a line network graph where $n$ denotes the size of the graph.

Without loss of generality, let us assume that $n = 2t + 1$ where $t$ is a positive integer. Let $f_{max} = \max\limits_{1 \leq i \leq 2t+1} f_i$, then another interesting problem is to compute the ratio

$$\frac{f_{max}}{\sum\limits_{i} f_i} = \frac{\binom{2t}{t}}{2^{2t}}, \tag{2.3}$$

which is equivalent to the probability that the most central vertex as the most probable spreading source is indeed the correct guess in this line graph. As shown in the next section, this quantifies the performance of any estimator in the contagion source detection problem. In fact, for this particular line graph special case, we can obtain more insights into the performance of this estimator as the graph grows larger by leveraging some bounds to the central binomial coefficient [78]. For example, we have

$$\frac{4^t}{\sqrt{4t}} \leq \binom{2t}{t} \leq \frac{4^t}{\sqrt{3t+1}}. \tag{2.4}$$

Applying (2.4) to (2.3), we obtain the following result

$$\frac{1}{\sqrt{4t}} \leq \frac{\binom{2t}{t}}{2^{2t}} \leq \frac{1}{\sqrt{3t+1}}, \tag{2.5}$$

which establishes the fact that using the most central vertex as the estimator, the probability of correct detection is of the order inversely proportional to the square root of the number of vertices in the line graph. Moreover, this detection probability diminishes to zero as the number of vertices in the graph becomes asymptotically large. Interestingly, another way to quantify this detection performance is a probabilistic proof given in [132] by viewing spreading in a line graph as two independent Poisson processes having the same rate beginning at the source and spreading in opposite directions. The estimator correctly detects the source with probability one if these two Poisson processes have exactly the same number of arrivals and with probability half if one of the Poisson processes is one less than the other [132].

### 2.3.1  Pólya Urn Probability Model

In this section, we introduce a classical probability model called Pólya Urn's model that captures a random sampling process where items are repeatedly added to and drawn from an urn and the probability of each item being sampled depends on the current composition of the urn.[1] Initially, there are R red balls and B blue balls in an urn. A ball is uniformly, with probability $\frac{1}{R+B}$, drawn from the urn. The ball is then returned to the urn with m additional balls of the same color. Suppose the first drawn ball is a red ball; then there are $R + m$ red balls and B blue balls in the urn after the first draw. The problem of interest is the number of red balls and blue balls after $n$ draws. In particular, the ratio between the number of red balls (or black balls) and the total number of balls in the urn.

We say a random process $\{M_n, n \geq 1\}$ is a martingale if $\mathbf{E}[|M_n|] < \infty$ and $\mathbf{E}[M_{n+1}|M_1, \ldots, M_n] = M_n$. For simplicity, assume that initially there are one red ball and one black ball in the urn and m = 1. Let $R_i$ and $B_i$ denote the number of red balls and black balls after the $i$th trial respectively, i.e., initially, we have $R_0 = 1$ and $B_0 = 1$. Assume that $R_n = k + 1$, then we have $B_n = n - k + 1$. Consider the expectation of ratio $\frac{R_{n+1}}{R_{n+1} + B_{n+1}}$ after $n + 1$ trial given the previous results, then we have

$$\mathbf{E}\left[\frac{R_{n+1}}{R_{n+1} + B_{n+1}} | R_n = k + 1, B_n = n - k + 1\right]$$
$$= \frac{k + 2}{n + 3} \cdot \frac{k + 1}{n + 2} + \frac{k + 1}{n + 3} \cdot \frac{n - k + 1}{n + 2}$$
$$= \frac{k + 1}{n + 2} = \frac{R_n}{R_n + B_n},$$

which implies that the ratio between the number of red balls (or black balls) and the total number of balls in the urn satisfies the property of a martingale.

**Theorem 2.1.** Let $M_i$ be a martingale for $i \geq 1$. Then if $\mathbf{E}[M_i]$ is finite, then there is a random variable $M_\infty$ such that

$$\lim_{i \to \infty} M_i = M_\infty.$$

We can infer from Theorem 2.1 that the ratio between the number of red balls (or black balls) and the total number of balls will finally converge to some random variables. Note that Pólya Urn's model is similar to a disease-spreading process on a tree network. Assume that the spreading process starts from the tree root, and in each time period, we randomly select a tree branch (a color) to spread the disease. Then, the number of susceptible individuals in the chosen branch will increase by m, which makes this branch (color) more probable to be selected again in the future. For more details about the application of Pólya Urn's model to contagion source detection problems, we refer the readers to Section 5.2.

---

[1] Pólya Urn's model, which is introduced by George Pólya, is able to capture the "rich get richer" phenomenon in the real world.

### 2.3.2 Generating Function

In this section, we introduce the idea of using analytical approaches to analyze the long-term behaviors of a combinatorial structure. Generating functions are one of the ideal tools to deal with combinatorial structures and their application to counting the number of increasing trees which is related to the number of spreading orders (cf. Section 3.1) starting from the spreading source (the tree root) on a given tree structure.

Suppose we are given a problem whose answer is an integer sequence $a_1$, $a_2$, ..., $a_n$. For example, how many different two-element subsets are there in an $n$-element set $S = \{1, 2, \ldots, n\}$? To give a formula of $a_n = \frac{n(n-1)}{2}$ may be the best way to present the solution. However, suppose we consider the problem of finding the $n$th Fibonacci number given $a_0 = 0$ and $a_1 = 1$. Then, the formula of $a_n$ may not be as simple as the first question we asked. Instead of providing the formula of $a_n$, generating functions gives us a sum of a power series whose coefficients are the sequence we are interested in. For instance, the $n$th Fibonacci number is the coefficient of $x^n$ in the power series expansion about $x = 0$ of the function $\frac{x}{1-x-x^2}$.

In the following, we use a simple example to illustrate how to find the generating function of the Fibonacci number by a basic technique in the method of generating functions.

**Example 2.1.** Suppose $F_{n+1} = F_n + F_{n-1}$ is the $(n+1)$th Fibonacci number where $F_0 = 0$ and $F_1 = 1$.

**Definition 2.1.** A function $f(x)$ is an ordinary power series generating function (OGF) of a sequence $\{f_n\}_{n \geq 0}$ if

$$f(x) = \sum_{n \geq 0} f_n x^n.$$

Let $F(x)$ be the OGF of the Fibonacci sequence. Then by the definition of $F_0$ and $F_1$, we have

$$F(x) = 0 + x + F_2 x^2 + F_3 x^3 + \ldots,$$

which implies

$$\frac{F(x) - x}{x} = F_2 x + F_3 x^2 + + F_4 x^3 \ldots. \qquad (2.6)$$

Note that

$$
\begin{aligned}
F(x) + x F(x) &= F_0 x^0 + (F_0 + F_1)x^1 + (F_1 + F_2)x^2 + (F_2 + F_3)x^3 + \ldots \\
&= F_0 x^0 + F_2 x^1 + F_3 x^2 + F_4 x^3 + \ldots \\
&= \text{the right hand side of (2.6)}.
\end{aligned}
$$

Hence, we have

$$\frac{F(x) - x}{x} = F(x) + x F(x),$$

which implies

$$F(x) = \frac{x}{1 - x - x^2}.$$

Finally, $\frac{x}{1-x-x^2}$ can be written as the difference of two geometric series which is shown as follows,

$$
\begin{aligned}
\frac{x}{1 - x - x^2} &= \frac{1}{\sqrt{5}} \left( \sum_{n \geq 0} a_1^n x^n - \sum_{n \geq 0} a_2^n x^n \right) \\
&= \frac{1}{\sqrt{5}} \sum_{n \geq 0} (a_1^n - a_2^n) x^n,
\end{aligned}
$$

where $a_1$ and $a_2$ are the roots of the quadratic function $1 - x - x^2$ and $a_1 > a_2$. We can conclude that $F_n = \frac{1}{\sqrt{5}}(a_1^n - a_2^n)$.

**Definition 2.2.** Let $f(x)$ be a series in powers of $x$. Then we denote the coefficient of $x^n$ in $f(x)$ by $[x^n]f(x)$.

In the previous example, $[x^n]F(x) = \frac{1}{\sqrt{5}}(a_1^n - a_2^n)$. Next, we introduce a combinatorial structure which is called *labeled tree*, and a subclass of the labeled tree, which is called *increasing tree*. A rumor spread in a given network is similar to the structure of an increasing tree, where the rumor source is the root of the increasing tree.

**Definition 2.3.** A labeled tree with $n$ vertices is a tree structure where each vertex in the tree is assigned a unique number from 1 to $n$, i.e., there is a bijection function from the vertex set of the tree to the integer set $\{1, 2, \ldots n\}$.

**Definition 2.4.** An increasing tree is a labeled tree such that the sequence of labels along any branch starting at the root is increasing.

The problem of interest is the number of different $n$-vertex increasing trees on a fixed underlying topology. For example, how many different $n$-vertex increasing trees are there if the underlying topology is a binary tree? It is well known that a one-to-one relation exists between an $n$-vertex binary tree and a permutation of $n$ objects. Hence, there are $n!$ different $n$-vertex increasing trees when the underlying topology is a binary tree. In the following, we consider the general case that the underlying topology is a $d$-ary tree.

**Definition 2.5.** A function $f(x)$ is an exponential generating function (EGF) of a sequence $\{f_n\}_{n \geq 0}$ if

$$f(x) = \sum_{n \geq 0} f_n \frac{x^n}{n!}.$$

**Definition 2.6.** Let $\{s_r\}_{r \geq 0}$ be a non-negative sequence. Let $s_r$ be defined as the number of different rooted trees of size $s_r + 1$ where $v$ is the root and the outdegree of $v$ is $r$.

For example, suppose $v$ is on a binary tree, then we have $s_0 = 1$, $s_1 = 2$, $s_2 = 2$ and $s_i = 0$ for all $i > 2$. If $v$ is on a strictly binary tree (the outdegree is either 0 or 2), then we have $s_0 = 1$, $s_1 = 0$, $s_2 = 2$ and $s_i = 0$ for all $i > 2$.

**Definition 2.7.** The degree function $\phi(w)$ of a variety of trees associated with $\{s_r\}_{r \geq 0}$ is defined by

$$\phi(w) = \sum_{r \geq 0} s_r w^r.$$

For example, the degree function $\phi(w)$ associated with the binary tree is $1 + 2w + w^2$, and $1 + w^2$ for the strictly binary tree. Let $T_n$ denote the number of $n$ vertices increasing trees on a $d$-ary tree. Then the following theorem states that the exponential generating function of $T_n$ can be expressed implicitly by leveraging the degree function of the $d$-ary tree.

**Theorem 2.2** ([9]). *The exponential generating function $T(z)$ of a variety of trees defined by the degree function $\phi(w)$ is given implicitly by*

$$\int_0^{T(z)} \frac{dw}{\phi(w)} = z.$$

In the following, we use the binary tree example to illustrate how to find the number of $n$ vertices increasing trees on a binary tree structure.

**Example 2.2.** We have $\phi(w) = 1 + 2w + w^2$ when the underlying structure is a binary tree. Let $T_n$ and $T(z)$ be defined as above, then we have

$$\int_0^{T(z)} \frac{dw}{1 + 2w + w^2} = z,$$

which implies

$$T(z) = T_0 \frac{z^0}{0!} + T_1 \frac{z^1}{1!} + T_2 \frac{z^2}{2!} + T_3 \frac{z^3}{3!} \cdots$$
$$= \frac{z}{1 - z}$$
$$= z + z^2 + z^3 + z^4 + \ldots$$

Hence, we have $T_0 = 0$ and $T_n = n!$ for $n \geq 1$, which is exactly the number of different $n$-object permutations. In Section 5.1, we can apply Theorem 2.2 to compute the exponential generating function of the number of different increasing trees on a given underlying network structure.

Lastly, we introduce a key theorem in [48], which plays an important role in analyzing the asymptotic behavior of the number of increasing trees.

**Theorem 2.3** (Theorem VI.1 in [48])**.** For $\alpha \in \mathbb{C} \setminus \mathbb{Z}_{\leq 0}$ set

$$f(z) := (1 - z)^{-\alpha}.$$

Then, as $n \to \infty$,

$$[z^n]f(z) \sim \frac{n^{\alpha-1}}{\Gamma(\alpha)} \left( 1 + \sum_{k=1}^{\infty} \frac{e_k(\alpha)}{n^k} \right),$$

where $e_k(\alpha)$ is a polynomial of degree $2k$.

### 2.3.3 Graph Algorithms

In this section, we introduce basic graph algorithms such as *Breadth-First Search*, *Depth-First Search*, and *Message Passing Algorithm*. A graph search algorithm provides a systematical way to traverse all the vertices on the given graph $G$ by leveraging the data structure in which the graph is stored. In the two graph-searching algorithms, we use three colors, say, white, gray, and black, to indicate the states of a vertex. A vertex is white if it is "undiscovered", and gray if it is "discovered" but some of its neighbors are still undiscovered. Finally, a black vertex is a discovered vertex, as well as all its neighbors. In the beginning, all vertices in $G$ are colored white and will be colored black after we apply a graph-searching algorithm on $G$ if $G$ is connected.

A graph $G = (V, E)$ can either be stored as an *adjacency matrix* or a collection of *adjacency lists*.

**Definition 2.8.** Given a graph $G = (V, E)$, the adjacency matrix $A(G) = \{a_{i,j}\}_{N \times N}$ of $G$ is called the adjacency matrix of $G$ if $a_{i,j}$ is defined as follows:

$$a_{i,j} = \begin{cases} 1 & \text{if } (i, j) \in E, \\ 0 & \text{else.} \end{cases}$$

For example, if we consider the induced subgraph $G$ with $V = \{v_1, v_2, v_3\}$ of the graph illustrated in Figure 2.3.

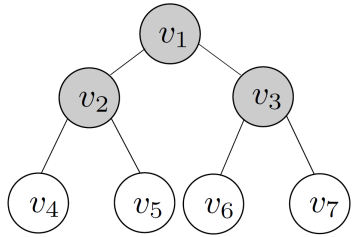

**Figure 2.3:** An example to illustrate how to store a graph as a $0 - 1$ matrix.

$$A(G) = \begin{bmatrix} 0 & 1 & 1 \\ 1 & 0 & 0 \\ 1 & 0 & 0 \end{bmatrix}$$

The other way is that for each vertex $u \in V$, we construct an adjacency list $adj(u)$ for $u$ such that $adj(u) = \{v | v \in N(u)\}$. If $G$ is a directed graph, then $adj(u)$ only contains the edge $(u, j) \in E$ where $j \in V$, i.e., the outdegree of $u$. In the following, we only consider the case where $G$ is an undirected graph. As a remark, these two searching algorithms can be applied to directed graphs as well.

**Breadth-first Search**

Assume that a connected graph $G = (V, E)$ is stored as a collection of adjacency lists. To start a breadth-first search traversal on $G$, we first select a starting vertex $r \in G$ and add $r$ into an empty list $L_{visited}$; therefore, $r$ becomes a gray vertex. Next, we add all vertices in $adj(r)$ into $L_{visited}$. Now, $r$ becomes a black vertex, and all neighbors of $r$ are colored in gray. For each gray vertex in $L_{visited}$, we recursively add their undiscovered neighbors into $L_{visited}$ according to their order when they are added into $L_{visited}$, i.e., the list $L_{visited}$ is a first-in-first-out queue. We can construct a BFS tree $G'$ of $G$ starting from the root $r$ by initially setting $V(G') = V$ and $E(G') = \emptyset$. Each time a vertex $v$ is discovered by its neighbor $u$ through an edge $(u, v) \in E$, we add the edge $(u, v)$ to $E(G')$. Then, $G'$ becomes a BFS tree of $G$ with root $r$ when all vertices are visited and the queue is empty. We can observe that for each vertex $v$ in $G'$, the path from $r$ to $v$ in $G'$ corresponds to one of the shortest paths from $r$ to $v$ in the original graph $G$. In the breadth-first search, the vertices which are closer to $r$ will be visited earlier than those farther vertices.

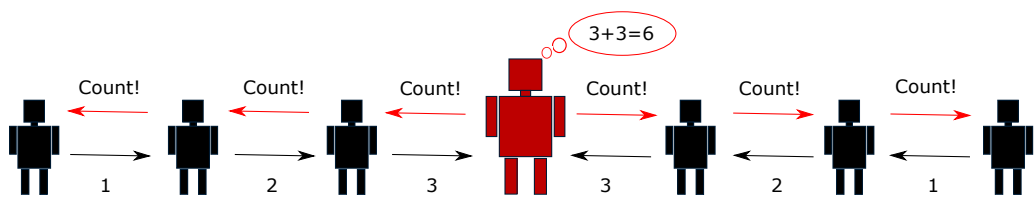

**Figure 2.4:** An example of counting the number of students in a line by passing a message between any two neighboring students.

### Depth-first Search

In the depth-first search algorithm, we again assume that $G$ is stored as a collection of adjacency lists and the search starts with a vertex $r$. Unlike the breadth-first search, the depth-first search tends to search as deep as possible in $G$. Hence, $L_{visited}$ is a first-in-last-out data structure.

For example, let $v_{new}$ denote the most recently discovered vertex. Then, initially we have $v_{new} = r$ and $L_{visited} = \{r\}$. Next, suppose $v \in adj(r)$ is the first neighbor "explored" along the edge $(r, v)$, then we have $v_{new} = v$ and $v$ is added into $L_{visited}$. According to the strategy of searching as deep as possible, we update $v_{new}$ each time a new vertex is added into $L_{visited}$. When all the neighbors of $v_{new}$ have been added into $L_{visited}$, we backtrack to the second newest vertex in $L_{visited}$ and recursively add vertices into $L_{visited}$ until all vertices are discovered. For more details about the vertex traversal algorithms, please refer to [27].

### Message Passing

Given a line network graph as in Figure 2.2, assume that each graph vertex is a single computing processor that can only exchange information with its neighboring processors. We consider a basic distributed computing problem in the following: "How many vertices are there in this line graph?" We can solve this problem using the *message passing* algorithm [113]. Following the example of a line of soldiers counting themselves in [113], imagine that one of the vertices is a teacher, and the rest of the vertices are all students in a line, illustrated in Figure 2.4. For any two neighboring students $u$ and $v$ on the line, if $u$ is closer to the teacher than $v$, then we say $u$ is in the front of $v$, and $v$ is at the back of $v$.

To find out how many students there are in the line, we can set the following rules for messages to be exchanged and passed between vertices in the line:

- The message starts with the teacher and ends with the teacher. Initially, the message sent from the teacher is "Count!".

- For each student, if the received message is "Count!" from the person in the front, then send the same message "Count!" to the person in the back. If there is no person at the back, then send a number "1" to the person in the front.

- For each student, if the received message is a number "$i$" from the person at the back, then send the number "$i + 1$" to the person in the front.

Lastly, the number of students in the line graph is the summation of all messages received from all neighboring students of the teacher. Note that the computation of the message-passing algorithm is performed locally, however, the solution is a global one when the algorithm converges. We refer the readers to [113] for an in-depth analysis of the message-passing algorithm and its convergence in general tree graphs.

### 2.3.4 Random Walk on Graphs

A random walk on a graph is a stochastic process where a "walker" moves from one node to another in a graph based on a set of rules [29], [104]. At each step, the walker moves to a neighboring node with a certain probability distribution, which may be uniform or non-uniform. The walker's position at any given time is a random variable, and the movement pattern can be analyzed to understand various properties of the graph, such as its connectivity, centrality, and other structural features [29], [104].

Graph random walks and Markov Chains are closely related [104]. Graph random walk can be viewed as a time-reversible finite Markov chain. On the other hand, every Markov Chain can be defined as a random walk on a weighted and directed graph [104]. In the following, we give an example of a uniform random walk on a given graph. Given a connected undirected graph $G(V, E)$, we begin the process by placing a random walker on an arbitrary node $v_i$, where $v_i$ is from some initial distribution and assume that at each stage, the walker must move from the current position to an adjacent node $v_j \in N(v_i)$. We denote the probability that the walker moves from $v_i$ to $v_j$ by $p_{i,j}$ and define $p_{i,j}$ as follows:

$$p_{i,j} = \begin{cases} \frac{1}{d(v_i)} & v_i \neq v_j \text{ and } (v_i, v_j) \in E \\ \\ 0 & v_i = v_j. \end{cases} \quad (2.7)$$

For example, if the random walker is at $v_6$ in Figure 2.5, then in the next stage, the walker will move to $v_7$ or $v_4$ with the same probability $1/2$. Regardless of the initial distribution of $v_i$, if $G$ is not a bipartite graph, then this uniform random walk will converge to a stationary distribution $\pi(v_k) = \frac{d(v_k)}{2|E|}$ eventually. Let $\tilde{\pi}$ be a distribution over $V$ and $P$ a transition matrix on $G$. We say $\tilde{\pi}$ and $P$ satisfy the *detailed balanced condition* if $\tilde{\pi}(v_i) \cdot p_{i,j} = \tilde{\pi}(v_j) \cdot p_{j,i}$ for all $v_i, v_j \in V$. Note that the stationary distribution $\pi(v)$ and the transition probability in (2.7) satisfy the detailed balanced condition [104].

**Theorem 2.4.** Let $P$ be the transition matrix for a random walk on a finite connected graph $G(V, E)$ and $\pi$ be a distribution over $V$. If $P$ and $\pi$ satisfy the detailed balanced condition, then $\pi$ is the stationary distribution of this Markov Chain.

The detailed balance condition can be applied to design the transition probability matrix in Markov Chain-based methods [104]. For example, the Ruelle-Bowen random walk can be used to construct a Markov Chain to maximize the entropy rate of the walk on an unweighted graph that is related to the Pagerank centrality [31].

## 2.4 Network Centrality

Network centrality can be seen as a function that maps a given graph vertex to a real number, i.e., we can assign a numerical score to each vertex based on the topological properties of the graph [59], [145]. The score of a vertex can be interpreted as the "influence" of this vertex in the whole graph; hence, we can rank all the vertices based on their scores to find the most influential one. In fact, Google's search engine is driven by such an idea through its *PageRank* algorithm that uses the *eigenvector centrality* to rank webpages on the Internet and also finds applications in big data analytics [26], [57], [124], [136], [145]. There are many other definitions of network centrality depending on the structural object of interest in the graph. We introduce some other commonly-known notions of network centrality that will be encountered in the next few sections. As a remark, the notion of network centrality only applies to the vertices in the same connected component of a graph.

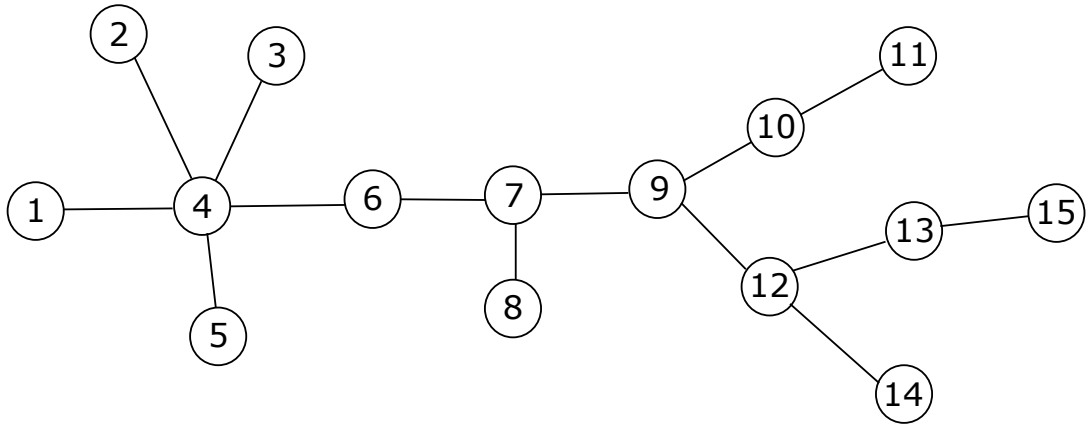

**Figure 2.5:** An example graph $G = (V, E)$, where $V = \{v_1, v_2, \ldots, v_{15}\}$, to illustrate different definition of network centrality centrality. For example, $v_4$ is the degree center, $v_7$ is both the distance center and centroid, and $v_9$ is the betweenness center.

**Degree Centrality**

Let $G = (V, E)$ be a simple undirected graph. The degree centrality of a vertex $v \in V$ is simply defined as the degree of $v$. Hence, for $v \in V$ we denote the degree centrality of $v$ as its degree $d(v)$. We call the vertex with the maximum degree the *degree center* of $G$.

For example, we have $d(v_4) = 5$, $d(v_1) = 1$, $d(v_6) = 2$ in Figure 2.5 and $v_4$ is the degree center. For a $d$-regular graph, the degree centrality for every non-leaf vertex is obviously the same, i.e., the degree centrality is not informative enough to capture the influence of any particular non-leaf vertex in the graph. In directed graphs, we can also define the degree centrality based on the in-degrees and out-degrees of the nodes.

**Distance Centrality**

To better understand the structure of the whole graph, we may use distance centrality, whose definition is the summation of the number of hops from a particular vertex to every other vertex (i.e., geodesic distance) in the graph

$$DisC(v, G) = \sum_{u \in V} d(v, u) \tag{2.8}$$

where $d(v, u)$ denotes the shortest path distance from $v$ to $u$. The vertex with the minimum distance centrality is the *distance center* of $G$. For example, we have $DisC(v_7, G) = 34$, $DisC(v_6, G) = 37$, and $DisC(v_9, G) = 35$ and $v_7$ is the distance center in Figure 2.5. Note that, for any two adjacent vertices $u$ and $v$ in a tree graph, if $DisC(u, G)$ is known then we can compute $DisC(v, G)$ in constant time. We have the following relationship,

$$DisC(v, G) = DisC(u, G) - t_v^u + t_u^v, \tag{2.9}$$

where $t_v^u$ and $t_u^v$ are defined in Section 2.2. Observe that the distance centrality along the path from a leaf to the distance center is a monotone decreasing sequence which shows that the distance centrality can capture the geodesic structural property of the graph. As a remark, the *closeness centrality* of a vertex can be defined as the reciprocal of its distance centrality.

**Graph Centroid and Branch Weight on Tree Graphs**

The branch weight centrality is a special kind of network centrality that is applicable only to a tree graph. Suppose $G$ is a tree graph, and we define the branch weight $\mathsf{weight}(v, G)$ of a $v \in G$ by

$$\mathsf{weight}(v, G) = \max_{c \in \mathsf{child}(v)} t_c^v. \tag{2.10}$$

We call the vertex with the minimum branch weight the *centroid*. As an example, using the line network in Figure 2.2, the centroid is the vertex in the middle of the line graph (with a branch weight $\mathsf{weight}(v_{(n+1)/2}, G) = (n-1)/2$ when $n$ is odd. Otherwise, when $n$ is even, the centroid is either $v_{n/2}$ or $v_{n/2+1}$ (with the same branch weight $\mathsf{weight}(v_{n/2}, G) = n/2$). Using Figure 2.5 as another example, we have $\mathsf{weight}(v_6, G) = 9$, $\mathsf{weight}(v_7, G) = 7$ and $\mathsf{weight}(v_4, G) = 10$, and $v_7$ is thus the centroid.

**Betweenness Centrality**

Consider the problem that for each vertex $v$ in a graph $G(V, E)$, we want to find the number of times that $v$ acts as a "bridge" to all shortest $s - t$ paths, where $s, t \in V$. In other words, how many shortest paths from $s$ to $t$ will have to go through $v$? This structural property can be captured by the betweenness centrality. Let $\rho_{su}$ denote the total number of shortest paths from $s$ to $u$ and $\rho_{su}(v)$ denote the number of those paths which pass through $v$. Then the betweenness centrality $\mathcal{B}(v, G)$ of $v$ can be defined as

$$\mathcal{B}(v, G) = \sum_{s \neq u \neq v} \frac{\rho_{su}(v)}{\rho_{su}}. \tag{2.11}$$

The vertex $v_{\mathcal{B}}$ with the maximum betweenness centrality is the *betweenness center*. Using the line network (when $n = 4$) in Figure 2.2 as an example, there is only one path between any pair of nodes, and hence that path is also the shortest path. Considering vertex 2, the $s - t$ paths pairs to be considered are $(v_1, v_3)$, $(v_1, v_4)$ and $(v_3, v_4)$. The number of shortest paths passing between $v_1$ and $v_3$ is one (i.e., $\rho_{v_1 v_3} = 1$), which also passes through $v_2$ (i.e., $\rho_{v_1 v_3}(v_2) = 1$), thus we have the ratio $\frac{\rho_{v_1 v_3}(v_2)}{\rho_{v_1 v_3}} = 1$. Likewise, the ratio $\frac{\rho_{v_1 v_4}(v_2)}{\rho_{v_1 v_4}} = 1$. On the other hand, the number of shortest paths passing between $v_3$ and $v_4$ is one (i.e., $\rho_{v_3 v_4} = 1$), which does not pass through $v_2$ (i.e., $\rho_{v_3 v_4}(v_2) = 0$), thus we have the ratio $\frac{\rho_{v_3 v_4}(v_2)}{\rho_{v_3 v_4}} = 0$. Adding up these three ratios, we have $\mathcal{B}(v_2, G) = 2$. Similarly, $\mathcal{B}(v_1, G) = \mathcal{B}(v_4, G) = 0$, as either of the two end vertices in the line graph does not lie on any of the shortest $s - t$ path pairs. Clearly, the betweenness center is either $v_2$ or $v_3$. Using another bigger example in Figure 2.5, we have $\mathcal{B}(v_6, G) = 45$, $\mathcal{B}(v_7, G) = 55$, $\mathcal{B}(v_4, G) = 46$ and $\mathcal{B}(v_9, G) = 56$ and hence $v_9$ is the betweenness center. Note that the betweenness centrality of all the leaves in $G$ is zero since no path will pass through any leaf except itself.

**Rumor Centrality and Epidemic Centrality**

From the above example of the line graph, observe that the spreading order from any node in the graph (also known as permitted permutation in [131], [132]) plays a role. The definition of the spreading order is given as follows [131], [132].

**Definition 2.9.** Given a connected tree $G(V, E)$ and a source node $v \in V$, consider any permutation $\sigma : V \to \{1, \ldots, |V|\}$ of its nodes where $\sigma(u)$ denotes the position of node $u$ in the permutation $\sigma$. We call $\sigma$ a *spreading order* for tree $G(V, E)$ with source node $v$ if

1. $\sigma(v) = 1$,

2. For any edge $(u, u') \in E$, if $d(v, u) < d(v, u')$, then $\sigma(u) < \sigma(u')$.

For brevity, we denote a permutation satisfying the above two conditions as $\sigma^v$. Let $M(v, G_n) = \{\sigma_1^v, \sigma_2^v, \ldots\}$ be the set of all spreading orders from a vertex $v$ in a given graph $G_n$ with $n$ vertices. For example, in the line graph example, the set $M(v_2, G_4)$ contains three possible spreading orders, which are $\sigma_1^{v_2} = (v_2, v_1, v_3, v_4)$, $\sigma_2^{v_2} = (v_2, v_3, v_1, v_4)$ and $\sigma_3^{v_2} = (v_2, v_3, v_4, v_1)$. The size of this set is exactly the binomial coefficient $f_i = \binom{n-1}{n-i}$ for $i = 1, 2, \ldots, n$ as given earlier. Observe that each of these three possible occurs with an equal probability. From a maximum-likelihood estimation perspective, the optimal estimator for the source is the vertex $\hat{v}$ that maximizes the likelihood probability $P(G_n|v)$ given by

$$P(G_n|v) = \sum_{\sigma_i^v \in M(v, G_n)} P(\sigma_i^v|v). \qquad (2.12)$$

Now, consider the above line graph example for any $G_n$, observe that all possible spreading orders for every single node occur with an equal probability, and hence the optimal solution of (3.3) is equivalent to

$$\hat{v} = \arg\max_{v \in G_n} |M(v, G_n)|,$$

which leads to $\hat{v}$ being $v_{\frac{n+1}{2}}$ when $n$ is odd or otherwise $\hat{v}$ being either $v_{\frac{n}{2}}$ or $v_{\frac{n}{2}+1}$ when $n$ is even.

The size of $M(v, G_n)$ is known as the *rumor centrality* of the vertex $v$ in $G_n$ whose *rumor center* is $\hat{v}$ [131], [132]. It is, however, important to note that all the possible spreading orders for every single node occur with an equal probability only for the special case of a degree-regular tree graph (e.g., the line graph in the illustrative example is a 2-regular tree). In general, each spreading order in a given general tree graph can occur with a different probability. Hence, the rumor center may not necessarily be the optimal maximum-likelihood estimator for a general tree graph. There are, however, many interesting properties associated with rumor centrality. When $G_n$ is a tree graph, the rumor centrality can be connected to the distance and branch weight centrality. As illustrated by the line graph example, the rumor center of a tree graph is equivalent to the distance center and the graph centroid, as should be the case (cf. Theorem 3.3 in subsequent sections).

Note that the rumor center in [132] is only defined on a spanning tree (e.g., breadth-first-search tree) of $G_n$. For the case when $G_n$ is any general graph, even those with cycles, we follow [165] and consider a generalized version of the rumor centrality that applies to general graphs. To illustrate the idea of computing the rumor centrality on a general graph, we use a triangle graph $G_3$ with three vertices $V = \{v_1, v_2, v_3\}$ as an example. With $v_1$ as the source, a spreading order $\sigma = (v_1, v_2, v_3)$ on $G_3$ can be due to two spreading orders on the spanning trees of $G_3$. One spanning tree is the tree with edge set $\{(v_1, v_2), (v_2, v_3)\}$, and the other one is the tree with edge set $\{(v_1, v_2), (v_1, v_3)\}$. Hence, we can compute the spreading order starting from $v \in V$ on a general graph $G_n$ by summing up the rumor centrality of $v$ on all spanning trees $T_i$ of $G_n$. We continue to use the notation $M(v, G_n)$ as the set of all spreading orders starting from $v$ on $G_n$. Then for a general graph $G_n$ and $v \in V$, we have

$$|M(v, G_n)| = \sum_{1 \le i \le h} |M(v, T_i)|,$$

where $T_i$ is a spanning tree of $G_n$, and $h$ is the number of spanning trees of $G_n$. We call $|M(v, G_n)|$ the *epidemic centrality* of $v$, and $\hat{v}$ is an *epidemic center* of $G_n$ if

$$|M(\hat{v}, G_n)| = \max_{v \in G_n} |M(v, G_n)|.$$

Besides tracking the probability of spreading orders, *computing the epidemic centrality efficiently to find the epidemic center* is an object of interest to solving the contagion source detection problem.

As we shall see in subsequent sections, we will delve into the fascinating connection between various measures of network centrality introduced in this section and their interpretation as optimal solutions to relevant optimization problems.

# 3

# Contagion Source Problem and Degree-regular Tree Case

This section focuses on the contagion source detection problem in graphs, which we formulate as a maximum likelihood estimation problem. We will explore the optimal solutions to this problem, specifically in the case of degree-regular tree graphs, using the concept of rumor centrality introduced in the seminal work of [131], [132]. Furthermore, we will delve into the relationships between rumor centrality and other network centralities, such as distance centrality and graph-theoretic branch weight centrality. By doing so, we can better understand the role of rumor centrality in network analysis. We will also explore the use of message-passing algorithms, similar to those used in inferential statistics and Markov chain algorithms, in efficiently computing the optimal solution for the contagion source detection problem.

## 3.1 Maximum-likelihood Estimation Problem

The problem of identifying the initial source can be formulated as a maximum likelihood estimation problem. Let $G$ be the underlying network (assumed to be an infinite graph) and $G_n$ be the infection graph. Let $P(v = v^\star | G_n)$ be the probability of the event that the vertex

$v$ in $G_n$ is exactly the rumor source $v^\star$ when $G_n$ is observed. By Bayes' theorem, we have

$$P(v = v^\star | G_n) = \frac{P(G_n | v = v^\star) \cdot P(v = v^\star)}{\sum_{i \in G_n} [P(G_n | i = v^\star) \cdot P(i = v^\star)]}.$$

We assume that each vertex in the graph is equally likely to be the source, that is

$$P(v^\star = v_i) = P(v^\star = v_j) \quad \forall \, v_i, v_j \in G_n,$$

and thus, we have

$$P(v = v^\star | G_n) = \frac{P(G_n | v = v^\star)}{\sum_{i \in G_n} P(G_n | i = v^\star)}. \tag{3.1}$$

For simplicity, we denote $P(v = v^\star | G_n)$ as $P(v|G_n)$ and $P(G_n | v = v^\star)$ as $P(G_n|v)$ in the following. From (3.1), we can deduce that $P(v|G_n)$ is proportional to $P(G_n|v)$.

Hence, we aim to solve the contagion source detection problem given as follows:

$$\begin{aligned} \underset{v \in G_n}{\text{maximize}} \quad & P(G_n|v) \\ \text{subject to} \quad & G_n \subset G. \end{aligned} \tag{3.2}$$

**Definition 3.1.** For a given $G_n$ over the underlying graph $G$, $\hat{v}$ is a maximum likelihood estimator for the source in $G_n$, i.e., $P(G_n|\hat{v}) = \max_{v_i \in G_n} P(G_n|v_i)$.

The maximum likelihood estimator is the most probable vertex to be the actual rumor source. From Definition 3.1, the vertex $\hat{v}$ has the maximum probability $P(G_n|\hat{v})$, so we need to consider $P(G_n|v)$, the probability that $v$ is the actual rumor source that leads to observing $G_n$. Recall that we denote a possible spreading order starting from $v$ as $\sigma_i^v$, and let $M(v, G_n)$ be the set of all $\sigma_i^v$ in $G_n$. For example, the set $M(v_4, G_4)$ contains two possible spreading orders which are $\sigma_1^v = (v_4, v_1, v_2, v_3)$ and $\sigma_2^v = (v_4, v_1, v_3, v_2)$ in Figure 3.1. Let $G_k(\sigma)$ be the infected subgraph following the spreading order $\sigma$ with $k$ vertices. Then we have

$$P(G_n|v) = \sum_{\sigma_i^v \in M(v, G_n)} P(\sigma|v). \tag{3.3}$$

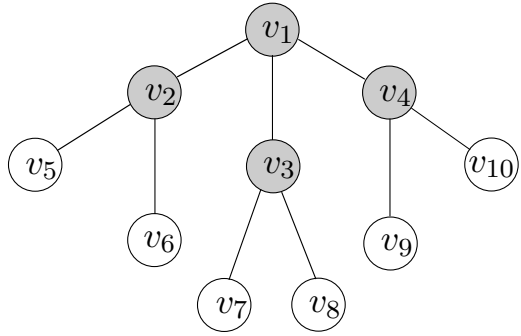

**Figure 3.1:** Example of $G_4$ as an infected subgrpah with infected vertices $\{v_1, v_2, v_3, v_4\}$.

In particular,

$$P(\sigma_i^v | v) = \prod_{k=1}^{n-1} \frac{1}{\sum\limits_{v_j \in G_k(\sigma_i^v)} d(v_j) - 2(k-1)}. \tag{3.4}$$

When $G$ is an infinitely large (each leaf of $G_n$ always has at least one susceptible neighbor) $d$-regular tree, we have

$$P(\sigma_i^v | v) = \prod_{k=1}^{n-1} \frac{1}{dk - 2(k-1)}. \tag{3.5}$$

The following example illustrates the computation for the probability $P(G_5 | v_4)$ in Figure 3.1. We have

$$\begin{aligned}
P(G_4 | v_4) &= P((v_4, v_1, v_2, v_3) | v_4) + P((v_4, v_1, v_3, v_2) | v_4) \\
&= 2 \cdot (\frac{1}{3} \cdot \frac{1}{4} \cdot \frac{1}{5}) \\
&= \frac{1}{30}.
\end{aligned}$$

Also, we have $P(G_4 | v_2) = 1/30$, $P(G_4 | v_3) = 1/30$ and $P(G_4 | v_1) = 1/10$. In this example, $v_1$ is the maximum likelihood estimator. Observe that, when $G$ is a $d$-regular tree, $P(\sigma_i^v | v) = P(\sigma_j^v | v)$ for all $\sigma_i^v, \sigma_j^v \in M(v, G_n)$, which means that $P(G_n | v)$ is proportional to $|M(v, G_n)|$. When $G_n$ is degree-regular, we call $|M(v, G_n)|$ the *rumor centrality* of $v$, and $v$ is a *rumor center* of $G_n$ if $|M(v, G_n)| = \max\limits_{v_i \in G_n} |M(v_i, G_n)|$. If $P(v|G_n)$ is proportional to $|M(v, G_n)|$, then the rumor center is the maximum likelihood estimator.

## 3.2   Rumor Centrality

Let the underlying network $G$ be a $d$-regular tree rooted at $v$ and $G_n$ is an infected subtree on $G$, then we have

$$
\begin{aligned}
|M(v, G_n)| &= (n-1)! \prod_{u \in child(v)} \frac{|M(u, t_u^v)|}{t_u^v!} \\
&= (n-1)! \prod_{u \in child(v)} \left( \frac{(t_u^v - 1)!}{t_u^v!} \cdot \prod_{w \in child(u)} \frac{|M(w, t_w^u)|}{t_w^u!} \right) \\
&= (n-1)! \prod_{u \in child(v)} \frac{1}{t_u^v} \cdot \prod_{w \in child(u)} \frac{|M(w, t_w^u)|}{t_w^u!} \\
&= (n-1)! \prod_{u \in G_n \setminus \{v\}} \frac{1}{t_u^v}.
\end{aligned}
$$

(3.6)

For example, consider $|M(v_4, G_n)|$ in Figure 3.1, we have

$$
\begin{aligned}
|M(v_4, G_n)| &= 4! \cdot \frac{1}{4 \cdot 2 \cdot 1 \cdot 1} \\
&= 3.
\end{aligned}
$$

The above closed-form formula for computing $M(v, G_n)$ on a tree graph was proposed in [132]. Now, suppose $G$ is a degree regular tree, and let $u$ and $v$ be two adjacent vertices on $G_n$. Consider the ratio

$$
\begin{aligned}
\frac{P(v|G_n)}{P(u|G_n)} &= \frac{P(G_n|v)}{P(G_n|u)} \\
&= \frac{|M(v, G_n)|}{|M(u, G_n)|} \\
&= \frac{t_v^u}{t_u^v} = \frac{t_v^u}{n - t_v^u}
\end{aligned}
$$

We have shown that the likelihood ratio is equivalent to the ratio $t_v^u / t_u^v$. This result leads to the following theorem.

**Theorem 3.1.** Given an $n$ vertices tree $G_n$. $v \in G_n$ is a rumor center if and only if

$$t_u^v \leq \frac{n}{2}$$

for all $u \in G_n - \{v\}$.

Theorem 3.1 characterize the topological feature for the rumor center in $G_n$, moreover, from Theorem 3.1, we can deduce that there are at most two rumor centers in a given tree $G_n$.

**Corollary 3.2.** Let $G$ and $G_n$ be defined as above, then we have

$$P(\hat{v}|G_n) \leq \frac{1}{2}.$$

The following theorem states the relation between *rumor center, centroid* and *distance center*.

**Theorem 3.3.** Let $G_n$ be a general tree graph and $v$ is a vertex in $G_n$. Then, the following statements are equivalent:

1. $v$ is a *distance center* of $G_n$.

2. $v$ is a rumor center of $G_n$.

3. $v$ is a graph centroid of $G_n$.

*Proof.* Let $G_N$ be a tree of size $N$ and $v \in G_N$. We prove ($1 \Rightarrow 2$) first by considering the contrapositive argument. Suppose $v$ is not a rumor center, by Theorem 3.1 there is a branch of $v$, say $T_u^v$, with size greater than $N/2$ and $u$ is adjacent to $v$. Now, we need a relationship between $\sum_{s \in G_N} d(v, s)$ and $\sum_{s \in G_N} d(u, s)$ as described by

$$\sum_{s \in G_N} d(v, s) = \sum_{s \in G_N} d(u, s) + (t_u^v - 1) - (t_v^u - 1).$$

We have $\sum_{s \in G_N} d(v, s) > \sum_{s \in G_N} d(u, s)$, since $t_u^v > t_u^v$. This implies that $v$ is not a distance center.

Next, let us prove ($2 \Rightarrow 3$): First, we need the following fact: If all $v$'s branches are of size $\leq N/2$, then $v$ is the centroid. Again, by contrapositive argument, suppose $v$ is not a centroid, then there exists a branch of $v$ whose size $> N/2$ by Theorem 3.1.

Lastly, let us prove $(3 \Rightarrow 1)$: Suppose $v$ is a centroid, then each of all its branches is of size $\leq N/2$. This implies that $v$ is a rumor center. Let $u \in G_N$, if $u$ is adjacent to $v$, then $\sum_{s \in G_N} d(v, s) < \sum_{s \in G_N} d(u, s)$ and we finish the proof. If $u$ is not adjacent to $v$, then we can partition all the nodes in $G_N$ into three sets. The first one is $T_v^u$, the second one is $T_u^v$ and the last one contains all the nodes not in $T_v^u$ and $T_u^v$, say $R$. Let $l$ denote $d(u, v)$. Now, consider $\sum_{s \in G_N} d(v, s) - \sum_{s \in G_N} d(u, s) = (\sum_{s \in T_v^u} d(v, s) + \sum_{s \in T_u^v} d(v, s) + \sum_{s \in R} d(v, s)) - (\sum_{s \in T_v^u} d(u, s) + \sum_{s \in T_u^v} d(u, s) + \sum_{s \in R} d(u, s))$.

Since $v$ is the centroid, we have:

(1) $|R| + t_u^v \leq N/2$, and $t_v^u > N/2$;

(2) $(\sum_{s \in T_u^v} d(v, s) + \sum_{s \in T_v^u} d(v, s)) - (\sum_{s \in T_u^v} d(u, s) + \sum_{s \in T_v^u} d(u, s)) = l \cdot (t_u^v - t_v^u)$;

(3) $|\sum_{s \in R} d(v, s) - \sum_{s \in R} d(u, s)| \leq l \cdot |R|$.

Combining these three properties, we conclude that $\sum_{s \in G_N} d(v, s) - \sum_{s \in G_N} d(u, s) < 0$, for any $u \in G_N$, that is, $v$ is the distance center.    □

From the perspective of the message-passing algorithm, the following theorem enables us to compare the above-mentioned centralities of any two adjacent vertices and extend the result in Theorem 3.3.

**Theorem 3.4.** *Let $G_n$ be a general tree with $n$ vertices and $u, v \in G_n$ are two adjacent nodes (neither $u$ nor $v$ needs to be the centroid). Then, the following statements are equivalent:*

1. $R(v, G_n) \geq R(u, G_n)$ .

2. $DisC(v, G_n) \leq DisC(u, G_n)$.

3. $\mathsf{weight}(v, G_n) \leq \mathsf{weight}(u, G_n)$.

*Proof.* Let $G_n$ be a tree of size $n$, and $u, v \in G_n$. Observe the following directions. Let us prove $(1 \Rightarrow 2)$: Suppose $R(v, G_n) \geq R(u, G_n)$, we have $DisC(v, G_n) = Disc(u, G_n) - t_u^v + t_v^u$ and $t_v^u \geq t_u^v$, and so we conclude that $DisC(v, G_n) \leq DisC(u, G_n)$.

Next, let us prove $(2 \Rightarrow 3)$: Suppose $DisC(v, G_n) \leq DisC(u, G_n)$, we have $DisC(v, G_n) - DisC(u, G_n) = t_u^v - t_v^u \leq 0$. This implies that $t_u^v \leq t_v^u$. We claim that $\mathsf{weight}(u, G_n) = t_v^u$. If not, then there is a branch of $u$ with its size larger than $t_v^u$, thereby implying $t_u^v \geq t_v^u$, which is a contradiction. Hence, we have $\mathsf{weight}(u, G_n) = t_v^u \geq \mathsf{weight}(v, G_n)$.

Lastly, let us prove $(3 \Rightarrow 1)$: Suppose $\mathsf{weight}(v, G_n) \leq \mathsf{weight}(u, G_n)$, and note that $\mathsf{weight}(u, G_n) = t_v^u$. Since $u$ is not the rumor center, we have $t_v^u > n/2$ and so $t_u^v \leq n/2$, this implies that $R(v, G_n) \geq R(u, G_n)$.
$\square$

## 3.3 Algorithms for Tree Networks

In this section, we present two algorithms to find the rumor center (i.e., the globally optimal solution of the maximum likelihood estimation problem) on tree networks. The first method is a message-passing algorithm to compute the graph centroid (which is equivalent to the rumor center) with linear time complexity, and the second one is a random-walk-based algorithm where the centrality is in the form of the stationary distribution of a specific graph random walk.

### 3.3.1 Message Passing Algorithm

Let $M^{i \to j}$ denote the message from node $i$ to node $j$. To calculate the weight of all nodes in $G_N$, we need to assign each $M^{i \to j}$ a number for all $(i, j) \in E(G_N)$. We set the value of $M^{i \to j}$ to be the size of $t_i^j$. So, we have $M^{i \to j} + M^{j \to i} = N$. And also for any node $v \in V(G_N)$, we have $\mathsf{weight}(v, G_N) = \max\{M^{i \to v} | \forall i \text{ is adjacent to } v\}$. In Algorithm 1 below, we first find all $M^{i \to j}$, and then use Theorem 3.5 to locate the *centroid*, finally, we set the weight to all nodes. Let $\mathsf{Diff}(i, j)$ be defined by $\mathsf{Diff}(i, j) = |M^{i \to j} - M^{j \to i}|$.

**Theorem 3.5.** *Given a tree $G_N$ with $N$ nodes, $\tilde{v} \in G_N$ is the centroid if and only if $\forall v$ adjacent to $\tilde{v}$ and $v_i, v_j \in V(G_N)$, $\min_{(v, \tilde{v}) \in E(G_N)} \{\mathsf{Diff}(\tilde{v}, v)\} \leq \{\mathsf{Diff}(v_i, v_j)\}$. Moreover, for any $u \in G_N$, on the path from $\tilde{v}$ to $u$ say $(v_1, v_2, ..., v_D)$, where $v_1 = \tilde{v}$ and $v_D = u$. The sequence of $\mathsf{Diff}(v_i, v_{i+1})$ for $i = 1, 2...D$ is increasing.*

Now, a practical implication of Theorem 3.5 is that this centroid for a given tree $G_N$ can be found using graph algorithms (thereby providing alternative algorithms to the contagion source detection in [132]). Using graph-theoretic analysis, new algorithms can be designed with computational time complexity $O(N)$, where $N$ is the size of the input graph. Now, a key algorithmic design in statistical learning is the message-passing algorithm framework (also known as belief propagation [50], [87], [113]) whereby simple messages are exchanged between neighboring nodes in a graph and these local operations converge to the solution of a global problem iteratively [113]. We next present such a new message passing algorithm to find the graph centroid in Algorithm 1.

---

**Algorithm 1** Message Passing Algorithm to compute the Centroid of a Graph [163]

---

Input a tree $G$

Choose a root $v_r$ from $G$

**for** $u$ in T **do**

    **if** $u$ is a leaf **then**

        $M^{u \to \mathsf{parent}(u)} = 1$

    **else**

        **if** $u \neq v_r$ **then**

            $M^{u \to \mathsf{parent}(u)} = \Sigma_{j \in \mathsf{child}(u)} M^{j \to u} + 1$

        **end if**

    **else**

        $N = \Sigma_{j \in \mathsf{child}(v_r)} M^{j \to v_r}$

    **end if**

**end for**

Find the longest path $P = (p_1, .., p_{k-1}, p_k)$ starting from $v_r$, such that $\mathsf{Diff}(p_i, p_j)$ is decreasing along $P$, where $p_1 = v_r$ and $\mathsf{Diff}(v_i, v_j) = |2M^{v_i \to v_j} - N|$.

$v^k$ is the centroid if $M^{v^k \to v^{k-1}} > M^{v^{k-1} \to v^k}$ else $v^{k-1}$ is the centroid.

---

The first part in Algorithm 1 is message passing, where each node exchanges messages with its neighboring nodes that are updated in a recursive manner. As these messages are passed from the leaf nodes to their parent nodes who in turn aggregate the messages collected from their children nodes and pass the aggregated result to their parent nodes, this process iterates until each node computes their individual network centrality. The number of iterations is the same as the number of nodes in the graph, i.e., the computational complexity of this part is $O(N)$. The second part is to find the centroid by leveraging Theorem 3.5. Since local optimality implies global optimality in the branch weight centrality, the length of the path in the second part is at most $N/2$, which implies the computational complexity is $O(N)$. Hence, the computational complexity for Algorithm 1 is $O(N)$. In the following, we highlight the key advantage of Algorithm 1 over that in [132]. The algorithm in [132] needs to compute the factorial of $N$, which is relatively larger than $N$. In contrast, the largest number that appears in Algorithm 1 is $N$. Furthermore, we can add a new node to the network by updating the nodes on the path from the newly-added node to the current centroid, and this modification complexity is at most the height of the tree. Hence, Algorithm 1 is scalable and adaptive to streaming data. In Figure 3.2, we use an example to illustrate two steps in Algorithm 1.

$$n = 12 + 1 + 2 + 1 = 16$$

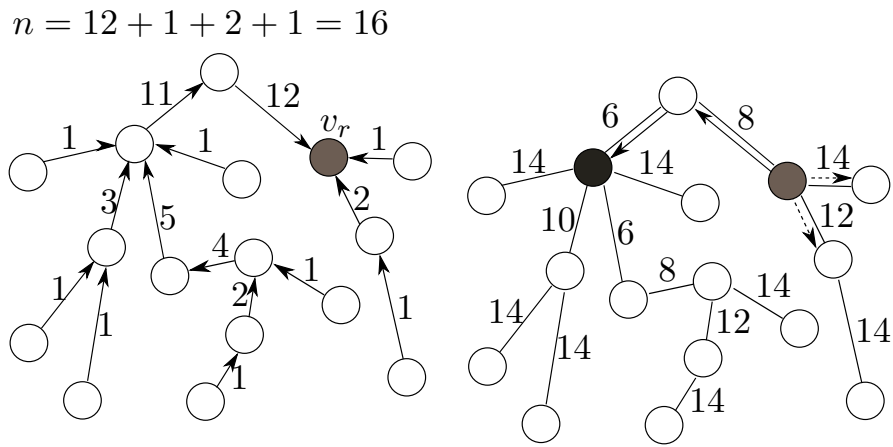

**Figure 3.2:** Example of how Algorithm 1 works on a given tree $G_{16}$. The figure on the left is the first part of Algorithm 1. We first randomly pick a node as the root, which is the node in the grey color. The process of message passing starts from the leaves until the root receives messages from all of its children. The figure on the right computes $\mathsf{Diff}(v_i, v_j)$ for all $v_i, v_j \in G_{16}$. Finally, find the longest decreasing path $P$, where $P = (8 \to 6)$. Note that if $P = (8 \to 6 \to 6)$, we can still find the centroid by the last line of Algorithm 1.

Suppose the tree $G_N$ is given, the message-passing Algorithm 1 ranks the importance of each node in terms of relative tree branch weight (or, equivalently, the number of linear extensions). The ranking of the nodes makes use of the relative centrality measure between adjacent nodes that is similar in measure to the rumor centrality, distance centrality, or the branch weight centrality that we characterized in the previous sections.

### 3.3.2   Rumor Centrality As a Stationary Distribution

It is also worth mentioning that there is a Markov Chain representation of the rumor centrality that can find the maximum-likelihood estimator on trees through a random walk approach [166]. We can define a specific transition probability matrix for a graph random walk such that the stationary distribution of this Markov chain is proportional to the rumor centrality for each node. Let $c$ be a constant such that $c > 1$.

$$
P_{i,j}^{rumor} = \begin{cases} \dfrac{T_j^i}{c(n-1)} & i \neq j \text{ and } (i,j) \in E \\[2mm] \dfrac{c-1}{c} & i = j \\[2mm] 0 & else \end{cases} \tag{3.7}
$$

The above Markov chain $((X_1, X_2, \ldots), P_{i,j}^{rumor})$ is aperiodic since $c > 1$. Note that this Markov chain is also irreducible since the corresponding graph is connected and the transition probability defined on each edge is non-zero.

Let $\pi \in R^n$ be a probability distribution over all states in $\Omega$ where

$$
\pi_i = k \cdot R(i, G_n),
$$

for $i = 1, 2, \ldots n$ and $k$ is a normalized constant. Then we have

$$
\begin{aligned}
\pi_i \cdot P_{i,j}^{rumor} &= k \cdot R(i, G_n) \cdot \frac{T_j^i}{c(n-1)} \\
&= k \cdot R(j, G_n) \frac{T_i^j}{T_j^i} \cdot \frac{T_j^i}{c(n-1)} \\
&= k \cdot R(j, G_n) \cdot \frac{T_i^j}{c(n-1)} \\
&= \pi_j \cdot P_{j,i}^{rumor}.
\end{aligned}
$$

By the Fundamental Theorem of Markov Chains and Theorem 2.4, this Markov chain has a unique stationary distribution which is the unique left eigenvector of $P$ with eigenvalue 1 and we denote this eigenvector as $\pi^\star$. We can deduce that $\pi = \pi^\star$, moreover, $\pi_i$ is proportional to the rumor centrality for each $i = 1, 2, \ldots, n$.

## 3.4 Conclusions and Remarks

In this section, we first showed that the contagion source detection problem could be formulated as a maximum likelihood estimation problem when considering a discrete SI spreading model with an underlying infinite graph. Based on Bayes's Theorem, a network centrality called "rumor centrality" was introduced in [132], [166] to optimally solve the maximum-likelihood estimation problem for infection networks that are degree-regular tree graphs. We have characterized the equivalence between the rumor center, the distance center and the graph centroid. We have also presented several efficient algorithms including a practical message-passing algorithm to compute the graph centroid (and therefore the rumor center) and computation by random walks on a Markov chain. We refer the reader to [114] on characterizing the probability distribution of the distance between the true source and the maximum-likelihood estimator in regular tree graphs and [37], [73] on characterizing the long-term behavior of graph centroids on increasing trees.

# 4

## Estimation and Detection in Graphs with Irregularities

In the previous section, we focused on the contagion source detection problem in the context of spreading over an infinite graph, where the resulting infection graphs were degree-regular tree types, and rumor centrality was able to determine the optimal solution for maximum-likelihood estimation. However, real-world scenarios often involve irregularities in the underlying graphs, such as finite boundaries or cycles, which require different approaches as rumor centrality may no longer be optimal. In this section, we delve into these irregularities to gain a better understanding of their effects.

We start with an illustrative example that highlights how limited graph size can introduce irregularities that impact maximum likelihood estimation. Subsequently, we investigate two specific cases: finite-size graphs with boundaries and graphs with cycles, both of which require tailored approaches for optimal maximum likelihood estimation. This exploration leads us to introduce the concept of epidemic centrality, which generalizes rumor centrality and offers new insights. By leveraging epidemic centrality-based techniques, we can develop heuristics that yield near-optimal solutions for maximum likelihood estimation in general graphs.

## 4.1 Epidemic Centrality for Trees with a Single End Vertex

The maximum likelihood estimation problem in the previous section has a constraint that the problem can be solved optimally only when the underlying network is an infinite-size degree-regular tree. Now we consider the case when $G$ is a regular tree that is finite, e.g., there are leaf vertices, each with degree one, and $G_n$ contains some of these leaf vertices. If a leaf vertex $v_e$ of $G_n$ is also a leaf of $G$, then we call $v_e$ an *end vertex* in $G_n$. Figure 4.1 gives an example of how a single end vertex in $G_n$ can affect the maximum-likelihood estimation performance.

**Example 4.1.** Consider $G$ as a finite 3-regular tree and $G_5 \subseteq G$ as shown in Figure 4.1. Consider $P(G_5|v_1)$ and with a spreading order $\sigma_1^{v_1} = (v_1, v_2, v_5, v_3, v_4)$, we have $P(\sigma_1^{v_1}|v_1) = \frac{1}{3} \cdot \frac{1}{4} \cdot \frac{1}{3} \cdot \frac{1}{4}$. Had $v_5$ not been the end vertex, then $P(\sigma_1^{v_1}|v_1) = \frac{1}{3} \cdot \frac{1}{4} \cdot \frac{1}{5} \cdot \frac{1}{6}$. This demonstrates that the order in which the rumor spreads to the end vertex $v_5$ is important when computing $P(\sigma_i^{v_1}|v_1)$. Table 4.1 lists down three possible values of $P(\sigma_i^{v_1}|v_1)$ according to the position of $v_5$. In particular, $P(G_5|v_1) = \frac{34}{720}$. By symmetry, we also have $P(G_5|v_4) = P(G_5|v_3) = \frac{7}{720}$, and $P(G_5|v_2) = \frac{40}{720}$. Note that $v_1$ is the *rumor center*, but $P(G_5|v_1) < P(G_5|v_2)$, and thus $\hat{v}$ is not $v_1$.

In particular, we compare this single end vertex special case with a naive prediction that assumes an underlying *infinite graph*. This illustrates that ignoring the boundary effect in the finite graph ultimately leads to a wrong estimate and thus requires an in-depth analysis and new contagion source detection algorithm design for the general case of *finite graphs*.

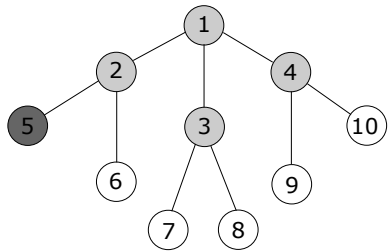

**Figure 4.1:** Example of $G$ as a finite 3-regular tree and $G_n$ as a subtree with a single end vertex $v_e = v_5$. The maximum likelihood estimate $\hat{v}$ is $v_2$, while a naive application of the rumor centrality in [133], i.e., rumor center $v_c$ of $G_n$, yields $v_1$.

**Table 4.1:** Numerical Example of $P(G_5|\sigma_i^{v_1})$ using $G_5$ in Figure 4.1.

| $\sigma_i^{v_1}$ | spreading order | $P(G_5|\sigma_i^{v_1})$ | $\sigma_i^{v_1}$ | spreading order | $P(G_5|\sigma_i^{v_1})$ |
|---|---|---|---|---|---|
| $\sigma_1^{v_1}$ | $v_1, v_2, v_5, v_3, v_4$ | $\frac{1}{144}$ | $\sigma_7^{v_1}$ | $v_1, v_2, v_3, v_4, v_5$ | $\frac{1}{360}$ |
| $\sigma_2^{v_1}$ | $v_1, v_2, v_5, v_4, v_3$ | $\frac{1}{144}$ | $\sigma_8^{v_1}$ | $v_1, v_2, v_4, v_3, v_5$ | $\frac{1}{360}$ |
| $\sigma_3^{v_1}$ | $v_1, v_3, v_2, v_5, v_4$ | $\frac{1}{240}$ | $\sigma_9^{v_1}$ | $v_1, v_3, v_2, v_4, v_5$ | $\frac{1}{360}$ |
| $\sigma_4^{v_1}$ | $v_1, v_4, v_2, v_5, v_3$ | $\frac{1}{240}$ | $\sigma_{10}^{v_1}$ | $v_1, v_3, v_4, v_2, v_5$ | $\frac{1}{360}$ |
| $\sigma_5^{v_1}$ | $v_1, v_2, v_3, v_5, v_4$ | $\frac{1}{240}$ | $\sigma_{11}^{v_1}$ | $v_1, v_4, v_2, v_3, v_5$ | $\frac{1}{360}$ |
| $\sigma_6^{v_1}$ | $v_1, v_2, v_4, v_5, v_3$ | $\frac{1}{240}$ | $\sigma_{12}^{v_1}$ | $v_1, v_4, v_3, v_2, v_5$ | $\frac{1}{360}$ |

### 4.1.1 Impact of Boundary Effects On $P(G_n|v)$

Example 4.1 reveals some interesting properties of boundary effects due to even a single end vertex:

- $P(\sigma_i^v|v)$ increases with how soon the end vertex appears in $\sigma_i^v$ (as ordered from left to right of $\sigma_i^v$).

- When there is at least one end vertex in $G_n$, then $P(G_n|v)$ is no longer proportional to $|M(v, G_n)|$.

This means that $P(\sigma_i^v|v)$ is no longer a constant for each $i$, and is dependent on the *end vertex* position in each spreading order. We proceed to compute $P(\sigma_i^v|v)$ as follows. For brevity of notation, let $v_e$ be the end vertex and define

$$M_v^{v_e}(G_n, k) = \{\sigma^v | v_e \text{ is on the } k\text{th position of } \sigma^v\};$$
$$P_v^{v_e}(G_n, k) = P(\sigma^v|v), \text{ for } \sigma^v \in M_v^{v_e}(G_n, k),$$

where $M_v^{v_e}(G_n, k)$ is the set of all the spreading orders starting from $v$ and with $v_e$ at the $k$th position, and its size is the combinatorial object of interest:

$$m_v^{v_e}(G_n, k) = |M_v^{v_e}(G_n, k)|. \tag{4.1}$$

Let $D$ be the distance (in terms of the number of hops) from $v$ to $v_e$. Then we have

$$|M(v, G_n)| = \sum_{k=D+1}^{n} m_v^{v_e}(G_n, k). \tag{4.2}$$

Now, (4.2) shows that $M(v, G_n)$ can be decomposed into $M_v^{v_e}(G_n, k)$ for $k = D+1, D+2, \ldots, n$. This decomposition allows us to handle the boundary effect due to the different positions of the end vertex in each spreading order. Let $P_v^{v_e}(G_n, k)$ be the corresponding probability for each $k$. We can rewrite $P(G_n|v)$ for the finite tree graph as:

$$P(G_n|v) = \sum_{k=D+1}^{n} m_v^{v_e}(G_n, k) \cdot P_v^{v_e}(G_n, k). \tag{4.3}$$

Thus, the contagion source detection problem is to find the vertex $\hat{v}$ that solves

$$P(G_n|\hat{v}) = \max_{v_i \in G_n} P(G_n|v_i). \tag{4.4}$$

Since $P(G_n|v)$ is no longer proportional to $|M(v, G_n)|$, we now describe how to compute $P(G_n|v)$ in $G_n$ over an underlying $d$-regular graph $G$. First, consider $P_v^{v_e}(G_n, k)$ and let $z_d(i) = (i-1)(d-2)$, then

$$P_v^{v_e}(G_n, k) = \prod_{i=1}^{k-1} \frac{1}{d + z_d(i)} \cdot \prod_{i=k-1}^{n-2} \frac{1}{d + z_d(i) - 1}, \tag{4.5}$$

where the first factor of $P_v^{v_e}(G_n, k)$ in (4.5) is the probability that $k$ vertices are infected once the rumor reaches the end vertex, i.e., $v_e$ is the $k$th vertex infected in $G_n$, and the second factor is the probability that all remaining $n - k$ vertices are infected thereafter. On the other hand, the value of $m_v^{v_e}(G_n, k)$ in (4.1) is dependent on the network topology, and thus there is no closed-form expression in general (though when $G_n$ is a line, a closed-form expression for $m_v^{v_e}(G_n, k)$ is given in (4.6)). In the following, we provide a two-step algorithm based on the vertex contraction to compute the value of $m_v^{v_e}(G_n, k)$.

Let $\mathbb{G}_v^{k-1}$ be the set of all $(k-1)$-subtrees rooted at $v$ and containing the parent vertex of $v_e$, for example, vertex $v_2$ is the parent vertex of $v_5$ in Figure 4.2. We decompose the computation of $m_v^{v_e}(G_n, k)$ into two parts to ensure that $v_e$ will be on the $k$-th position of the spreading order. The first part is to compute how many spreading orders are there before the rumor reaches $v_e$, and the second part is to compute the number of spreading order after the rumor reach $v_e$.

In line 3, we can traverse all $(k-1)$-subtrees in $\mathbb{G}_v^{k-1}$ by an algorithm in [157] with a slight change as follows. Suppose $d(v, v_e) = D$, and we contract all the vertices on the path from $v$ to $parent(v_e)$ to make sure that the $(k - 1)$-subtrees contain the $parent(v_e)$. Then we have a new vertex $v'$, which is the contraction of $D$ vertices, and a new graph $G'_{n-D+1}$. Now, we can apply the algorithm in [157] to find all $(k - D)$-subtrees rooted at $v'$ on $G'_{n-D+1}$, then each $(k - D)$-subtree on $G'_{n-D+1}$ is corresponding to a $(k - 1)$-subtree which contains $parent(v_e)$ on $G_n$. The traversal of $(k - 1)$-subtrees cost $O(n)$ per $(k - 1)$-subtree. As a remark, the algorithm in [157] goes through all the $(k-1)$-subtrees, even the $(k - 1)$-subtrees are not rooted at $v$. Hence, we only need a part of the algorithm in [157].

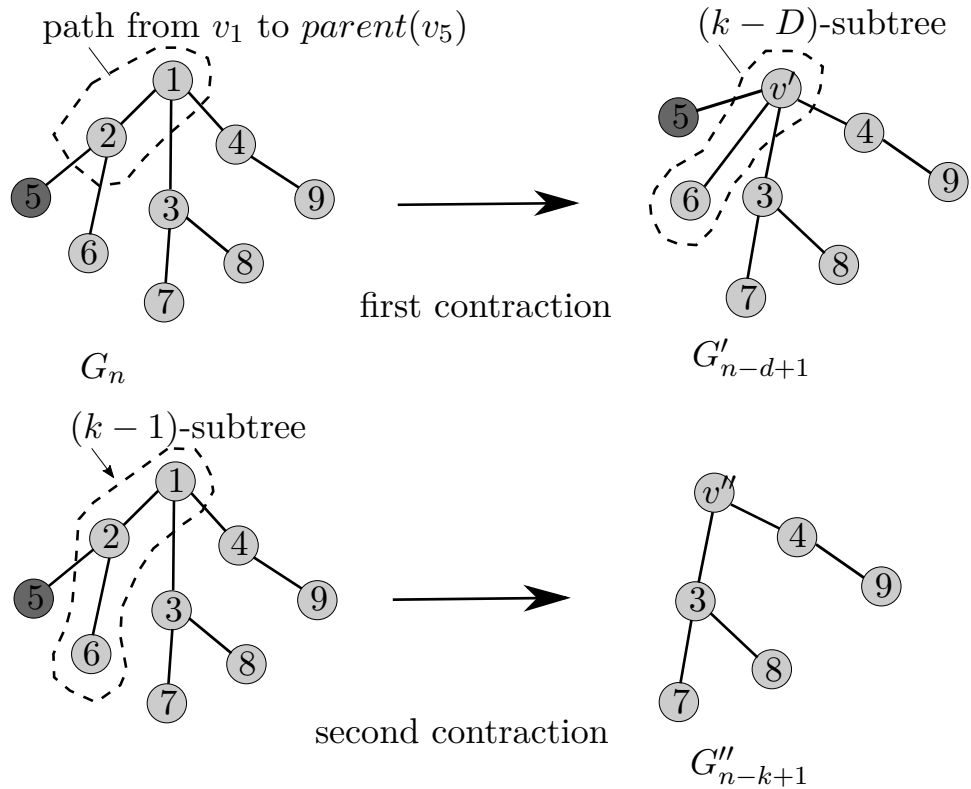

**Figure 4.2:** Illustration of Algorithm 2 for computing $m_{v_1}^{v_5}(G_n, k)$ where $n = 9$, $k = 4$, and $D = 2$. After the first contraction on $v_1$ and $v_2$, we apply the algorithm in [120] to find all $(k - D)$-subtrees starting from $v'$ on $G'_{n-d+1}$. Each resultant graph is a $(k - 1)$-subtree containing $v_1, v_2, v_6$. Lastly, the graph on the lower right is the resultant graph after contracting $v_1, v_2, v_6$ and $v_e$.

Back to Algorithm 2, $P_1$ represents the number of spreading orders starting from $v$ on the $(k - 1)$-subtree denoted by $G_{k-1}$, which corre-

---

**Algorithm 2** Computing $m_v^{v_e}(G_n, k)$ in (4.1)

---

1: **Input:** $G_n$, $k$, $v$
2: $m_v^{v_e}(G_n, k) = 0$
3: **for** $G_{k-1} \in \mathbb{G}_v^{k-1}$ **do**
4:    $P_1 = |M(v, G_{k-1})|$
5:    **for** $v_i \in G_k$ **do**
6:       **if** $v_i \neq v$ **then**
7:          **for** $v_j \in neighbor(v_i)$ **do**
8:             $E(G_n) = E(G_n) \cup (v_j, v)$
9:          **end for**
10:         delete $v_i$ from $G_n$
11:       **end if**
12:       $P_2 = |M(v, G''_{n-k+1})|$
13:    **end for**
14:    $m_v^{v_e}(G_n, k) = m_v^{v_e}(G_n, k) + P_1 \cdot P_2$
15: **end for**
16: Output: $m_v^{v_e}(G_n, k)$

---

sponds to the first part. The for-loop in line 5 is the vertex contraction that contracts all the infected vertices in $(k-1) - subtree$ and the end vertex $v_e$ to obtain a new vertex, say $v''$. The second part is to compute $P_2$, which is the number of spreading orders starting from $v''$ on $G''_{n-k+1}$. We can compute $P_1$ and $P_2$ by applying the algorithm in [131]. Lastly, the value of $P_1 \cdot P_2$ is the number of all possible spreading orders on $G_n$ for a given $(k-1)$-subtree. The time complexity of Algorithm 2 heavily depends on $\mathbb{G}_v^{k-1}$, since line 4 and line 12 cost $O(n)$ computing time [132].

For example, consider $m_{v_1}^{v_5}(G_9, 4)$ on Figure 4.2. For the 3-subtree containing $\{v_1, v_2, v_6\}$, line 4 calculates the number of spreading order starting from $v_1$ on this 3-subtree, which yields $P_1 = 1$. Line 5 to line 11. contracts $v_1, v_2, v_6$ and $v_5$ to yield $v''$ and a new graph $G''_6$. Line 12 computes the number of spreading order starting from $v''$ on $G''_6$, which yield $P_2 = 20$. Thus, the output is $1 \cdot 20$. After going through all 3-subtrees that contains $v_2$, the computation of $m_{v_1}^{v_5}(G_9, 4)$ is completed. The complexity of this task is dependent on the network topology. When $G_n$ is an arbitrary tree on $G$, then going through all $(k-1)$-subtrees on $G_n$ is computationally intensive, especially when $k$ is large.

### 4.1.2   Analytical Characterization of Likelihood Function: Trees with a Single End Vertex

Suppose $G$ is a finite degree-regular tree and $G_n$ is a line graph with a single end vertex. Without loss of generality, suppose $n$ is odd (to ensure a unique $v_c$) and $n = 2t + 1$ for some $t$. Label all the vertices in $G_n$ as shown in Figure 4.3. To compute $P(G_n|v_i)$ for $v_i \in G_n$, from (4.3) and (4.5), we already have $P_{v_i}^{v_e}(G_n, k)$, so we need to compute $m_{v_i}^{v_e}(G_n, k)$. The enumeration of $m_{v_i}^{v_e}(G_n, k)$ can be accomplished in polynomial-time complexity with a path-counting message-passing algorithm (see, e.g., Chapter 16 in [113]). In particular, we have a closed-form expression for $m_v^{v_e}(G_n, k)$ given by:

$$m_{v_i}^{v_e}(G_n, k) = \binom{k-2}{k-n+i-1}, \tag{4.6}$$

when $i \neq n$. Figure 4.4 shows how to derive the above equation by counting the number of paths from the upper left corner to the lower right corner.

Thus, we have the following analytical formula for $P(G_n|v_i)$: $P(G_n|v_i) =$

$$\begin{cases} \displaystyle\prod_{l=1}^{n-1} \frac{1}{z_d(l) + 1}, & i = n; \\ \displaystyle\sum_{k=n-i+1}^{n} \binom{k-2}{k-n+i-1} \cdot P_{v_i}^{v_e}(G_n, k), & \text{otherwise,} \end{cases} \tag{4.7}$$

where $P_{v_i}^{v_e}(G_n, k)$ is given in (4.5).

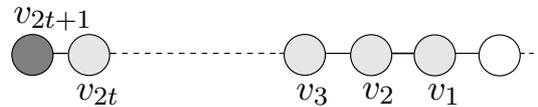

**Figure 4.3:** $G_n$ as a line graph with a single end vertex $v_e = v_{2t+1}$.

In (4.7), we suppose that $n$ is odd. Using (4.7), let us numerically compute $P(G_n|v_i)$ for all $v_i$ in Figure 4.5, where $G$ is a 4-regular tree and $G_n$ is a line graph with a single end vertex $v_e = v_n$ as boundary for different values of $n = 7, 8, 9, 10$. The $x$-axis is the vertex $v_i$ where $i = 1, 2, \ldots, 10$, and the $y$-axis plots $P(G_n | v_i = v^\star)$. As shown in Figure 4.5, the influence due to the end vertex on $P(G_n | v_i = v^\star)$ dominates that of rumor center $v_c$ when $n = 7, 8, 9$. However, when $n = 10$, the influence due to $v_c$ on $P(G_n | v_i)$ dominates that of the end vertex $v_e$.

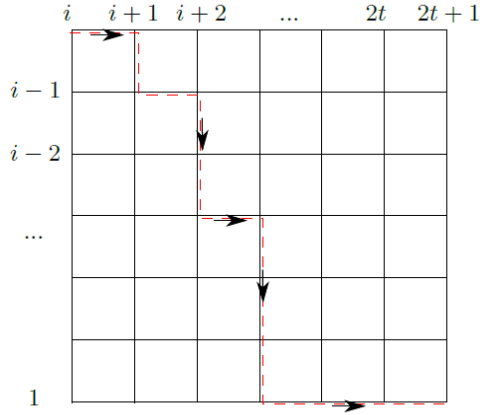

**Figure 4.4:** The dashed line on the grid corresponds to a spreading order $\sigma^i = (i, i+1, i-1, i+2, i-2, ..., 1, 2t-1, 2t, 2t+1)$ starting from upper left corner $i$ and ending at lower right corner $2t+1$.

**Theorem 4.1.** Suppose $G$ is a $d$-regular graph $(d > 2)$ with finite size. If $G_n$ is a line-graph with a single end vertex, then $\exists j$ such that $P(G_n | v_c) > P(G_n | v_e)$ when $n > j$.

As a remark, observe that when $n$ increases, i.e., the distance between $v_c$ and $v_e$ increases, the location of $\hat{v}$ in $G_n$ converges to the neighborhood of $v_c$.

*Proof.* In this proof, we use the fact that

$$P(v_c | G_n) > P_{v_c}^{v_e}(G_n, n) \cdot m_{v_c}^{v_e}(G_n, n),$$

and consider the ratio between the lower bound of $P(v_c | G_n)$ and $P(v_e | G_n)$ to simplify the proof. Let $G$ and $G_n$ be defined as in Theorem 1, and without loss of generality, suppose $n = 2t + 1$ and $d \geq 3$. For $v_e$, we have

$$P(v_e|G_n) = m_{v_e}^{v_e}(G_n, 1) \cdot P_{v_e}^{v_e}(G_n, 1)$$
$$= 1 \cdot P_{v_e}^{v_e}(G_n, 1)$$
$$= \prod_{i=0}^{2t-1} \frac{1}{1+i(d-2)} = \prod_{i=0}^{n-2} \frac{1}{1+i(d-2)}.$$

For $v_c$, it is simpler to consider the last term of (4.3) only, that is, $P_{v_c}^{v_e}(G_n, n) \cdot m_{v_c}^{v_e}(G_n, n)$. Note that $m_{v_c}^{v_e}(G_n, n) = |M(v_e, G'_{n-1})|$ where $G'_{n-1} = G_n \setminus \{v_e\}$. We have

$$P_{v_c}^{v_e}(G_n, n) \cdot m_{v_c}^{v_e}(G_n, n)$$
$$= \left[ \prod_{i=0}^{2t-1} \frac{1}{d+i(d-2)} \right] \cdot \frac{(2t)!}{2t(t-1)!t!}$$
$$= \frac{(n-1)!}{(n-1)(\frac{n-3}{2})!(\frac{n-1}{2})!} \cdot \prod_{i=0}^{n-2} \frac{1}{d+i(d-2)}.$$

Now, let us consider the ratio given by

$$\frac{P_{v_c}^{v_e}(G_n, n) \cdot m_{v_c}^{v_e}(G_n, n)}{P(v_e|G_n)} = \frac{(n-1)!}{(n-1)(\frac{n-3}{2})!(\frac{n-1}{2})!} \cdot \frac{\prod_{i=0}^{n-2} \frac{1}{d+i(d-2)}}{\prod_{i=0}^{n-2} \frac{1}{1+i(d-2)}}$$
$$= \frac{(n-2)!}{(\frac{n-3}{2})!(\frac{n-1}{2})!} \cdot \prod_{i=0}^{n-2} \frac{1+i(d-2)}{d+i(d-2)}$$
$$= \frac{\Gamma(n-1)}{\Gamma(\frac{n-1}{2})\Gamma(\frac{n+1}{2})} \cdot \frac{\Gamma(\frac{d}{d-2}+1)\Gamma(n+\frac{1}{d-2}-1)}{d \cdot \Gamma(\frac{1}{d-2}+1)\Gamma(n+\frac{d}{d-2}-1)}$$
$$= c_1 \cdot \frac{\Gamma(n-1)}{\Gamma(\frac{n-1}{2})\Gamma(\frac{n+1}{2})} \cdot \frac{\Gamma(n+\frac{1}{d-2}-1)}{\Gamma(n+\frac{d}{d-2}-1)}$$
$$= c_1 \cdot \frac{2}{(n-1) \cdot B(\frac{n-1}{2}, \frac{n-1}{2})} \cdot \frac{\Gamma(n+\frac{1}{d-2}-1)}{\Gamma(n+\frac{d}{d-2}-1)}$$
$$\approx c_1 \cdot \frac{2^{n-1}}{\sqrt{2\pi}} \cdot n^{-c_2},$$

where $c_1$ and $c_2$ are some positive values with respect to $d$. The approximation is given by using Stirling's formula. The above result shows

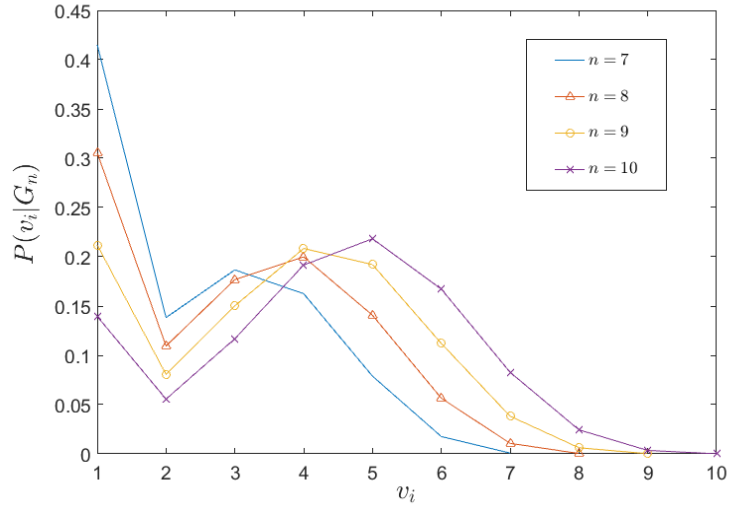

**Figure 4.5:** $P(G_n|v)$, where $G_n$ is a line graph with a single end vertex $v_1$ over an underlying 4-regular finite graph. Observe that $v_1$ in this figure is corresponding to $v_{2t+1}$ in Figure 4.3.

that the ratio becomes larger than 1 when $d$ is fixed and $n$ is sufficiently large enough. This leads to

$$\frac{P(v_c|G_n)}{P(v_e|G_n)} > \frac{P_{v_c}^{v_e}(G_n, n) \cdot m_{v_c}^{v_e}(G_n, n)}{P(v_e|G_n)} > 1,$$

when $n$ is sufficiently large. □

**Example 4.2.** To verify Theorem 4.1, we plot $P(G_n|v_i)$ for an example of a line $G_n$ with $G$ being a finite 4-regular graph in Figure 4.5. Clearly, we have $j = 9$.

Theorem 4.1 implies that, for any $d$-regular underlying graph, when $G_n$ is a line with a single end vertex, the influence of the end vertex $v_e$ on $P(v_i|G_n)$ decreases monotonically as $n$ grows. In fact, this reduces to the special case in [133], when $n$ goes to infinity asymptotically, i.e., $\hat{v}$ is rumor center.

**Optimality Characterization of the Maximum Likelihood Estimation**

From (4.3), we observe that, in addition to the spreading order, the distance (number of hops) between the end vertex and $v$ also affects the likelihood probability $P(G_n|v)$. Let $v_p$ be a neighbor of $v_c$ which has the largest rumor centrality among the neighbors of $v_c$ and satisfies $d(v_p, v_e) = D + 1$ where $D$ is the distance from $v_c$ to $v_e$. Since $v_c$ is the rumor center, we have $|M(v_c, G_n)| > |M(v_p, G_n)|$ by its definition. Moreover, the end vertex $v_e$ is closer to $v_c$ than $v_p$. Hence, these two assumptions may lead us to $P(G_n|v_c) > P(G_n|v_p)$. We formalize this first optimality result that characterizes the probabilistic inference performance between any two adjacent vertices and the location of $\hat{v}$ in $G_n$ with a single end vertex $v_e$ in the theorem. We first prove Lemma 4.2 and Lemma 4.3, which are tools to help us prove Theorem 4.4.

**Lemma 4.2.** Let $G$ be a tree and $v_a$, $v_b \in G$. If $d(v_a, v_b) = 2$ and $M(v_a, G) > M(v_b, G)$, then $t^{v_b}_{v_a} > t^{v_a}_{v_b}$.

*Proof.* Let $v_m$ denote the vertex on the path from $v_a$ to $v_b$. Then we can express $M(v_a, G)$ as

$$M(v_a, G) = \frac{t^{v_m}_{v_a}}{t^{v_a}_{v_m}} \cdot M(v_m, G),$$

and $M(v_b, G)$ can be expressed in the same form. Since $M(v_a, G) > M(v_b, G)$, we have

$$\frac{t^{v_m}_{v_a}}{t^{v_a}_{v_m}} > \frac{t^{v_m}_{v_b}}{t^{v_b}_{v_m}}.$$

Note that $t^{v_m}_{v_a} = t^{v_b}_{v_a}$ and $t^{v_m}_{v_b} = t^{v_a}_{v_b}$, we can rewrite the above inequality as

$$\frac{t^{v_b}_{v_a}}{t^{v_a}_{v_m}} > \frac{t^{v_a}_{v_b}}{t^{v_b}_{v_m}}.$$

Moreover, we can leverage the fact $n = t^{v_a}_{v_m} + t^{v_b}_{v_a} = t^{v_b}_{v_m} + t^{v_a}_{v_b}$ to replace $t^{v_a}_{v_m}$ and $t^{v_b}_{v_m}$ in the above inequality. Hence, we have

$$\frac{t^{v_b}_{v_a}}{n - t^{v_b}_{v_a}} > \frac{t^{v_a}_{v_b}}{n - t^{v_a}_{v_b}},$$

which implies $t^{v_b}_{v_a} > t^{v_a}_{v_b}$.

$\square$

**Lemma 4.3.** Let $G$ be a tree of size $n$, for any three vertices say $v_a$, $v_b$ and $v_{ir}$, if one of the following conditions is satisfied

1. $|M(v_a, G_n)| \geq |M(v_b, G_n)|$ and $d(v_a, v_{ir}) < d(v_b, v_{ir})$

2. $|M(v_a, G_n)| > |M(v_b, G_n)|$ and $d(v_a, v_{ir}) \leq d(v_b, v_{ir})$,

then we have

$$\sum_{i=d(v_a, v_{ir})+1}^{k} m_{v_a}^{v_{ir}}(G_n, i) \geq \sum_{i=d(v_a, v_{ir})+1}^{k} m_{v_b}^{v_{ir}}(G_n, i),$$

for all possible $k$.

*Proof.* To prove Lemma 4.3, we first consider the second condition, i.e., the case when $d(v_a, v_{ir}) = d(v_b, v_{ir}) = 1$. Note that in this case, $v_a$ is not on the shortest path from $v_b$ to $v_{ir}$ and vice versa.

We consider the ratio of $m_{v_a}^{v_{ir}}(G_n, i)$ to $m_{v_b}^{v_{ir}}(G_n, i)$ for all possible $i$. We can express $m_v^{v_{ir}}(G_n, i)$ as

$$m_v^{v_{ir}}(G_n, i) = \binom{n-i}{t_{v_{ir}}^v - (i-1)} M(v, t_{v_a}^{v_{ir}}) \cdot M(v_{ir}, t_{v_{ir}}^v),$$

where $v = v_a, v_b$. Since $M(v_a, t_{v_a}^{v_{ir}})$, $M(v_{ir}, t_{v_{ir}}^{v_a})$, $M(v_b, t_{v_b}^{v_{ir}})$ and $M(v_{ir}, t_{v_{ir}}^{v_b})$ are fixed when $G_n$ is given, we have

$$\frac{m_{v_a}^{v_{ir}}(G_n, i)}{m_{v_b}^{v_{ir}}(G_n, i)} \propto \frac{\binom{n-i}{t_{v_a}^{v_{ir}} - (i-1)}}{\binom{n-i}{t_{v_b}^{v_{ir}} - (i-1)}} = \frac{\binom{n-i}{n - t_{v_a}^{v_{ir}} - 1}}{\binom{n-i}{n - t_{v_b}^{v_{ir}} - 1}}.$$

Since $G_n$ is a tree and $d(v_a, v_{ir}) = d(v_b, v_{ir}) = 1$, we have $d(v_a, v_b) = 2$. By Lemma 4.2, we have $t_{v_a}^{v_b} > t_{v_b}^{v_a}$, which implies $t_{v_a}^{v_{ir}} > t_{v_b}^{v_{ir}}$. Hence, the ratio $\frac{m_{v_a}^{v_{ir}}(G_n, i)}{m_{v_b}^{v_{ir}}(G_n, i)}$ is an increasing sequence with respect to $i$.

Since $m_{v_a}^{v_{ir}}(G_n, 2) = m_{v_b}^{v_{ir}}(G_n, 2)$ and $\frac{m_{v_a}^{v_{ir}}(G_n, i)}{m_{v_b}^{v_{ir}}(G_n, i)}$ is increasing with respect to $i$, we can conclude that $m_{v_a}^{v_{ir}}(G_n, i) \geq m_{v_b}^{v_{ir}}(G_n, i)$ which leads to

$$\sum_{i=d(v_a, v_{ir})+1}^{k} m_{v_a}^{v_{ir}}(G_n, i) \geq \sum_{i=d(v_a, v_{ir})+1}^{k} m_{v_b}^{v_{ir}}(G_n, i).$$

On the other hand, we consider the case when $v_a$ is on the path from $v_b$ to $v_{ir}$, i.e., $\{v_a, v_b\} \in E(G)$ and $d(v_a, v_{ir}) = d(v_b, v_{ir}) + 1$.

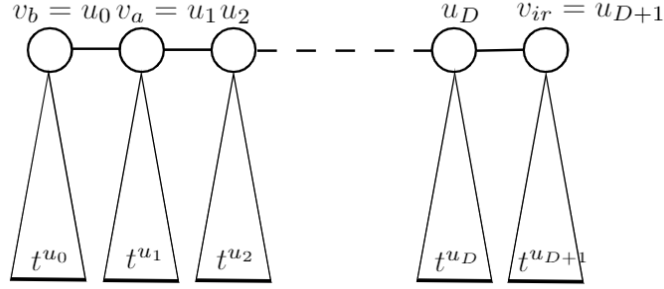

**Figure 4.6:** Each triangle labeled $t^{u_i}$ is a branch rooted at $u_i$.

For brevity, we denote the distance from $v_a$ to $v_{ir}$ as $D$, and relabel the vertices on the path from $v_p$ to $v_e$ as $u_0$ to $u_{D+1}$ which are shown in Figure 4.6.

Again, we consider the ratio of $m_{u_1}^{u_{D+1}}(G_n, i)$ to $m_{u_0}^{u_{D+1}}(G_n, i)$ for all possible $i$. We can express $m_{u_1}^{u_{D+1}}(G_n, i)$ and $m_{u_0}^{u_{D+1}}(G_n, i)$ as follows:

$$m_{u_1}^{u_{D+1}}(G_n, i) =$$

$$\sum_{1 < i_1 < \ldots < i_{D-1} < i} \Big[ \prod_{l=1}^{D-1} \binom{n - i_l - t^{u_1}_{u_{l+1}}}{t^{u_l} - 1} \Big] \binom{n - i}{t^{u_1}_{u_{D+1}} - 1} \mathbf{M}_a,$$

where

$$\mathbf{M}_a = M(u_1, t^{u_{D+1}}_{u_1}) \cdot \prod_{l=2}^{D+1} M(u_l, t^{u_l}).$$

We can express $m_{u_0}^{u_{D+1}}(G_n, i)$ in the same way. Since $\mathbf{M}_a$ is fixed when $G_n$ is given, we have

$$\frac{m_{u_1}^{u_{D+1}}(G_n, i)}{m_{u_0}^{u_{D+1}}(G_n, i)} \propto \frac{\sum\limits_{1 < i_1 < \ldots < i_{D-1} < i} \Big[ \prod_{l=1}^{D-1} \binom{n - i_l - t^{u_1}_{u_{l+1}}}{t^{u_l} - 1} \Big]}{\sum\limits_{1 < i_1 < \ldots < i_D < i} \Big[ \prod_{l=1}^{D} \binom{n - i_l - t^{u_1}_{u_{l+1}}}{t^{u_l} - 1} \Big]},$$

which is a decreasing sequence with respect to $i$.

Note that when $i = D + 1$, we have $m_{u_1}^{u_{D+1}}(G_n, D+1) > m_{u_0}^{u_{D+1}}(G_n, D+1)$.

Now, we can prove the main statement of Lemma 4.3. To contrary, suppose there is an integer $k$ such that

$$\sum_{i=D+1}^{k} m_{v_a}^{v_{ir}}(G_n, i) < \sum_{i=D+1}^{k} m_{v_b}^{v_{ir}}(G_n, i).$$

Since the ratio of $m_{u_1}^{u_{D+1}}(G_n, i)$ to $m_{u_0}^{u_{D+1}}(G_n, i)$ is a decreasing sequence, we have for all $k' > k$,

$$\sum_{i=D+1}^{k'} m_{v_a}^{v_{ir}}(G_n, i) < \sum_{i=D+1}^{k'} m_{v_b}^{v_{ir}}(G_n, i).$$

This leads to $M(u_1, G_n) < M(u_0, G_n)$, which is a contradiction to the assumption. Hence, we can conclude that $\forall k$,

$$\sum_{i=D+1}^{k} m_{v_a}^{v_{ir}}(G_n, i) \geq \sum_{i=D+1}^{k} m_{v_b}^{v_{ir}}(G_n, i).$$

$\square$

**Theorem 4.4.** Let $G$ be a $d-$regular tree with a single irregular vertex and $G_n \subseteq G$ be a subtree of $G$ with a single irregular vertex $v_{ir} \in G_n$. where $deg(v_{ir}) < d$. Then, the maximum likelihood estimator $\hat{v}$ with maximum probability $P(G_n|v)$ is on the path from the $v_c$ to $v_{ir}$.

In conclusion, given any $G_n$ with a single end vertex, in order to find $\hat{v}$, we first apply Theorem 4.4 to simplify our task by focusing the search for $\hat{v}$ to only those vertices on the path from $v_c$ to $v_e$, and then apply Algorithm 2 to only those vertices on this path. We refer the readers to Section 6 for simulation results.

**Likelihood Ratio Between Rumor Center and End Vertex on Different Network Topology**

Let us further elaborate on how the graph distance, i.e., the network topology, enables the search for $\hat{v}$ to narrow down to either $v_c$ or $v_e$. Observe that for any two infected subgraphs $G_n$ and $G'_n$ with the same underlying graph $G$, if $G_n$ is not isomorphic to $G'_n$, then the probability $P(G_n|v)$ for each $v$ is not the same. In other words, the topology of $G_n$ affects the probability $P(G_n|v)$. For $G_n$ and $G'_n$, if there exists an axis of symmetry of $G_n$ such that $G'_n$ can be obtained by rotating $G_n$ along this axis, then for any $v_i \in G_n$ and its corresponding vertex $v'_i \in G_n$, we have $P(G_n|v_i) = P(G_n|v'_i)$.

In the following, we consider the likelihood ratio between $v_c$ and $v_e$,

$$
\begin{aligned}
\frac{P(G_n|v_c)}{P(G_n|v_e)} &= \frac{\sum_{k=2}^{n} m_{v_c}^{v_e}(G_n, k) \cdot P_{v_c}^{v_e}(G_n, k)}{|M(v_e, G_n)| \cdot P_{v_e}^{v_e}(G_n, 1)} \\
&> \frac{P_{v_c}^{v_e}(G_n, n) \cdot \sum_{k=2}^{n} m_{v_c}^{v_e}(G_n, k)}{|M(v_e, G_n)| \cdot P_{v_e}^{v_e}(G_n, 1)} \\
&= \frac{P_{v_c}^{v_e}(G_n, n) \cdot |M(v_c, G_n)|}{P_{v_e}^{v_e}(G_n, 1) \cdot |M(v_e, G_n)|}.
\end{aligned}
$$

Assume that $n$ and $d$ is fixed, then $P_{v_c}^{v_e}(G_n, n)/P_{v_e}^{v_e}(G_n, 1)$ is a constant, and the lower bound of the likelihood ratio can be written as

$$
\frac{P(G_n|v_c)}{P(G_n|v_e)} > \mu \cdot \frac{|M(v_c, G_n)|}{|M(v_e, G_n)|}, \tag{4.8}
$$

where $\mu$ is a function of $d$ and $n$. From (4.8), we can conclude that $\hat{v} = v_c$ when $|M(v_c, G_n)|/|M(v_e, G_n)| > 1/\mu$. On the other hand, we have

$$
\frac{P(G_n|v_c)}{P(G_n|v_e)} < \tau \cdot \frac{|M(v_c, G_n)|}{|M(v_e, G_n)|}, \tag{4.9}
$$

where $\tau$ is a function of $d$ and $n$. This implies that $\hat{v} = v_e$ when $|M(v_c, G_n)|/|M(v_e, G_n)| < 1/\tau$. For simplicity, we fix $n$ and $d$ to analyze the effect of the network topology on $|M(v_c, G_n)|/|M(v_e, G_n)|$.

**Theorem 4.5.** Assume that the size $n$ and the degree are fixed. Let $\hat{G}_n$ be the graph with $d(v_c, v_e) = \lfloor (n-1)/2 \rfloor$, and $\hat{g}_n$ be the graph with $d(v_c, v_e) = 1$. Then we have

$$
\hat{G}_n = \underset{G_n}{\operatorname{argmax}} \frac{|M(v_c, G_n)|}{|M(v_e, G_n)|},
$$

$$
\hat{g}_n = \underset{G_n}{\operatorname{argmin}} \frac{|M(v_c, G_n)|}{|M(v_e, G_n)|},
$$

where $\hat{G}_n$ and $\hat{g}_n$ each represent a class of network topology that leads to the extreme value of the ratio.

$$
|M(v_c, G_n)|/|M(v_e, G_n)|.
$$

Theorem 4.5 reveals that when the distance $d(v_c, v_e)$ reaches its maximum value, then the probability that $\hat{v} = v_c$ is greater than the case when $d(v_c, v_e) < \lfloor (n-1)/2 \rfloor$. Note that in Theorem 4.1, we showed that there exist a threshold $j$ such that if $n > j$, then $P(G_n|v_c) > P(G_n|v_e)$. Assume that $G_n$ is a graph with maximum ratio $|M(v_c, G_n)|/|M(v_e, G_n)|$, and $G_n'$ is the graph with

$$|M(v_c', G_n')|/|M(v_e', G_n')| < |M(v_c, G_n)|/|M(v_e, G_n)|.$$

Combining Theorem 4.1 and Theorem 4.5, we can conclude that the threshold $j' > j$, where $j'$ is the threshold of $n$ such that $\hat{v}$ switches from $v_e'$ to $v_c'$.

**Theorem 4.6.** Let $G$ be a $d$-regular graph and $G_n$ is an infected subgraph such that $d(v_c, v_e) = 1$, then $\hat{v}$ is always $v_e$.

*Proof.* Let $G$ be a $d$-regular graph, where $d > 2$, and $G_n$ be the infected subgraph with one end vertex such that $d(v_c, v_e) = 1$. Then, we have $M(v_c, G_n) = (n-1)M(v_e, G_n)$. Note that, we partition $M(v_c, G_n)$ into $|M(v_e, G_n)|$ sets and each is defined by an unique spreading order $\sigma_i^{v_e} \in M(v_e, G_n)$. For example, given $\sigma_i^{v_e} = (v_e, v_c, v_{i_3}, v_{i_4}, \ldots, v_{i_n}) \in M(v_e, G_n)$, we can construct a set $S(\sigma_i^{v_e})$ of $(n-1)$ spreading orders in $M(v_c, G_n)$ by removing $v_e$ from the first position of $\sigma_i^{v_e}$ and insert into any position from 2 to $n$. Hence, for each $\sigma_i^{v_e} \in M(v_e, G_n)$, we can compare the probability $P(\sigma_i^{v_e}|v_e)$ with $\sum\limits_{\sigma_j^{v_c} \in S(\sigma_i^{v_e})} P(\sigma_j^{v_c}|v_c)$.

We have

$$P(\sigma_i^{v_e}|v_e) = \frac{1}{1} \cdot \frac{1}{d-1} \cdot \frac{1}{2d-3} \cdots \cdot \frac{1}{(n-1)d-2n+3},$$

and

$$\sum\limits_{\sigma_j^{v_c} \in S(\sigma_i^{v_e})} P(\sigma_j^{v_c}|v_c) = P(\sigma_i^{v_e}|v_e)(\frac{1}{d} + \frac{d-1}{d(2d-2)} + \frac{(d-1)(2d-3)}{d(2d-2)(3d-4)} + \ldots).$$

Note that when $d = 3$, the denominator of the irreducible fraction of each term in $(\frac{1}{d} + \frac{d-1}{d(2d-2)} + \frac{(d-1)(2d-3)}{d(2d-2)(3d-4)} + \ldots)$ form a sequence of triangular numbers, i.e., $(3, 6, 10, 15, 21, \ldots)$. Moreover, we have

$$\sum\limits_{x=2}^{\infty} \frac{2}{x^2+x} = 1,$$

which implies $\sum_{\sigma_j^{v_c} \in S(\sigma_i^{v_e})} P(\sigma_j^{v_c}|v_c) = P(\sigma_i^{v_e}|v_e)$, when $d = 3$. For $d \geq 4$, we have

$$(\frac{1}{d} + \frac{d-1}{d(2d-2)} + \frac{(d-1)(2d-3)}{d(2d-2)(3d-4)} + \ldots) < 1.$$

Hence, for all $d > 2$, $\sum_{\sigma_j^{v_c} \in S(\sigma_i^{v_e})} P(\sigma_j^{v_c}|v_c) \leq P(\sigma_i^{v_e}|v_e)$.

We can conclude that,

$$P(G_n|v_e) = \sum_{\sigma_i^{v_e} \in M(v_e, G_n)} P(\sigma_i^{v_e}|v_e)$$
$$\geq \sum_{\sigma_i^{v_e} \in M(v_e, G_n)} [\sum_{\sigma_j^{v_c} \in S(\sigma_i^{v_e})} P(\sigma_j^{v_c}|v_c)]$$
$$= P(G_n|v_c).$$

$\square$

Theorem 4.6 reveals how the distance $d(v_c, v_e)$ and the ratio $|M(v_c, G_n)|/|M(v_e, G_n)|$ affect the likelihood in the extreme case $d(v_c, v_e) = 1$. However, for the other extreme case $d(v_c, v_e) = \lfloor (n-1)/2 \rfloor$, the maximum-likelihood estimator may not be $v_c$. For example, $G_5$ is a line graph with five vertices and one end of the line is the end vertex, then $v_e$ is the maximum-likelihood estimator.

**Example 4.3.** Let $G_n$ be a star graph as shown in Figure 4.7 with the center vertex as rumor center $v_c$ and only one leaf of $G_n$ is an end vertex $v_e$. The likelihood ratio $|M(v_c, G_n)|/|M(v_e, G_n)|$ reaches the minimum possible value as compared to the line graph illustrated by Theorem 4.5. This means that, for a star graph, $P(G_n|v)$ is dominated by $v_e$, therefore, $\hat{v} = v_e$. However, if we keep adding new vertices to $G_n$ such that $n > d$, then $G_n$ is no longer a star graph, which means that $\hat{v}$ switches from $v_e$ to $v_c$ eventually when $n$ becomes sufficiently large.

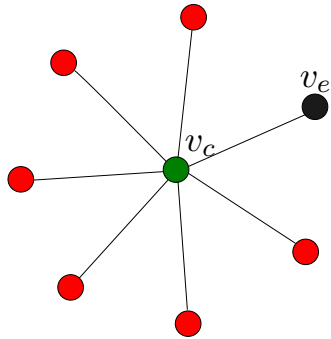

**Figure 4.7:** An example of a star graph with eight vertices.

## 4.2 Epidemic Centrality for Trees with Multiple End Vertices

In this section, we consider the case when $G_n$ has more than a single end vertex (naturally, this also means $d > 2$ in $G$ ruling out the trivial case of $G$ being a line). The key insight from the single end vertex analysis still holds: Once the rumor reaches an end vertex in $G$, $\hat{v}$ can be located near this very first infected end vertex. In addition, the algorithm design approach is to decompose the graph into subtrees to narrow the search for the maximum-likelihood estimate solution. To better understand the difficulty of solving the general case, we start with a special case: The entire finite underlying network is infected, i.e., $G_n = G$, then $P(G_n|v) = 1/n$ for each vertex in $G_n$, as each vertex is equally likely to spread the rumor to all the other vertices in $G$ to yield $G_n = G$. In this case, $P(G_n|\hat{v})$ is exactly the minimum detection probability. For example, consider a 3-regular tree underlying graph $G$ with 10 vertices, note that leaves of $G$ have degree 1. Figure 4.8 illustrates the maximum detection probability as the number of end vertices in $G_n$ increases with increasing $n$ as the rumor spreads. This means that the problem is harder to solve when the number of end vertices increases.

Note that when there is no end vertex, $G_n$ is composed of 4 vertices of the inner part of $G$. We can see that as the number of end vertices increases, $P(G_n|\hat{v})$ decreases to $1/10$. Therefore, when simulating the rumor spreading in a network, we will set an upper bound $n/k$ on the number of end vertices where $k$ is some integer greater than 1. Once the number of end vertices in $G_n$ reaches $n/k$, then we will stop the spreading process.

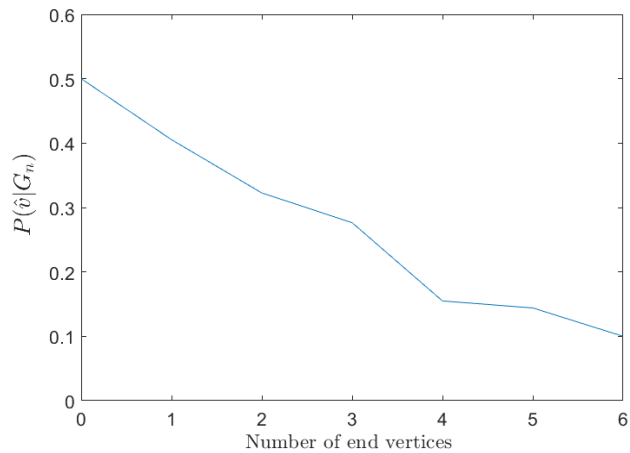

**Figure 4.8:** A numerical plot of $P(\hat{v}|G_n)$ versus the number of end vertices by using the example in Figure 4.1.

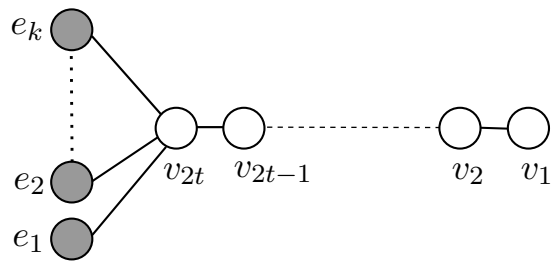

**Figure 4.9:** $G_n$ as a broom graph with $k$ star-like end vertices $e_1$ to $e_k$.

## Degree-Regular Tree ($d \geq 3$) Special Case: $G_n$ is Broom-Shaped

In Section 4.1, we have shown that, when $G_n$ is sufficiently large, the effect of the single end vertex on $P(v|G_n)$ for each vertex $v$ on the line graph $G_n$ is dominated by the rumor center. Now, we study the effect of multiple end vertices on a class of graphs whose topology is richer than the line graph in Section 4.1. In particular, as shown in Figure 4.9, we add end vertices to $v_{2t}$, so that when $G$ is $d$-regular, then there will be at most $d-1$ end vertices in $G_n$. We call this the *broom* graph. We can compute $P(v|G_n)$ by extending the result in Section 4.1. Let $P_{v_i}^{\{e_1, e_2, \dots, e_k\}}(\{h_1, h_2, \dots, h_k\}, G_n)$ be the probability of the spreading order starting from $v_i$ with the end vertex set $\{e_1, e_2, \dots, e_k\}$ and their position set $\{h_1, h_2, \dots, h_k\}$ in this spreading order. Note that we do not assume that $h_i$ is the position of $e_i$, as it can be the position of any end vertex in $G_n$. The probability $P_{v_i}^{\{e_1,e_2,\ldots,e_k\}}(\{h_1,h_2,\ldots,h_k\},G_n)$ can be obtained by the same analysis in (4.5). To compute $m_{v_c}^{\{e_1,e_2,\ldots,e_k\}}(\{h_1,h_2,\ldots,h_k\},G_n)$, we first consider the line-shaped part of $G_n$, i.e., the part $\{v_1,v_2,\ldots,v_{2t}\}$, say $G_n'$. From the previous discussion, we have

$$m_{v_i}^{v_{2t}}(G_n',j) = \binom{j+(2t+i-1)}{j},$$

and for each spreading order that $v_{2t}$ lies on the $j$th position, the end vertices $e_1,e_2,\ldots,e_k$ can be placed on any position after the $j$th position. So for each spreading order in $m_{v_i}^{v_{2t}}(G_n',j)$, there are $k! \cdot \binom{n-k-j+1}{k}$ corresponding spreading orders in $G_n$. Thus, we have

$$m_{v_i}^{\{e_1,\ldots,e_k\}}(G_n,\{h_1,\ldots,h_k\}) = k! \sum_{j=2t-i+1}^{h_1-1} \binom{j-2}{2t-i-1}. \qquad (4.10)$$

With $P_{v_i}^{\{e_1,e_2,\ldots,e_k\}}$ and $m_{v_i}^{\{e_1,e_2,\ldots,e_k\}}$, we can now compute the probability $P(v_i|G_n)$ by going through all possible $\{h_1,h_2,\ldots,h_k\}$. Figure 4.10 shows that even though there are five end vertices, the effect of the rumor center on $P(v|G_n)$ eventually dominates that of the end vertices as $n$ grows from 37 to 39. This result implies that: When there are more end vertices in $G_n$, $n$ needs to be sufficiently large to offset the effect of end vertices, i.e., for the transition phenomenon to take place. For other $d$ and $n$ in the *broom* graph, as shown in the proof of Theorem 4.1, we can prove this, in the same way. To conclude that, if we fix the number of end vertex, the probability $P(v_c|G_n)$ will be greater than $P(v_e|G_n)$ when $n$ is large enough.

## 4.3 Epidemic Centrality for Pseudo-Trees with a Cycle

In this section, we consider the special case where $G$ is a degree-regular graph, and $G_n$ has only a single cycle, i.e., $G_n$ is a pseudo-tree.

**Definition 4.1.** A pseudo-tree is a connected graph with an equal number of vertices and edges, i.e., a tree plus an edge that creates a cycle.

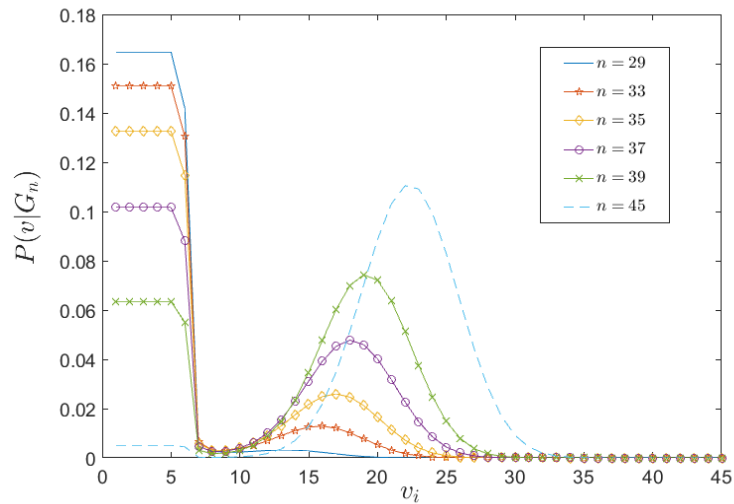

**Figure 4.10:** Probability distribution of each vertex on $G_n$ with five end vertices when $G$ is 6-regular, The $y$-axis plots the probability $P(v_i|G_n)$ and the $x$-axis plots the vertex $v_i$'s number $i$. In particular, $v_1, \ldots, v_5$ are the leaves (end vertices) corresponding to $e_1, \ldots, e_k$ in Figure 4.9, where $k = 5$. Observe that the transition phenomenon happens when $n$ grows from 37 to 39.

We denote the cycle as $C_h$ where $h$ is the number of vertices on the cycle. Here, we call those vertices on $C_h$ *cycle vertices*. Assuming $v$ is a cycle vertex, then we define $t_v$ to be the subtree rooted at $v$ in $G_n$. We take Figure 4.11 as an example, the subtree $t_{v_1}$ contains $v_1$, $v_4$, and $v_7$. In this section, we study how a cycle affects the probability $P(v|G_n)$ when $G_n$ contains a cycle $C_h$. To generalize the analysis in [131], we should intuitively assume that the probability of being infected is proportional to the number of infected neighborhoods. With this assumption, the analysis in [131] will not change, but we can consider the case that two infected vertices have a common susceptible neighborhood, i.e., there is a cycle in $G_n$.

### 4.3.1 Impact of a Single Cycle On $P(G_n|v)$

**Example 4.4.** Consider the infected subgraph $G_6 \subset G$ as shown in Figure 4.11, where $G_6 = \{v_1, v_2, v_3, v_4, v_5, v_7\}$ and there is a 3-cycle in $G_6$. Consider a spreading order $\sigma_1^{v_4} \in M(v_4, G_6)$, where $\sigma_1^{v_4} = (v_4, v_1, v_2, v_3, v_5, v_7)$. We have $P(\sigma_1^{v_4}|v_4) = (1/3) \cdot (1/4) \cdot (2/5) \cdot (1/4) \cdot (1/5) = 2/1200$. Note that when $v_1$ and $v_2$ are infected, $v_3$ has two infected neighborhoods which implies that the probability of $v_3$ be infected in the next time period is twice higher than $v_5, v_7$ and $v_8$. In particular, there are three possible values for $P(\sigma_i^{v_4}|v_4)$ as shown in Table 4.2, for all $\sigma_i^{v_4} \in M(v_4, G_6)$. Moreover, we observe that the denominators are different in Table 4.2, due to sharing a common neighbor in the presence of a cycle. We call this property the *cycle effect*.

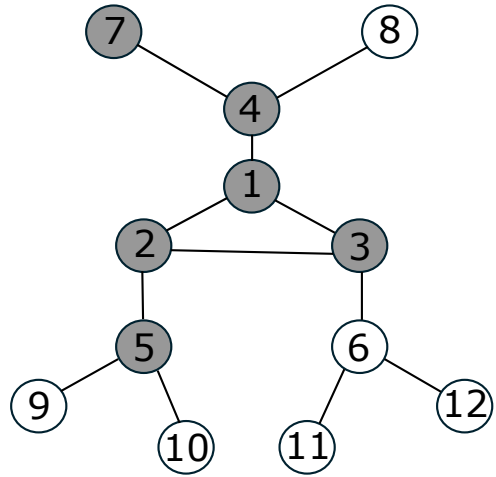

**Figure 4.11:** $G_6$ is an infected subgraph with a single cycle $C_3$ containing three cycle vertices $v_1$, $v_2$ and $v_3$. We can partition $G_6$ into three subtrees say $t_{v_1} = \{v_1, v_4, v_7\}$, $t_{v_2} = \{v_2, v_5\}$, $t_{v_3} = \{v_3\}$.

**Table 4.2:** Numerical Example of $P(\sigma_i^{v_4}|G_6)$ using $G_6$ in Figure 4.11

| $\sigma_i^{v_4}$ | spreading order | $P(\sigma_i^{v_4}|G_6)$ |
|---|---|---|
| $\sigma_1^{v_4}$ | $(v_4, v_1, v_3, v_2, v_5, v_7)$ | $\frac{2}{1200}$ |
| $\sigma_2^{v_4}$ | $(v_4, v_1, v_2, v_5, v_3, v_7)$ | $\frac{2}{1800}$ |
| $\sigma_3^{v_4}$ | $(v_4, v_1, v_2, v_5, v_7, v_3)$ | $\frac{2}{2520}$ |

Example 4.4 reveals some interesting properties of the cycle effect due to a single cycle:

1. $P(\sigma_i^v|v)$ increases with how soon the last cycle vertex appears in $\sigma_i^v$ (as ordered from left to right of $\sigma_i^v$). For example, the last cycle vertex on $\sigma_1^{v_4}$ is $v_2$, and is $v_3$ on $\sigma_2^{v_4}$ and $\sigma_3^{v_4}$.

2. When there is a cycle in $G_n$, then $P(G_n|v)$ is no longer proportional to $|M(v, G_n)|$.

3. For each $\sigma_i^v$, there are actually two corresponding permitted spreading orders due to the cycle.

The first property shows that $P(\sigma_i^v|v)$ is dependent on the position of the last cycle vertex in each spreading order. Note that the cycle effect is similar to the boundary effect, and the main difference lies in that all cycle vertices may cause the cycle effect instead of only one end vertex may cause the boundary effect. We proceed to compute $P(\sigma_i^v|v)$ as follows. For brevity of notation, let $v_l$ denote the last cycle vertex.

**Definition 4.2.** We let the distance from a vertex $v$ to the cycle $C_h$ be denoted by $d(v, C_h)$ and defined by the minimum value of distances from $v$ to all cycle vertices on $C_h$. That is,

$$d(v, C_h) = \min_{v_i \in C_h} \{d(v, v_i)\}.$$

We take Figure 4.11 as an example. Let the cycle contains $v_1$, $v_2$ and $v_3$ be denoted as $C_3$, then $d(v_7, C_3) = d(v_7, v_1) = 2$ and $d(v_5, C_3) = 1$.

As a remark, for each $\sigma^v \in |M(v, G_n)|$, the last cycle vertex $v_l$ can be any vertex on the cycle $C_h$ except the vertex $v'$ with the distance $d(v, v') = d(v, C_h)$. Hence, there are $h - 1$ possible choices $v_l$.

From previous observations, we have

$$|M(v, G_n)| = 2 \cdot \sum_{k=d(v,C_h)+h}^{n-t_{v_l}+1} m_v^{v_l}(G_n, k) \qquad (4.11)$$

since the position of $v_l$ on the spreading order ranges from $d(v, C_h) + h$ to $n - t_{v_l} + 1$. For example, in Table 4.2, we can see that $v_l = v_2$ is the 4th element on $\sigma_1^{v_4}$ and $v_l = v_3$ is the 6th element on $\sigma_3^{v_4}$. Finally, the multiplication with 2 is due to the third property.

Now, we can rewrite $P(G_n|v)$ for $G_n$ with a cycle as:

$$P(G_n|v) = \sum_{k=d(v,C_h)+h}^{n-t_{v_l}+1} m_v^{v_l}(G_n, k) \cdot P_v^{v_l}(G_n, k), \qquad (4.12)$$

and our goal is to find the vertex $\hat{v}$ that achieves

$$P(G_n|\hat{v}) = \max_{v_i \in G_n} P(G_n|v_i). \tag{4.13}$$

Since $P(G_n|v)$ is not proportional to $|M(v, G_n)|$, we should compute $P(G_n|v)$ by considering each part $m_v^{v_l}(G_n, k)$ and their corresponding probability $P_v^{v_l}(G_n, k)$. Let $z_d(i) = (i-1)(d-2)$, then

$$P_v^{v_l}(G_n, k) = 2 \cdot \prod_{i=1}^{k-1} \frac{1}{d + z_d(i)} \cdot \prod_{i=k-1}^{n-2} \frac{1}{d + z_d(i) - 1}. \tag{4.14}$$

The first factor in (4.14) is the probability that $k$ vertices are infected where the $k$th infected vertex is $v_l$, and the second factor is the probability of that all remaining $n-k$ vertices being infected thereafter. The $-1$ in the denominator of the second factor and the coefficient 2 at the front are due to the common neighbor in a cycle. Note that multiplying by 2 at the front makes no difference when computing $P(G_n|v)$ for each $v \in G_n$. From (4.12), we see that the number of spreading orders and the corresponding position of $v_l$ affect $P(G_n|v)$.

### 4.3.2 Analytical Characterization of Likelihood Function: Pseudo-Trees with a Cycle

This section focuses on computing $|M(v, G_n)|$. To compute $|M(v, G_n)|$, we can leverage the message-passing algorithm in [132] if $G_n$ is a tree. Each infected vertex in $G_n$ is infected by one of its infected neighbors (even if it has two infected neighbors), so the actual infecting route is a spanning tree of $G_n$ instead of a graph with a cycle. Hence, the number of all spreading orders on a graph $G_n$ with a cycle can be computed as

$$|M(v, G_n)| = \sum_{1 \le i \le h} |M(v, T_i)|, \tag{4.15}$$

where $T_i$ is a spanning tree of $G_n$, for $i = 1, 2, ..., h$. If $G_n$ contains a $C_h$, then $G_n$ has $h$ spanning trees. For each spanning tree $T_i$, we can apply the message-passing algorithm in [132] with $O(n)$ time complexity to compute $|M(v, T_i)|$. Hence, the time complexity of this approach is $O(hn)$.

To simplify the computation, we can leverage some analytical results to find $v_c$ in $O(n+h)$ time instead of $O(hn)$. In the following, we present a theorem and a lemma to characterize the location of the epidemic center in $G_n$. Let $t_i = t_{v_i}$ be defined as above, and slightly abusing the notation of the subtree size $|t_i|$ as $t_i$.

**Theorem 4.7.** Let $G$ be a degree regular graph, and $G_n$ be a subgraph of $G$ with a single cycle $C_h = \{v_1, v_2, ..., v_h\}$. The epidemic center $v_c$ of $G_n$ satisfies one of the following conditions:

1. Each connected component of $G_n \backslash \{v_c\}$ is of size less or equal to $n/2$.

2. $v_c$ is a cycle vertex and $t_i \leq n/2$ for $i = 1, 2, ..., h$.

As a remark, observe that $v$ being the rumor center of $G_n$ does not mean that each connected component of $G_n \backslash \{v\}$ is of size less or equal to $n/2$. (Had $G_n$ been a tree, then this is true [132].)

*Proof.* Let $v$ be a non-cycle vertex in $G_n$, and each connected component of $G_n \backslash \{v\}$ is of size less or equal to $n/2$. Consider any given spanning tree $T_j$ of $G_n$, assuming that $v$ is not the rumor center of $T_j$. Then, there is a vertex $u$ such that $t_u^v > t_v^u$ on $T_j$. Since $t_u^v + t_v^u = n$, we can conclude that the size of the connected component of $G_n \backslash \{v\}$ containing $u$ is greater than $n/2$, which contradicts the assumption. Hence, we have the fact that $v$ is the epidemic center on each spanning tree $T_j$, for $j = 1, 2, \ldots, h$, and $v$ is also the epidemic center of $G_n$ form (4.15). □

We can locate $v_c$ of $G_n$ by the first condition in Theorem 4.7. However, if the first condition is not satisfied, then $v_c$ is on the cycle. In the following, we proceed to consider the case if $v_c$ is a cycle vertex. Let $T_j$ denote the spanning tree of $G_n$ which is constructed by $G_n \backslash (v_j, v_{j+1})$, for $j = 1, 2, ..., h-1$ and $T_h = G_n \backslash (v_h, v_1)$. Note that $(v_h, v_1)$ and $(v_j, v_j + 1)$ for $j = 1, 2, ..., h-1$ are cycle edges of $C_h$.

**Proposition 4.1.** Let $v_i$ be a cycle vertex and assume that $|M(v_i, T_p)| = r$, where $r$ and $p$ are integers and $1 \leq p \leq h$. Then for $1 \leq q \leq h$, we have:

$$\frac{|M(v_i, T_q)|}{|M(v_i, T_p)|} = \frac{\prod\limits_{j \in C_h, j \neq i} T_{p,j}^i}{\prod\limits_{k \in C_h, k \neq i} T_{q,k}^i},\tag{4.16}$$

where $T_{p,j}^i$ is the subtree $T_j^i$ of the spanning tree $T_p$.

From [132], we know that the ratio of $|M(v_i, T_p)|/|M(v_j, T_p)|$ is proportional to their branch size in $T_p$ if $v_i$ and $v_j$ are adjacent. Now, for the same vertex $v_i$, but in different spanning tree say $T_p$ and $T_x$, can also derive the ratio $|M(v_i, T_p)|/|M(v_i, T_q)|$. Hence, if we assume $|M(v_1, T_1)| = r$, then we can derive $|M(v, T)|$ for all $v \in C_h$ and $T$ is a spanning tree of $G_n$ in terms of $r$ and $t_i$, where $t_i$ is shown in Figure 4.12. We give an algorithm in the following to find $v_c$ with time complexity $O(n + h^2)$ in the worst case. The $h^2$ term is the complexity to construct the table as shown in Table 4.3 and Table 4.4.

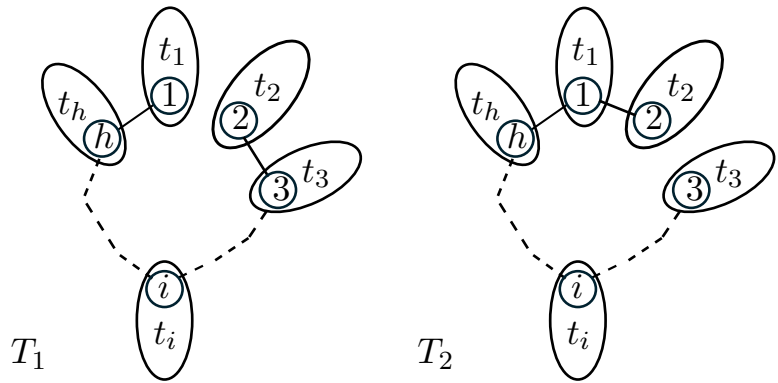

**Figure 4.12:** $C_h$ is constructed by $v_1, v_2, ..., v_h$, and $t_i$ is a subtree rooted at $v_i$.

**Example 4.5.** Take $C_3$ for example, $|M(v_1, T_1)|/|M(v_1, T_2)| = t_3/(t_2 + t_3)$. Assume $|M(v_1, T_1)| = r$, then Table 4.3 shows the ratio of $|M(v_i, T_j)|$, for all $1 \leq i, j \leq 3$. With Table 4.3, we can conclude that

$$|M(v_1, G_n)| = \frac{2(t_2 + t_3)}{t_3}r;$$
$$|M(v_2, G_n)| = \frac{2t_2(t_2 + t_3)}{t_1 t_3}r;$$
$$|M(v_3, G_n)| = \frac{2(t_2 + t_3)}{t_1}r,$$

which implies

$$|M(v_1, G_n)| : |M(v_2, G_n)| : |M(v_3, G_n)| = t_1 : t_2 : t_3.$$

---

**Algorithm 3** Compute the rumor center $v_c$ for unicyclic $G_n$ [162]

---

**Input:** $G_n$

Step 1: If there is a vertex $v_c$ satisfying the condition in Theorem 4.7, then $v_c$ is rumor center. If no such vertex exists, go to Step 2.

Step 2: Given a spanning tree $T^*$ and a cycle vertex $v'$ on $G_n$, compute $|M(v', T^*)|$.

Step 3: Construct a table (e.g. Table 4.3) of $|M(v,T)|$, where $v$ is a cycle vertex and $T$ is a spanning tree of $G_n$, starting from $|M(v', T^*)|$. For each row, i.e., on the same spanning tree $T$, if $v_i$ and $v_j$ are adjacent then we have $|M(v_i,T)| : |M(v_j,T)| = T_i^j : T_j^i$. For each column, i.e., on the same vertex $v$ but a different spanning tree, we can use the ratio in (4.16) to complete the column.

Step 4: Sum over each column, and $v_c = \underset{v \in G_n}{\operatorname{argmax}} \sum_{i=1}^{h} |M(v,T_i)|$.

---

**Table 4.3:** Table of $|M(v_i, T_i)|$, when $G_n$ contains a $C_3$

|       | $|M(v_1, T_i)|$      | $|M(v_2, T_i)|$                    | $|M(v_3, T_i)|$                    |
|-------|---------------------|-----------------------------------|-----------------------------------|
| $T_1$ | $r$                 | $\frac{t_2(t_2+t_3)}{t_1(t_1+t_3)}r$ | $\frac{t_2+t_3}{t_1}r$            |
| $T_2$ | $\frac{t_2+t_3}{t_3}r$ | $\frac{t_2(t_2+t_3)}{t_3(t_1+t_3)}r$ | $\frac{t_2+t_3}{t_1+t_2}r$      |
| $T_3$ | $\frac{t_2}{t_3}r$  | $\frac{t_2(t_2+t_3)}{t_1 t_3}r$   | $\frac{t_2(t_2+t_3)}{t_1(t_1+t_2)}r$ |

By Theorem 4.7 and 4.1, we conclude that if $G_n$ contains a $C_3$, then $v_c$ of $G_n$ is either a vertex that satisfies the condition in Theorem 4.7 or a vertex $v_i$ on $C_3$ with $t_i = \underset{1 \leq j \leq 3}{\max} t_j$.

We can use the same approach to find the rumor center for any regular graph with a single cycle, Table 4.4 shows the ratio of $|M(v_i, T_j)|$ between any $1 \leq i, j \leq 4$ when $G_n$ contains a $C_4$. From Table 4.4, we have

$$\frac{|M(v_i, G_n)|}{|M(v_j, G_n)|} = \frac{t_i(t_{i-1}+t_{i+1})(n-t_j)}{t_j(t_{j-1}+t_{j+1})(n-t_i)},$$

if $v_i$ and $v_j$ are adjacent to each other.

**Table 4.4:** Table of $|M(v_i, T_i)|$, when $G_n$ contains a $C_4$

|  | $|M(v_1,T_i)|$ | $|M(v_2,T_i)|$ |
|---|---|---|
| $T_1$ | $r$ | $\frac{(t_2+t_3+t_4)(t_2+t_3)t_2}{(t_1+t_3+t_4)(t_1+t_4)t_1}r$ |
| $T_2$ | $\frac{(t_2+t_3+t_4)(t_2+t_3)}{(t_3+t_4)t_3}r$ | $\frac{(t_2+t_3+t_4)(t_2+t_3)t_2}{(t_1+t_3+t_4)(t_3+t_4)t_3}r$ |
| $T_3$ | $\frac{(t_2+t_3+t_4)t_2}{t_3t_4}r$ | $\frac{(t_2+t_3+t_4)(t_2+t_3)t_2}{(t_1+t_4)t_3t_4}r$ |
| $T_4$ | $\frac{(t_2+t_3)t_2}{(t_3+t_4)t_4}r$ | $\frac{(t_2+t_3+t_4)(t_2+t_3)t_2}{(t_3+t_4)t_1t_4}r$ |
|  | $|M(v_3,T_i)|$ | $|M(v_4,T_i)|$ |
| $T_1$ | $\frac{(t_2+t_3+t_4)(t_2+t_3)}{(t_1+t_4)t_1}$ | $\frac{t_2+t_3+t_4}{t_1}r$ |
| $T_2$ | $\frac{(t_2+t_3+t_4)(t_2+t_3)}{(t_1+t_2+t_4)(t_1+t_2)}r$ | $\frac{(t_2+t_3+t_4)(t_2+t_3)}{(t_1+t_2)t_3}r$ |
| $T_3$ | $\frac{(t_2+t_3+t_4)(t_2+t_3)t_2}{(t_1+t_2+t_4)(t_1+t_4)t_4}r$ | $\frac{(t_2+t_3+t_4)t_2}{(t_1+t_2+t_3)t_3}r$ |
| $T_4$ | $\frac{(t_2+t_3+t_4)(t_2+t_3)t_2}{(t_1+t_2)t_1t_4}r$ | $\frac{(t_2+t_3+t_4)(t_2+t_3)t_2}{(t_1+t_2+t_3)(t_1+t_2)t_1}r$ |

Lastly, we characterize the location of the maximum likelihood estimator of the source on regular pseudo-trees by combining Lemma 4.3 and Theorem 4.7.

**Theorem 4.8.** Let $G$ and $G_n$ be defined as in Theorem 4.7. The optimal solution to (3.2) is either on the path from the epidemic center of $G_n$ to the cycle or on the cycle.

As a remark, Theorem 4.8 is a combination of Theorem 4.7 and Theorem 4.4. Besides, Theorem 4.8 generalizes the results in [161], [162].

## 4.4 Conclusions and Remarks

In this section, we characterized the impact of the boundary effect and a single cycle on a regular graph and reformulated the maximum likelihood estimation problem in a more general setting. Note that the impact on the likelihood caused by a single cycle is similar to the one caused by the graph boundary. Hence, a cycle in a graph can be treated as a "end vertex" in the inner part (not on the graph boundary) of the graph. On the other hand, an end vertex can be treated as a size 1 cycle on the graph boundary. From (4.3) and (4.12), we can observe that the likelihood of any given vertex is no longer proportional to its rumor centrality. Moreover, we showed that the exact computation of $m_v^{v_e}(G_n, k)$ in (4.3) and $m_v^{v_l}(G_n, k)$ in (4.12) is impractical [14], [157].

Besides the combinatorial approach in this section, there are many other interesting approaches to the contagion source detection problem for general graphs. Another widely used and promising approach is to adopt a probabilistic method to design computationally efficient algorithms. Some of these methods are the breadth-first search tree heuristic in [131], [132], the probabilistic sampling of the spanning tree of $G_n$ (e.g., the Gromov matrix approach in [68]), stochastic approximation techniques like the Markov chain Monte Carlo sampling with statistical confidence in [30], [85] and the class of estimators that is oblivious to the underlying stochastic spreading process [75], [89]. We expect that a jointly combinatorial approach with probabilistic reasoning can offer meaningful approaches to tackle the difficult problem of contagion source detection in general graphs. In fact, the next section will illustrate an application of *algebraic combinatorics* to analyze asymptotically large random graphs for this problem.

# 5

---

# Asymptotic Analysis and Pólya Urn Models

---

Epidemic and infodemic outbreaks can involve massive graphs representing complex networks of interactions between individuals, which makes it difficult to identify the origin of the outbreak. Asymptotic analysis provides a powerful tool for analyzing the behavior of algorithms in such large networks and deriving insights into the performance of source detection algorithm as the size of the network grows. In particular, we analyze the limiting performance of the maximum likelihood estimator when $n$ approaches infinity. The fundamental limits to performance are established using analytic combinatorics and Pólya Urn Model, which may be of interest in their own right. Asymptotic analysis also reveals interesting phase transition phenomenon such as the detection probability converging to some limits depending on the graph topological features.

## 5.1   Asymptotic Analysis

In this section, we assume that the underlying graph $G$ is a $d$-regular tree with an infinite number of nodes (cf. Section 2.2.1). We derived the probability $P(\hat{v}|G_n)$ for $G_n$ as $n$ grows infinitely large over $G$. The probability $P(\hat{v}|G_n)$ can be obtained by the basic counting method when $d = 2, 3$. But when $d \geq 4$, we formulate the original problem as an increasing tree counting problem and apply results in analytical combinatorics to compute $P(\hat{v}|G_n)$.

**Theorem 5.1.** Suppose $G$ is a 2-regular tree, and $G_n \subseteq G$ is an infected subtree of $G$ without end vertices, then

$$\lim_{n \longrightarrow \infty} P(\hat{v}|G_n) = 0.$$

*Proof.* This theorem can be immediately proved from the result in (2.5). □

In the following, we derive a closed-form formula for the detection probability in a $d$-regular tree, where $d > 2$. By applying Theorem 3.1, we define $A_d$ and $B_d$ as follows:

$$A_d = \{(a_1, a_2, ..., a_d)|0 \leq a_i \leq \frac{n}{2}, \sum_{i=1}^{d} a_i = n - 1\},$$

$$B_d = \{(b_1, b_2, ..., b_d)|b_i \in \mathbb{N} \cup \{0\}, \sum_{i=1}^{d} b_i = n - 1\}.$$

The above two sets correspond to all possible sequences of branch sizes of a vertex $v$. If the sequence of branch sizes of $v$ is in $A_d$, then we can conclude that $v$ is the rumor center. To compute $|A_d|$, we first define new sets $S_k = \{(s_1, s_2, ..., s_d)|s_k > \frac{n}{2}, 0 \leq s_j < \frac{n}{2}, \sum_{i=1}^{d} s_i = n - 1\}$, for $k = 1, 2, \ldots, d$. Since $S_i \cap S_j = \phi$ for $i \neq j$, we have $|B_d| = \binom{n-1+d-1}{d-1}$, $|S_k| = \binom{d+\lceil \frac{n}{2} \rceil - 3}{d-1}$, and

$$|A_d| = |B_d| - \sum_{k=1}^{d} |S_k|$$

$$= \binom{n+d-2}{d-1} - d \cdot \binom{\lceil \frac{n}{2} \rceil + d - 3}{d-1}.$$

Given an infected subgraph $G_n \subseteq G$, and suppose $v$ is the source of $G_n$ with branch size $(t_{u_1}^v, t_{u_2}^v, ..., t_{u_d}^v)$, where $u_i$ is child of $v$ for $i = 1, 2, \ldots d$. According to (3.6), we first compute the spreading order in each branch of $v$. For each $u \in N(v)$, if $u$ is infected by $v$, then there are $\dfrac{-1}{d-3} \prod_{i=0}^{t_u^v}((d-2)(i-1)+1)$ possible spreading orders starting from $u$ in $T_u^v$, for $d \geq 4$. We shall discuss the case $d = 3$ separately in the next theorem since the fraction $\frac{-1}{d-3}$ is meaningless. Thus, given a sequence of branch size $(t_{u_1}^v, t_{u_2}^v, ..., t_{u_d}^v)$ of $v$, the number of spreading orders is

$$\frac{(n-1)!}{t_{u_1}^v! \cdot t_{u_2}^v! \cdot \ldots \cdot t_{u_{d(v)}}^v!} \cdot \prod_{k=1}^{d} [\frac{-1}{d-3} \prod_{i=0}^{t_{u_k}^v}((d-2)(i-1)+1)], \text{ for } d \geq 4.$$

The corrected detection occurs when the rumor center equals to source, so it occurs when $(t_{u_1}^v, t_{u_2}^v, ..., t_{u_d}^v) \in A_d$. Then we have

$$P(\hat{v}|G_n) = \frac{(n-1)! \displaystyle\sum_{(t_{u_1}^v, t_{u_2}^v, ..., t_{u_d}^v) \in A_d} (\displaystyle\prod_{k=1}^{d} \frac{\frac{-1}{d-3} \prod_{i=0}^{t_{u_k}^v}((d-2)(i-1)+1)}{t_{u_k}^v!})}{(n-1)! \displaystyle\sum_{(t_{u_1}^v, t_{u_2}^v, ..., t_{u_d}^v) \in B_d} (\displaystyle\prod_{k=1}^{d} \frac{\frac{-1}{d-3} \prod_{i=0}^{t_{u_k}^v}((d-2)(i-1)+1)}{t_{u_k}^v!})},$$

$$\tag{5.1}$$

for $d \geq 4$.

**Theorem 5.2.** Let $G$ be a 3-regular tree, then

$$\lim_{n \to \infty} P(\hat{v}|G_n) = \frac{1}{4}.$$

*Proof.* Let $A_3$ and $B_3$ as defined above. We should avoid the case when $i = 0$, that is, one of the branch sizes of $v$ is 0. Let $A_3^1$ be the subset of $A_3$ with at least a zero in $(t_{u_1}^v, t_{u_2}^v, v_{u_3})$ and $A_3^2 = A_3 \setminus A_3^1$. Also, $B_3^1$ and $B_3^2$ are defined respectively. Let $z_d(i) = (d-2)(i-1)+1$. Then we have

$$P(\hat{v}|G_n) = \frac{\displaystyle\sum_{(t_{u_1}^v, t_{u_2}^v, t_{u_3}^v) \in A_3^1} (\displaystyle\prod_{k=1}^{2} \frac{\prod_{i=1}^{t_{u_k}^v}(z_d(i))}{t_{u_k}^v!}) + \displaystyle\sum_{(t_{u_1}^v, t_{u_2}^v, t_{u_3}^v) \in A_3^2} (\displaystyle\prod_{k=1}^{3} \frac{\prod_{i=1}^{t_{u_k}^v}(z_d(i))}{t_{u_k}^v!})}{\displaystyle\sum_{(t_{u_1}^v, t_{u_2}^v, t_{u_3}^v) \in B_3^1} (\displaystyle\prod_{k=1}^{2} \frac{\prod_{i=1}^{t_{u_k}^v}(z_d(i))}{t_{u_k}^v!}) + \displaystyle\sum_{(t_{u_1}^v, t_{u_2}^v, t_{u_3}^v) \in B_3^2} (\displaystyle\prod_{k=1}^{3} \frac{\prod_{i=1}^{t_{u_k}^v}(z_d(i))}{t_{u_k}^v!})}$$

$$= \frac{\displaystyle\sum_{(t_{u_1}^v, t_{u_2}^v, t_{u_3}^v) \in A_3^1} 1 + \displaystyle\sum_{(t_{u_1}^v, t_{u_2}^v, t_{u_3}^v) \in A_3^2} 1}{\displaystyle\sum_{(t_{u_1}^v, t_{u_2}^v, t_{u_3}^v) \in B_3^1} 1 + \displaystyle\sum_{(t_{u_1}^v, t_{u_2}^v, t_{u_3}^v) \in B_3^2} 1} = \frac{|A_3^1| + |A_3^2|}{|B_3^1| + |B_3^2|}.$$

When $n$ is an even number, we have

$$\frac{|A_3^1| + |A_3^2|}{|B_3^1| + |B_3^2|} = \frac{6 + \binom{n-2}{2} - 3 \cdot \binom{\lceil \frac{n}{2} \rceil - 2}{3 - 1}}{3 + 3(n-2) + \binom{n-2}{2}} = \frac{n^2 + 10n}{4n^2 + 4n}.$$

When $n$ is an odd number, we have

$$\frac{|A_3^1| + |A_3^2|}{|B_3^1| + |B_3^2|} = \frac{3 + \binom{n-2}{2} - 3 \cdot \binom{\lceil \frac{n}{2} \rceil - 2}{3 - 1}}{3 + 3(n-2) + \binom{n-2}{2}} = \frac{n^2 + 4n + 3}{4n^2 + 4n}.$$

Thus, $\lim_{n \longrightarrow \infty} P(\hat{v}|G_n) = \frac{1}{4}$.

$\square$

**Theorem 5.3.** Let $G$ be a $d$-regular tree, where $d > 2$, then

$$\lim_{n \longrightarrow \infty} P(\hat{v}|G_n) = 1 - \frac{d}{2} + \frac{d \cdot \Gamma\left(\frac{d}{d-2}\right)}{2^{\frac{d}{d-2}} \cdot \Gamma\left(\frac{1}{d-2}\right) \Gamma\left(\frac{d-1}{d-2}\right)}.$$

*Proof.* We consider rumor spreading on $d$-regular graphs. First, we fix some notations.

Let $\tilde{\mathcal{T}}_n$ denote the tree after the rumor has spread to $n$ nodes. We give the vertices labels from the set $\{1, 2, \ldots, n\}$ according to the first time a vertex learns the rumor. Thus, the vertex with label 1 is the source and it is the only node with outdegree $d$ whereas all other nodes have outdegree $d - 1$ (if $\tilde{\mathcal{T}}_n$ is drawn as a rooted tree in the usual way).

By the definition of an increasing tree, a permitted permutation of a rumor spreading on a tree is equivalent to an increasing tree where the rumor source is the root of the increasing tree. Note that the $d$ subtrees of the source are $(d-1)$-ary increasing trees. Set

$$T_n = \text{number of } (d-1)\text{-ary increasing trees}, \qquad T(z) = \sum_{n \geq 1} T_n \frac{z^n}{n!}.$$

We can apply Theorem 2.2 to derive the explicit form of $T(z)$, we have

$$T(z) = -1 + (1 - (d-2)z)^{-1/(d-2)}. \tag{5.2}$$

Next, set

$$\tilde{T}_n = \text{number of possible } \tilde{\mathcal{T}}_n \quad and \quad \tilde{T}(z) = \sum_{n \geq 1} 2 \tilde{T}_n \frac{z^n}{n!}.$$

Then,

$$\tilde{T}_n = \frac{1}{2}\prod_{i=1}^{n}\left[2 + (d-2)(i-1)\right] \quad and \quad \tilde{T}(z) = -1 + (1-(d-2)z)^{-2/(d-2)} \tag{5.3}$$

We are interested in the event that the source of $\tilde{\mathcal{T}}_n$ is the rumor center. Thus, we have

$$P(\hat{v}|G_n) = 1 - d \cdot P(\text{size of one branch of the source of } \tilde{\mathcal{T}}_n \geq n/2).$$

We fix one branch of the source of $\tilde{\mathcal{T}}_n$ say $t_1^v$ and denote its size by $I$. Let $j_i$ represent the size of the $i$-th subtree of $\tilde{\mathcal{T}}_n$, then we have

$$
\begin{aligned}
P(I = j) &= \frac{1}{\tilde{T}_n}\sum_{j+j_2+j_3+\ldots+j_d=n-1}\binom{n-1}{j, j_2, j_3, \ldots, j_d}T_j T_{j_2} T_{j_3}\ldots T_{j_d}\\
&= \frac{(n-1)! T_j}{j!\tilde{T}_n}\sum_{j_2+j_3+\ldots+j_d=n-1-j}\frac{T_{j_2}}{j_2!}\frac{T_{j_3}}{j_3!}\ldots\frac{T_{j_d}}{j_d!}\\
&= \frac{(n-1)! T_j}{j!\tilde{T}_n}[z^{n-1-j}](1 + T(z))^{d-1}\\
&= \frac{(n-1)! T_j}{j!\tilde{T}_n}[z^{n-1-j}](1 - (d-2)z)^{-\frac{d-1}{d-2}}\\
&= \frac{(n-1)! T_j}{j!\tilde{T}_n}(d-2)^{n-1-j}[z^{n-1-j}](1 - z)^{-\frac{d-1}{d-2}}\\
&= \frac{(n-1)! T_j}{j!\tilde{T}_n}(d-2)^{n-1-j}\frac{(n-1-j)^{\frac{1}{d-2}}}{\Gamma(\frac{d-1}{d-2})}. \tag{5.4}
\end{aligned}
$$

In the sequel, we apply Theorem 2.3 to describe the asymptotic behavior of $T_n$ and $\tilde{T}_n$ as $n$ goes to infinity.

**Asymptotics.** The following computation is based on $n \to \infty$.

First, applying Theorem 2.3 to (5.2) gives

$$T_n \sim \frac{n! \cdot n^{\frac{2-d}{d-1}}}{\Gamma(\frac{1}{d-1})}(d-1)^n \qquad (n \to \infty).$$

Similarly, applying Theorem 2.3 to (5.3) yields

$$\tilde{T}_n \sim \frac{n! \cdot n^{\frac{4-d}{d-2}}}{2\Gamma(\frac{2}{d-2})}(d-2)^n \qquad (n \to \infty).$$

By Theorem 3.1, we need to compute

$$\sum_{n/2 \leq j \leq n-1} P(I = j),$$

where $P(I = j)$ is given by (5.4). Therefore, we again use Theorem 2.3 and the expansions for $T_n$ and $\tilde{T}_n$ from above. Let $d' = d - 2$, we have

$$\sum_{n/2 \leq j \leq n-1} P(I = j)$$

$$\sim \frac{(n-1)!}{\tilde{T}_n} \sum_{n/2 \leq j \leq n-1} \frac{T_j}{j! \Gamma(\frac{d-1}{d'})} (d')^{n-1-j} (n-1-j)^{\frac{1}{d'}}$$

$$\sim \frac{(n-1)! \cdot 2\Gamma(\frac{2}{d'})}{n! \cdot n^{\frac{4-d}{d'}} (d')^n} \sum_{n/2 \leq j \leq n-1} \frac{\frac{j! \cdot j^{\frac{3-d}{d'}}}{\Gamma(\frac{1}{d'})} (d')^j}{j! \Gamma(\frac{d-1}{d'})} (d')^{n-1-j} (n-1-j)^{\frac{1}{d'}}$$

$$\sim \frac{2\Gamma\frac{2}{d'}}{(d') \cdot n^{\frac{2}{d'}} \Gamma(\frac{1}{d'}) \Gamma(\frac{d-1}{d'})} \sum_{n/2 \leq j \leq n-1} j^{\frac{1}{d'}-1} (n-1-j)^{\frac{1}{d'}}$$

$$\sim \frac{2\Gamma\frac{2}{d'}}{(d') \cdot n^{\frac{2}{d'}} \Gamma(\frac{1}{d'}) \Gamma(\frac{d-1}{d'})} \cdot n^{\frac{2}{d'}-1} \sum_{n/2 \leq j \leq n-1} \left(\frac{j}{n}\right)^{\frac{1}{d'}-1} \left(\frac{n-1-j}{n}\right)^{\frac{1}{d'}}$$

$$\sim \frac{2\Gamma(\frac{2}{d'})}{(d') \Gamma(\frac{1}{d'}) \Gamma(\frac{d-1}{d'})} \int_{1/2}^1 x^{\frac{1}{d'}-1} (1-x)^{\frac{1}{d'}} dx \qquad (n \to \infty).$$

$$(5.5)$$

**Lemma 5.4.** For $\alpha > 0$,

$$\int_{1/2}^1 x^{\alpha-1} (1-x)^\alpha \mathrm{d}x = \frac{1}{2} \left( B(\alpha, \alpha+1) - \frac{1}{\alpha 2^{2\alpha}} \right),$$

where $B(\cdot, \cdot)$ denotes the beta function.

*Proof.* First, observe that

$$B(\alpha, \alpha+1) = \int_0^1 x^{\alpha-1} (1-x)^\alpha \mathrm{d}x$$

$$= \int_0^{1/2} x^{\alpha-1} (1-x)^\alpha \mathrm{d}x + \int_{1/2}^1 x^{\alpha-1} (1-x)^\alpha \mathrm{d}x.$$

Now, call the first and second integral on the right-hand side $L$ and $R$, respectively. By integration by parts and substitution, we have

$$L = \frac{1}{\alpha} x^\alpha (1-x)^\alpha \Big|_0^{1/2} + R.$$

Thus,

$$R = \frac{1}{2}\left(B(\alpha, \alpha+1) - \frac{1}{\alpha 2^{2\alpha}}\right)$$

which is the claimed result. □

Finally, combining everything yields for the detection probability of the source in $d$-regular trees is the following limit

$$\lim_{n \longrightarrow \infty} P(\hat{v}|G_n) = 1 - d \cdot \sum_{n/2 \le j \le n-1} P(I = j)$$

$$= 1 - \frac{d}{2} + \frac{(d-2)\cdot\Gamma\left(\frac{d}{d-2}\right)}{2^{\frac{d}{d-2}}\cdot\Gamma\left(\frac{1}{d-2}\right)\Gamma\left(\frac{d-1}{d-2}\right)}$$

□

**Lemma 5.5.** Let $k^{(d)} = \lim\limits_{n \longrightarrow \infty} P(\hat{v}|G_n)$ on a $d$-regular tree $G$, then we have

$$\lim_{d\to\infty} k^{(d)} = 1 - \ln 2.$$

Moreover, we have

$$1/4 \le k^{(d)} < 1 - \ln 2,$$

for $d \ge 3$.

This lemma shows that the detection probability is bounded by $1/4$ and $1 - ln2$ for $d \ge 3$.

*Proof.* By Stirling's formula for the gamma function, we have

$$k^{(d)} = 1 - \ln 2 + \frac{\pi^2/12 - 2\ln 2 + \ln^2 2}{d} + \mathcal{O}\left(\frac{1}{d^2}\right). \qquad (5.6)$$

This implies the first claim.

As for the second claim, note that

$$\frac{\pi^2}{12} - 2\ln 2 + \ln^2 2 = -0.08337431\cdots < 0.$$

From this, the second claim follows for all $d$ large enough. In order to show that the claim holds for all $d \geq 3$, we make the constant in the error term of (5.6) explicit. For this, we use Taylor expansions with error terms. For instance, we have

$$\Gamma(x) = 1 - \gamma(x-1) + \left(\frac{\pi^2}{12} + \frac{\gamma^2}{2}\right)(x-1)^2 + E_1(x),$$

where $\gamma$ is Euler's constant and

$$\frac{1}{2^x} = \frac{1}{2} - \frac{\ln 2}{2}(x-1) + \frac{\ln^2 2}{4}(x-1)^2 + E_2(x).$$

The error terms are bounded by

$$|E_1(x)| \leq \left(\frac{\zeta(3)}{3} + \frac{\pi^2 \gamma}{12} + \frac{\gamma^3}{6}\right)(x-1)^3, \qquad |E_2(x)| \leq \frac{\ln^3 2}{12}(x-1)^3$$

for $x \in [1,3]$, where $\zeta(x)$ denotes the Riemann zeta function. Plugging this into the expression of $k^{(d)}$ from (5.6) and using again Taylor expansion (several times), we obtain the rough bound

$$\left| k^{(d)} - 1 + \ln 2 - \frac{\pi^2/12 - 2\ln 2 + \ln^2 2}{d} \right| \leq \frac{100}{d^2} \qquad (10 \leq d). \quad (5.7)$$

From this, the claimed upper bounds are deduced for $d \geq 1200$. One then easily checks that the upper bound also holds for all smaller $d$. Likewise, for the lower bound, we obtain from (5.7) that it holds for $d \geq 43$. We showed that the bound holds for all $d > 1200$ and the finite remaining cases can be checked by computing software. $\qquad\square$

## 5.2  Contagion Source Detection via Pólya Urn Models

In this section, we first show that the spreading process on regular trees is equivalent to the ball drawing process in the Pólya's urn model, a connection first utilized in [133] for asymptotic analysis. Then, we leverage the convergence property of Pólya's urn model to derive the exact detection probability in the more general finite regime (recovering the asymptotic result as a special case).

Recall that in Pólya's urn model: Initially, the urn contains $b_j$ balls of color $C_j$ $(1 \leq j \leq d)$; at each uniform drawing of a single ball, the ball is returned together with $\mathsf{m}$ balls of the same color; after $n$ draws, the number $X_j$ is the number of times that the balls of color $C_j$ are drawn. Then the joint distribution of $\{X_j, 1 \leq j \leq d\}$ is given by

$$\mathbf{P_G} \left[ \bigcap_{j=1}^{d} (X_j = x_j) \right] = \frac{n!}{x_1! x_2! \cdots x_d!} \frac{\prod_{j=1}^{d} b_j (b_j + \mathsf{m}) \cdots (b_j + (x_j - 1)\mathsf{m})}{b(b + \mathsf{m}) \cdots (b + (n-1)\mathsf{m})}, \tag{5.8}$$

where $b = \sum_{j=1}^{d} b_j$ and $\sum_{j=1}^{d} x_j = n$.

As $n \to \infty$, the limiting joint distribution of the ratios $\{X_j/n, 1 \leq j \leq d\}$ converges to the Dirichlet distribution with a density function given by

$$\lim_{n \to \infty} \mathbf{P_G} \left[ \bigcap_{j=1}^{d} \left( \frac{X_j}{n} = y_j \right) \right] = \frac{\Gamma(\alpha)}{\prod_{j=1}^{d} \Gamma(\alpha_j)} \prod_{j=1}^{d} y_j^{\alpha_j - 1}, \tag{5.9}$$

where $\alpha_j = b_j/\mathsf{m}$, $\alpha = \sum_{j=1}^{d} \alpha_j$ and $\sum_{j=1}^{d} y_j = 1$. Here, $\Gamma(\alpha)$ is the gamma function with parameter $\alpha$.

Besides, the marginal distribution of $X_1$ is

$$\mathbf{P_G} (X_1 = x_1) = \frac{n!}{x_1! x_2'!} \frac{\prod_{j=1}^{2} b_j' (b_j' + \mathsf{m}) \cdots (b_j' + (x_j' - 1)\mathsf{m})}{b(b + \mathsf{m}) \cdots (b + (n-1)\mathsf{m})}, \tag{5.10}$$

where $b_1' = b_1$, $b_2' = b - b_1$, $x_1' = x_1$ and $x_2' = n - x_1$. In fact, it can be seen as a special case of the Pólya's urn model with two colors.

Due to the martingale property stated in Theorem 2.1, as $n \to \infty$, the limiting marginal distribution of the ratio $X_1/n$ converges to the Beta distribution with a density function given by

$$\lim_{n \to \infty} \mathbf{P_G} \left( \frac{X_1}{n} = y_1 \right) = \frac{\Gamma(\alpha_1' + \alpha_2')}{\Gamma(\alpha_1')\Gamma(\alpha_2')} y_1^{\alpha_1' - 1} (1 - y_1)^{\alpha_2' - 1}, \tag{5.11}$$

where $\alpha_1' = \alpha_1$ and $\alpha_2' = \alpha - \alpha_1$. The cumulative distribution function of the Beta distribution is

$$I_x(\alpha_1', \alpha_2') := \frac{\Gamma(\alpha_1' + \alpha_2')}{\Gamma(\alpha_1')\Gamma(\alpha_2')} \int_0^x y^{\alpha_1' - 1} (1 - y)^{\alpha_2' - 1} dy, \tag{5.12}$$

for all $x \in [0, 1]$. In particular, $I_x(\alpha_1', \alpha_2')$ is called the incomplete Beta function with parameters $\alpha_1'$ and $\alpha_2'$. Note that the (5.12) has the same structure as the last line of (5.5).

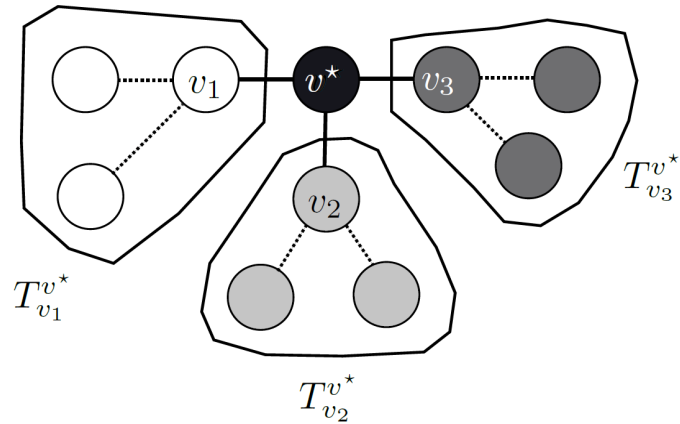

**Figure 5.1:** The black node $v^\star$ is the source and there are three colors ($d = 3$) of nodes connecting to $v^\star$. We have $x_1 = |T_{v_1}^{v^\star}|$, $x_2 = |T_{v_2}^{v^\star}|$, and $x_3 = |T_{v_3}^{v^\star}|$.

We are interested in $I_{1/2}(\alpha'_1, \alpha'_2)$ with parameters $\alpha'_1 = 1/(d-2)$ and $\alpha'_2 = (d-1)/(d-2)$, where $d$ is the node degree of a regular tree and $d \geq 3$. For the case when $d = 2$, we can simply apply the result in (2.5) to compute the desired limit.

### 5.2.1   Equivalence to the Pólya's urn model

Next, we show that the spreading on regular trees can be modeled by ball drawing in the Pólya's urn model, whose well-known distributions can be used to obtain a similar result in Theorem 5.3.

For a contagion source $v^\star$ with $d$ neighboring nodes $v_1, \ldots, v_d$, let $T_{v_j}^{v^\star}$ ($1 \leq j \leq d$) be the subtree rooted at node $v_j$ with node $v^\star$ as the source in $G_n$, and define a random variable $X_j$ as the number of nodes in $T_{v_j}^{v^\star}$; e.g., see Figure 5.1. We denote the set of susceptible neighbors of infected nodes as the *contagion boundary*, i.e., each vertex in the contagion boundary has a chance to be infected in the next time period. In the rumor spreading process, nodes in $G_n$ are infected sequentially, and thus we have the following: initially, $v^\star$ has one neighbor in each subtree $T_{v_j}^{v^\star}$ ($1 \leq j \leq d$) that belongs to the contagion boundary; after one of those nodes is infected, it introduces $d-1$ new nodes into the contagion boundary; finally, $n-1$ nodes are infected besides $v^\star$. Due to the memoryless and i.i.d. properties of exponentially distributed infection times $\{\tau_{ij}, (i, j) \in E\}$, in each step, the infected node is uniformly selected from the contagion boundary.

Now, the resulting infection $G_n$ with $X_j$ nodes in $T_{v_j}^{v^\star}$ $(1 \leq j \leq d)$ can be constructed in an equivalent way by the Pólya's urn model [74, Chapter 4]: initially, the urn has one ball for each color $C_j$; at each uniform drawing of a single ball, the ball is returned together with $\mathsf{m} = d - 2$ additional balls of the same color; after $n - 1$ draws, $X_j$ is the number of times that the balls of color $C_j$ are drawn.

Therefore, in the rumor spreading process, if we assume $v^\star$ to be the rumor source with $d$ neighbors $v_1, \ldots, v_d$ and observe $n$ infected nodes $G_n$ with $X_j$ nodes in each subtree $T_{v_j}^{v^\star}$ $(1 \leq j \leq d)$, then from (5.8), the joint distribution of $\{X_j, 1 \leq j \leq d\}$ is given by

$$\mathbf{P_G}\left[\bigcap_{j=1}^d (X_j = x_j)\right] = \frac{(n-1)!}{x_1! x_2! \cdots x_d!} \frac{\prod_{j=1}^d 1(1+\mathsf{m})\cdots(1+(x_j-1)\mathsf{m})}{d(d+\mathsf{m})\cdots(d+(n-2)\mathsf{m})}, \tag{5.13}$$

where $\sum_{j=1}^d x_j = n - 1$.

Besides, the marginal distribution of $X_1$ is

$$\mathbf{P_G}(X_1 = x_1) = \binom{n-1}{x_1} \frac{\prod_{j=1}^2 b_j'(b_j'+\mathsf{m})\cdots(b_j'+(x_j'-1)\mathsf{m})}{d(d+\mathsf{m})\cdots(d+(n-2)\mathsf{m})}, \tag{5.14}$$

where $b_1' = 1$, $b_2' = d - 1$, $x_1' = x_1$ and $x_2' = n - x_1 - 1$.

We are also interested in the limiting marginal distribution of the ratio $X_1/n$ as $n \to \infty$. From (5.11), we have

$$\lim_{n \to \infty} \mathbf{P_G}\left(\frac{X_1}{n} = y\right) = \frac{\Gamma(\alpha+\beta)}{\Gamma(\alpha)\Gamma(\beta)} y^{\alpha-1}(1-y)^{\beta-1}, \tag{5.15}$$

where $\alpha = 1/(d-2)$, $\beta = (d-1)/(d-2)$. We also refer the readers to [168] for a probabilistic characterization of the boundary of the infection graph that takes the effect of time to infection into account.

**Other Asymptotic Results on Infinite Trees** The method of generating function used in Theorem 5.3 can be further generalized to other increasing trees, such as recursive trees, $d$-ary trees, and plain-oriented recursive trees. For example, if we assume that the underlying network $G$ is an infinite $d$-ary tree, then we only need to change the generating function (5.2) to $T(z) = -1 + (1 - (d-1)z)^{-1/(d-1)}$. For more details of detection probability on increasing trees, please refer to [53].

Recall that in Theorem 3.3, we have shown that the rumor center is equivalent to the centroid on tree graphs. Hence, the above results can be seen as an analysis of the asymptotic behavior of the tree centroid. In addition to the above-mentioned results in the convergence of the detection probability, there are other results focused on the asymptotic behavior of the tree centroid. The Pólya's urn model is applied in [85] to show that the set of vertices with the maximum number of spreading orders form a confidence set on a $d$-regular tree. As the tree size $n$ goes to infinity, the probability of the source vertex not in the confidence set approaches zero, which generalizes the result on uniform attachment trees derived in [16]. We refer the readers to [72], [73], [85] for more details on the relationship between network centrality and the random growth process of infinitely large regular graphs.

## 5.3 Conclusions and Remarks

In this section, we showed that the correct detection probability on regular tree networks could be computed by constructing increasing trees for the spreading process. Note that we have introduced two more different approaches to compute the correct detection probability on regular trees so far. For the case with graph boundary, we characterize the position of the maximum-likelihood estimator on a regular tree with a single end vertex. We showed that the likelihood of any given vertex could be exactly computed in some special graphs such as line graphs and broom graphs. Moreover, if the distance from the end vertex to the rumor center is one then we showed that the end vertex is exactly the maximum-likelihood estimator by considering their likelihood ratio. For the case of pseudo-trees, we present a simple algorithm framework to find the rumor center on pseudo-trees and characterize the position of the maximum-likelihood estimator on regular pseudo-trees. Note that the approach we have used in pseudo-trees can be extended to a cactus graph, i.e., a class of graphs with multiple disjoint cycles.

We refer the reader to [37], [53] on extensions of computation of the correct detection probability on different increasing trees through a generating function approach (so-called *generatingfunctionology*) in [48] and other extensions of considering the specific class of graphs such as star graphs [84]. In the next section, we present practical algorithms to efficiently compute the maximum likelihood estimator and provide simulation results on different network topologies.

# 6

# Applications to COVID-19 Pandemic and Infodemics

In this section, we present the framework known as "*network centrality as statistical inference*" and demonstrate its applicability in the design of digital contact tracing algorithms that are relevant in the context of the COVID-19 pandemic. In addition, we showcase another potential application of this framework in detecting the source of rumors circulating within Twitter's online social networks. Through numerical examples, we aim to illustrate the practical implementation of this approach by adapting algorithms presented in previous sections to effectively address source attribution in various contagion problems.

## 6.1 Epidemic Centrality for Digital Contact Tracing

Upon the initial onslaught of a pandemic (e.g., COVID-19 pandemic or the future 'Disease X' pandemic), digital contact tracing will generate a tremendous amount of networked data that are essentially massive graphs modulated by stochastic processes. The main scientific question will be how best to analyze and leverage these networked data to develop effective contact-tracing strategies? There are several challenging unsolved problems in digital contact tracing [17], [81], [91], [139]. First, what is the fundamental relationship between infectiousness and the agility of contact tracing? How to quadruple the speed of contact tracing whenever infectiousness exceeds certain thresholds? What are efficient and scalable contact tracing strategies to find Patient Zero or superspreaders?

Tracing sources of spreading (i.e., *backward contact tracing*), as had been used in Japan and Australia during the COVID-19 pandemic, has proven effective as going backward can pick up infections that might otherwise be missed at superspreading events [12]. From an epidemiology perspective, all the infected persons in a contact tracing graph are potential candidates for tracking purposes as well as identification of Patient Zero or superspreaders [81], [139]. Designing effective backward contact tracing is part of robust predictive healthcare analytics that can prevent recurrent outbreaks and breakthrough transmissions as well as provide early warning of the arrival of future pandemics [91].

Let us generalize the results in the previous sections to address the case when $G_n$ is a general graph that will be applied to the digital contact tracing of superspreaders. Recall that computing the $|M(v, G_n)|$ of vertex $v$ in $G_n$ is crucial to solving the maximum-likelihood estimation problem. When $G_n$ is a tree, Lemma 4.3 reveals how the epidemic centrality connects to the distance centrality. Recall that, for the case when $G_n$ is a general graph, we call $|M(v, G_n)|$ the *epidemic centrality* of $v$, and $v$ is a *epidemic center* of $G_n$ if $|M(v, G_n)| = \max\limits_{v_i \in G_n} |M(v_i, G_n)|$. In the following, we present two message-passing algorithms to compute an approximation to the epidemic centrality based on the results in the previous sections. We also evaluate the performance for tracing the contagion source using real-world data.

### 6.1.1 Message-passing Algorithm for Trees with Multiple End Vertices

Let us describe a message-passing algorithm to find $\hat{v}$ on the finite regular tree $G$ by leveraging the key insights derived in Section 4. We summarize these features as follows:

1. If there is only a single end vertex $v_e$ in $G_n$, then $\hat{v}$ is located on the path from $v_c$ to $v_e$.

2. If $G_n = G$, then for all $v_i \in G_n$, $P(G_n|v_i) = 1/n$.

3. If $G_n$ has $q$ end vertices, then there exists an $n'$ such that, if $n > n'$, then $P(G_n|v_c) > \max_{1 \le i \le k} \{P(G_n|v_{e_i})\}$. Furthermore, $n'$ increases as $q$ increases.

4. If two vertices $v_1$ and $v_2$ are on the symmetric position of $G_n$, then $P(G_n|v_1) = P(G_n|v_2)$. For example, $v_3$ and $v_4$ are topologically symmetric in Figure 4.1.

In particular, Feature 1 is the optimality result pertaining to the decomposition of $G_n$ into subtrees to narrow the search for $\hat{v}$. The subtree $t_{ML}$ in $G_n$ corresponds to first finding the decomposed subtree containing the rumor center and the likelihood estimate needed for Theorem 4.4 to apply. Then, Features 3 and 4 identify $\hat{v}$ on a subtree $t_{ML}$ of $G_n$ as Theorem 4.4 only pinpoints the relative position of $\hat{v}$.

---

**Algorithm 4** Message-passing algorithm to compute $\hat{v}$ for $G_n$ with multiple end vertices [165]

---

Input: $G_n$, $\kappa = \{\}$

Step 1: Compute rumor center $v_c$ of $G_n$.

Step 2: Choose $v_c$ as the root of a tree and use a message-passing algorithm to count the number of end vertices on each branch of this rooted tree.

Step 3: Starting from $v_c$, and at each hop choose the child with the maximum number of end vertices (if there were more children with the same maximal number of end vertices, then choose all of them). This tree traversal yields a subtree $t_{ML}$ rooted at $v_c$.

**Output:** $\kappa = \{\text{parent vertices of leaves of } t_{ML}, v_c\}$

---

Algorithm 4 first finds the rumor center of $G_n$, and then determines the number of end vertices corresponding to each branch of the rumor center $v_c$. The final step is to collect vertices on the subtree where $\hat{v}$ is, and this leads to a subtree of $G_n$ denoted as $t_{ML}$. Observe that each step requires $O(n)$ computational time complexity. Observe that $t_{ML}$ in a graph with multiple end vertices is akin to the *path* from the rumor center to the end vertex in a rumor graph with a single end vertex in Section 4.1. Finally, we obtain a set $\kappa$ containing the parent vertices of the leaves of $t_{ML}$ and $v_c$.

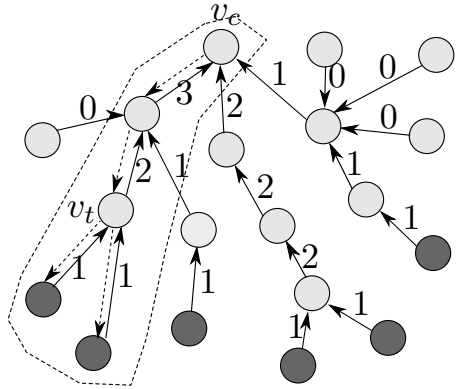

**Figure 6.1:** An illustration of how Algorithm 4 works on a tree graph rooted at $v_c$ with six end vertices (more shaded). Observed that $v_c$ branches out to three subtrees. Here, $t_{ML}$ is the subtree containing the five vertices within the dotted line. The numerical value on the edge indicates the message containing the number of end vertices being counted.

Now, let us use the example in Figure 6.1 to illustrate how Algorithm 4 runs. Let $G_{19}$ be the network in Figure 6.1 with the six end vertices depicted as more shaded. Suppose $v_c$ is determined by the end of Step 1. Then, Step 2 enumerates the number of end vertices at each branch of the subtrees connected to $v_c$, and these numbers are then passed iteratively from the leaves to $v_c$. These messages correspond to the numerical value on the edges in Figure 6.1. The message in Step 2 is an *upward* (leaf-to-root) message. Step 3 is a message passing procedure from $v_c$ back to the leaves, which is a *downward* message, and the message is the maximum number of end vertices in each branch. For example, the message from $v_c$ to $child(v_c)$ is $\max\{1, 2, 3\}$ which is 3. Lastly, the second part of Step 3 collects those vertices whose *upward message = downward message*. For example, the left-hand-side child of $v_c$ is first added to $t_{ML}$, and then $v_t$ is added to $t_{ML}$, and finally, the two leaves on the left-hand-side is added to $t_{ML}$. Observe that $t_{ML}$ must be connected.

**Simulation Results for Finite $d$-regular Tree Networks**

We simulate the rumor spreading in the degree regular tree network $G$ for $d = 3, 4, 5, 6$ with $|G| = 1000$ and $|G_n| = 100$. For each $d$, we simulate a thousand times the spread of a rumor on $G$ by picking $v^\star$ uniformly on $G$, and compare the average performance of Algorithm 4 and a naive heuristic that simply uses the rumor centrality approach in [131]. To fairly compare these two algorithms, when Algorithm 4 yields a set $\kappa$ with $|\kappa|$ vertices, then the naive heuristic finds a set of $|\kappa|$ vertices having the top $|\kappa|$ maximum $|M(v, G_n)|$ for all $v$ of $G_n$. Obviously, the size of the solution set $\kappa$ depends on the topology of $G_n$ in each run of the simulation and thus is not a constant in general over the thousand times. To quantify the performance of these two algorithms, let us define the error function of a vertex set $\eta$:

$$\text{error}(\eta) = min\{d(v, v^\star)|\forall v \in \eta\}.$$

This is simply the smallest number of hops between $v^\star$ and the nearest vertex in the set $\eta$. Figure 6.2 shows the distribution of these error hops for both algorithms when the underlying graph $G$ is 4-regular. This illustrates that Algorithm 4 can make a good guess for $P(\text{error}(\kappa) \leq 1 \text{ hop}) > 0.70$ in most cases, but there are occasions when the error is large. Table 6.1 shows the average of the error (number of hops) between the estimate and $v^\star$ for a thousand simulation runs. We can observe that the average error decreases as $d$ grows. The reason is that the number of infected vertices is fixed, and so as $d$ becomes larger, the diameter of the graph becomes smaller, and moreover, the Top-$k$ heuristic always chooses the set of vertices in the "center" of $G_n$. Hence, the average error decreases.

**Simulation Results for Finite General Tree Networks**

In this section, we evaluate the performance of Algorithm 4 for a finite general tree graph. In particular, the underlying graph $G$ is a tree satisfying the condition, that the degree of each vertex is less than or equal to $d_m$, where $d_m$ is a fixed positive integer. The construction of $G$ starts with a single vertex $v_1$, and we then randomly pick an integer, say $i$, from 0 to $d_m$ to be the degree of $v_1$, and then assign $v_2$ to $v_{i+1}$ to be the neighborhood of $v_1$. We recursively apply these steps until $G$ has one thousand vertices. The maximum degree in $G$ will be less than or equal to $d_m + 1$. The spreading model used is the same as in the previous simulation. We simulate the rumor spreading hundreds of times, particularly noting that $G$ is randomly generated and thus in each simulation $G$ is different. Note that, in the $d$-regular graph simulation, $G$ is however always the same. From Figure 6.2, we can observe that the error distribution is similar to the regular tree case in Figure 6.2, but with a smaller 1-hop error. The average error is roughly 1-hop larger than the regular case. The number of vertices in $\kappa$ is surprisingly small as compared to the regular case. Moreover, $|\kappa|$ is decreasing as $d_m$ increases.

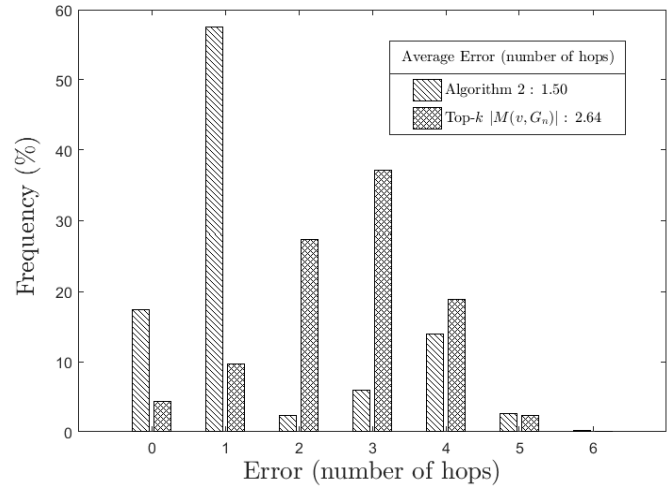

**Figure 6.2:** Comparing the error distribution (in the number of hops) between Algorithm 4 and the top-$k$ algorithm when $G$ is a 4-regular finite tree.

**Table 6.1:** Average error (in terms of number of hops) comparing Algorithm 4 and Top-$k$ Algorithm when $G$ is a $d$-regular graph, for $d = 3, 4, 5, 6$.

| $d$ | $|\kappa|$ | Algorithm 4 | Top-$k$ Algorithm |
|---|---|---|---|
| 3 | 6.34 | 1.44 | 3.26 |
| 4 | 5.65 | 1.50 | 2.64 |
| 5 | 4.05 | 1.48 | 2.36 |
| 6 | 3.72 | 1.40 | 2.32 |

### 6.1.2 Statistical Distance Based Algorithm for Finite Degree Regular Graph with Cycles

In this section, we present a statistical distance-based algorithm to solve the contagion source detection problem on a finite-degree regular graph with cycles. From Theorem 4.4 and 4.8, we can deduce that the likelihood of a vertex is greater if its distance to those end vertices and cycles is smaller. Hence, the maximum likelihood estimator should lie on the smallest induced subgraph containing three specific vertices, which are the rumor center, the vertex closest to all cycles, and the vertex closest to all end vertices. Note that an end vertex can be treated mathematically as an size-one cycle, hence we combine the boundary effect and the cycle effect together in our algorithm.

**Definition 6.1.** We say a cycle is a *minimum cycle* if there is no path between any two non-consecutive cycle vertices except the path along the cycle.

Since a vertex $v$ can be contained in multiple cycles with different sizes, we only take the minimum cycle that contains $v$ into consideration. Let $\mathcal{C}(v)$ denote the size of the minimum cycle containing $v$. If $v$ is not in any cycle and $deg(v) > 1$, then we set $\mathcal{C}(v) = \infty$, otherwise $\mathcal{C}(v) = 1$. Note that when $deg(v) = 1$, $v$ is regarded as a size 1 cycle.

Based on Lemma 4.3, Theorem 4.4 and Theorem 4.8, we heuristically define the weight $w_v$ of a vertex $v$ as

$$w_v = \frac{\mathcal{C}(v)}{\mathcal{C}(v) + 1}. \tag{6.1}$$

Since we define the distance center to be the vertex with minimum distance centrality (cf. Equation (2.8)), we can design a weight such that the location of the maximum likelihood estimator tends to close to vertices with "small weights". This is also motivated by the fact that the likelihood of a vertex $v$ being the source is greater if $v$ has a larger epidemic centrality and is closer to those irregular vertices or cycles (cf. Theorem 4.4 and Theorem 4.8). We shall call this distance-based centrality the *statistical distance centrality* (SDC) denoted as $\mathsf{SDC}(v, G)$ for $v \in G$. We define $\mathsf{SDC}(v, G)$ as the weighted-distance sum from $v$ to all the other vertices in $G$ (cf. Definition 6.2). Furthermore, the definition of $w_u$ reveals that the irregular effect caused by a small cycle is greater than that caused by a large cycle which can be observed from Table 4.2 and Equation (4.12). This definition implies that a vertex within a smaller cycle has a smaller weight which contributes "more" to $\mathsf{SDC}(v, G)$ while a vertex not in any cycle has $weight = 1$ which contributes "less" to $\mathsf{SDC}(v, G)$. Figure 6.3 illustrates such an example of a regular graph $G$ and the infected subgraph $G_{10}$ containing two different-sizes cycles, say $C_3$ and $C_4$. Note that, $v_{10}$ and $v_8$ have the same epidemic centrality, however, we have $P(G_{10}|v_{10}) > P(G_{10}|v_8)$ since $v_{10}$ is closer to the smaller cycle than $v_8$.

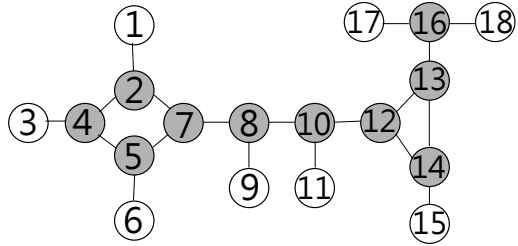

**Figure 6.3:** The infected subgraph $G_{10}$ contains ten nodes colored in grey. The epidemic centralities of $v_8$ and $v_{10}$ are the same, however, $v_{10}$ is the maximum-likelihood estimator of the true source in $G_{10}$. Since the irregular effect caused by a small cycle is greater than that of a large cycle.

**Definition 6.2.** Given a $d$-regular graph $G$ and vertex $v$ of $G$, we define the *statistical distance centrality* of $v$, $\mathsf{SDC}(v, G)$ as the summation of the weighted distance from $v$ to all other vertices in $G$. Hence, the statistical distance centrality of $v$ in $G$ is defined by

$$\mathsf{SDC}(v, G) = \sum_{u \in G} w_u \cdot d(v, u). \qquad (6.2)$$

The vertex $v_s$ with the minimum value for $\mathsf{SDC}(v, G)$ is called the *statistical distance center*.

Algorithm 5 is based on the idea of message passing. Let $lv(v)$ denote the level of $v$ in a BFS tree. In Step 2, for a given root $v_r$, we start a message-passing procedure in a BFS traversal to send a downward message containing level information to other vertices in the BFS tree. Upon receipt of this information, each leaf $v_l$ sends back an upward message containing $w_{v_l} \cdot lv(v_l)$ to its parent. Each internal vertex $v_{in}$

---

**Algorithm 5** Statistical Distance-based Contact Tracing (SCT)[165]

---

**Input:** $G_n$

Step 1: For each vertex $v$, compute the size $\mathcal{C}(v)$ of the minimum cycle containing $v$, and set $w_v = \frac{\mathcal{C}(v)}{\mathcal{C}(v)+1}$.

Step 2: For each vertex $v$, compute $\mathsf{SDC}(v, G_n)$.

Step 3: Let $\hat{v} = \underset{v \in G_n}{\operatorname{argmin}} \, \mathsf{SDC}(v, G_n)$.

---

sends an upward message, containing the summation of all message from its children plus $w_{v_{in}} \cdot lv(v_{in})$, to its parent.

In the following, we provide a time complexity analysis of Algorithm 5. For Step 1, the worst case time complexity is $O(|\mathcal{C}_{min}| \cdot |E(G_n)|)$ [146], where $|\mathcal{C}_{min}|$ is the number of all minimum cycles in $G_n$. Since the underlying graph is $d$-regular, each vertex in $G_n$ is contained in at most $d$ minimum cycles which implies $\mathcal{C}_{min} \leq d \cdot n$. The worst case time complexity for the Step 2 in the Algorithm 5 is $O(n^3)$, since the $\mathsf{BFS}$ traversal for each vertex takes $O(n + |E(G_n)|)$. Hence, the worst case time complexity of Algorithm 5 is $O(d \cdot n^3)$. In comparison with the $\mathsf{BFS}$ heuristic approach in [132], it applies the $\mathsf{BFS}$ traversal for each vertex and computes their epidemic centrality which ends up with worst time complexity $O(n^3)$.

**Simulation Results for Synthetic Regular Graphs with Cycles**

We provide simulation results on two finite regular graphs with cycles. The first simulation is conducted on a finite-size grid graph which is a 4-regular graph except for vertices on the boundary. The second simulation is conducted on a *circulant graph*. In our simulation, we pick the statistical distance center to be $\hat{v}$. We assume that the boundary and cycle effects dominate the rumor centrality. Furthermore, we simply set $w_{h_k} = \frac{1}{|C_{h_k}|}$ for each minimum cycle $C_{h_k}$.

**Grid Graph**    Disease spreading on grid graphs is often considered under different spreading rules and models [7]. Hence, we select grid graphs to be one of the testing synthetic networks. Simulation results are in Table 6.2 and one of the error distributions is in Figure 6.4. We can observe from Table 6.2 that the statistical distance based algorithm outperforms the BFS heuristic in [131]. Moreover, the average error increases as the number of end vertices are increasing which again reveals the fact shown in Figure 4.8 that the likelihood is evened out to those end vertices.

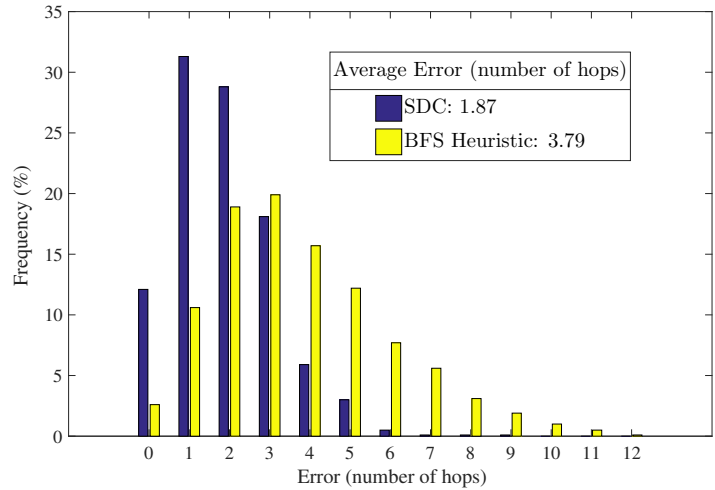

**Figure 6.4:** Comparing the error distribution (in the number of hops) between the SDC algorithm and the BFS heuristic [131] in a finite grid graph with $|G| = 10000$ and $|G_n| = 150$. In particular, the rate of the correct detection, i.e., $error = 0$, is 12.1% for the SDC and 2.6% for the BFS heuristic.

**Table 6.2:** Average error (in terms of number of hops) comparing SDC and BFS heuristic in [131] when $G$ is a $100 \times 100$ grid graph with different sizes of $G_n$.

| n | $|C_4|$ | $|v_e|$ | SDC | BFS heuristic [131] |
|-----|-------|-------|------|---------------------|
| 150 | 85.5  | 3.5   | 1.87 | 3.79                |
| 300 | 199.5 | 7.5   | 2.33 | 6.11                |
| 500 | 364.2 | 12.2  | 3.14 | 8.37                |
| 800 | 625.0 | 19.1  | 4.23 | 11.71               |

**Circulant Graph**  A circulant graph $G = (N, \mathsf{S})$ with $N$ vertices is a class of graphs that can be defined by its vertex set $V(G)$ and a set $\mathsf{S}$ of integers. The edge set is defined by $E(G) = \{(v_i, v_j)| \text{ if } |i - j| \in \mathsf{S}\}$. Hence, a circulant graph $G = (N, \mathsf{S})$ is a $|\mathsf{S}|$-regular graph. Note that a circulant graph $G$ is connected if and only if $\mathsf{S}$ generates the integer group $Z_N$ and we only consider the connected circulant graph. In the simulation, we fix $N$ and $n$ and randomly choose integers from the interval $[1, n/2]$ to form the set $\mathbf{S}$. Simulation results are summarized in Table 6.3 and one of the error distributions is shown in Figure 6.5.

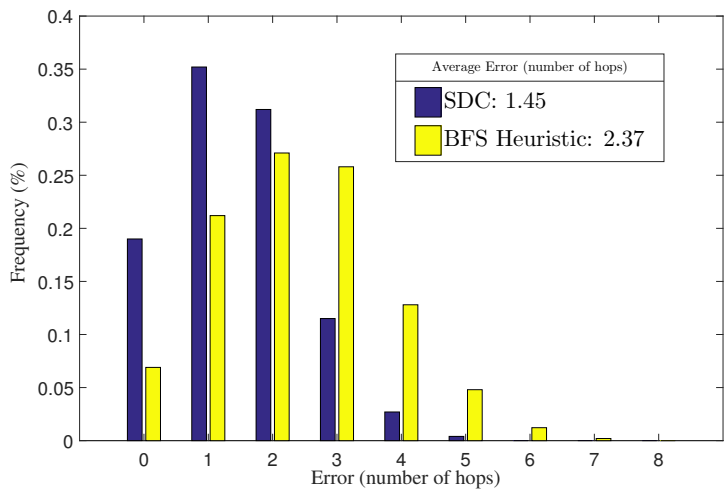

**Figure 6.5:** Comparing the error distribution (in the number of hops) between the SDC algorithm and the BFS heuristic in [131] on a circulant graph with $|G| = 6000$ and $|G_n| = 400$. In particular, the rate of the correct detection, i.e., $error = 0$, is 19.1% for the SDC and 7.0% for the BFS heuristic.

**Simulation Results for Real-World Networks**

We conduct the other four experiments on real-world SARS-CoV2003 and COVID-19 contact tracing networks in Singapore and Taiwan. If we can identify the connection between any two confirmed cases in real-world contact tracing networks, we denote the connection (or contact) as an edge. However, when the number of confirmed cases is too large to record details of contact information, we can only have information about the geographical footprint of some confirmed cases. In this situation, we also denote those visited places as vertices, and we add an edge between a confirmed case and a place if the confirmed case has visited the place.

Since $G$ is unknown in practice and contact tracing networks are infected subgraphs $G_n$, we assume that $G$ is a regular graph with a few irregular vertices and apply Algorithm SCT to the contact tracing networks to compute the source estimator. We use the graph distance from the actual source to the estimator to evaluate its performance.

**Table 6.3:** Average error (in terms of number of hops) comparing the SDC algorithm and BFS heuristic in [131] when $|G| = 6000$ is a random $d$-regular circulant graph with $|G_n| = 400$.

| $d$ | #(cycles) | SDC | BFS heuristic [131] |
|---|---|---|---|
| 3 | 255.6 | 1.67 | 2.75 |
| 4 | 198.0 | 1.45 | 2.37 |
| 5 | 167.1 | 1.33 | 2.07 |
| 6 | 147.7 | 1.24 | 1.97 |

**SARS-CoV2003 Contact Tracing Network in Taiwan** We reconstruct the contact tracing network data of SARS-CoV2003 Taiwan from a graph, which indicates potential bridges among hospitals and households, in [23]. In the original data, there are four types of nodes which represent the confirmed case, suspected case, hospital, and area, respectively. Since cities or countries provide no information for personal contact, we delete all area nodes from the original data. In addition, we also delete all the nodes that represent suspected cases. We apply Algorithm SCT on this infected network and correctly identify the first place, Taipei Municipal Heping Hospital (now Taipei City Hospital Heping Branch), of cluster infection in April 2003 in Taiwan. In addition, the BFS heuristic approach chooses the red vertex, which represents a confirmed case (not the first case) who had been to Taipei Municipal Heping Hospital. The network graph is shown in Figure 6.6, and the orange vertex is the statistical distance center representing the Taipei Municipal Heping Hospital.

**COVID-19 Contact Tracing Network in Singapore, 2020 Mar-Apr** The contact tracing network is an unconnected network due to the asymptomatic carriers, so we focus on the largest connected subgraph (cluster), including several worker dormitories and a construction site. We apply Algorithm SCT on subgraphs of the contact tracing network in Singapore that were reconstructed from the publicly-available news highlights provided by the Singapore Ministry of Health in [134]. In our computation, each vertex represents either an infected person or a place that the person had visited. An edge between two vertices implies that either a person has visited a place or two places have at least one common visitor. Here we omit the edge of person-person contact since most of the contact history can only be traced back to a place, not a single person. Hence, we treat each person-vertex as a leaf vertex connecting to a place.

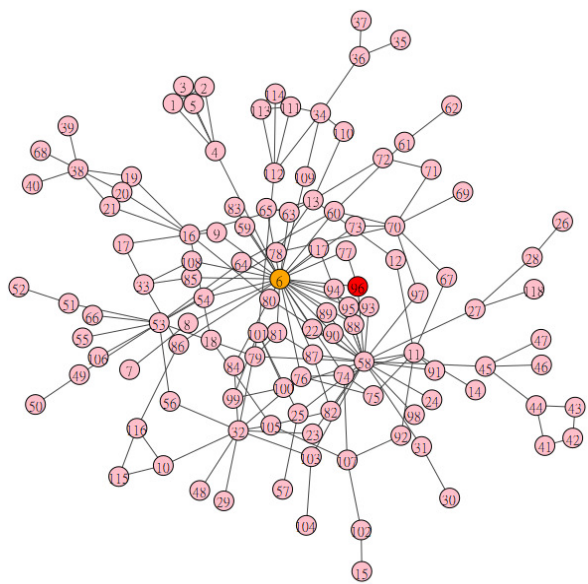

**Figure 6.6:** SARS-CoV2003 Contact Tracing Network in Taiwan. Each vertex represents either a confirmed case or a hospital. The orange and red vertices represent the source estimator determined by SDC and the BFS heuristic respectively. The orange vertex is the Taipei City Hospital Heping Branch where the major outbreak occurred and the red vertex is a confirmed case (not the first case) who had been to this hospital.

The first massive outbreak occurred at the beginning of April and peaked on April 20. Hence, we consider the infected subgraph after April 1. We define the source in the connected subgraph to be the first case in this connected subgraph. As far as we know, Case 655 attached to Westlite Toh Guan (WTG) is the first case found in this subgraph on March 26. On April 3, the connected subgraph formed a 4-cycle, which is shown in Figure 6.7. We can apply Algorithm SCT, which shows that WTG is the source estimator in this subgraph. After April 10, S11 Dorm (S11) becomes the new epidemic center due to the link between STL and S11 Dorm (S11). Note that S11 contains the second earliest case in this cluster and becomes the largest cluster in Singapore, which has more than two thousand cases confirmed in the middle of May.

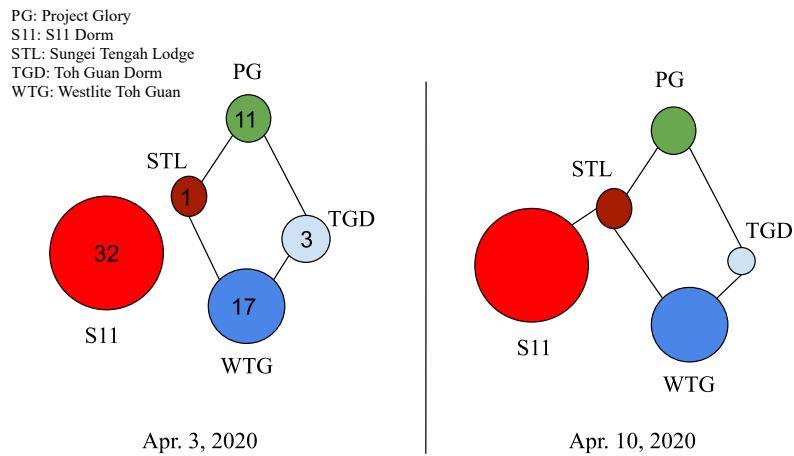

**Figure 6.7:** Each vertex is a cluster (place), and the number on each vertex is the total amount of infected people who have visited the place. On April 3, four places form a cycle, and all vertices are on the cycle. Algorithm SCT suggests that the source estimator is WTG where the first case in this subgraph comes from. On April 10, the epidemic center is S11 due to the link between S11 and STL.

**COVID-19 Contact Tracing Network in Taiwan, 2021 Feb, and 2021 May** We conduct experiments on two cluster infections in Taiwan recently. The first cluster infection originated in a northern Taiwan hospital in February. This network contains 18 tractable domestic cases, shown in Figure 6.8. The reason we select this networked data is that the contact network is public information provided by Central Epidemic Command Center in Taiwan [141] and the relation between cases in this network is clearly defined. Note that if we apply the BFS heuristic, then both case 838 and case 856 have the same possibility of being the source estimator. However, case 838 is the vertex with maximum statistical distance centrality, i.e., Algorithm SCT correctly identifies the first domestic case.

The second cluster infection is the latest cluster infection found at the beginning of May 2021. As the source of this cluster is unknown, we let the source be the person with the earliest symptom onset. We collect the data before May 14, 2021 to build the subgraphs using publicly available press releases provided by the Taiwan Centers of Disease Control in [141], and apply the 2-mode network model [23] to this cluster. Each vertex in this graph is either a workplace or a confirmed case. Both Algorithm SCT and the BFS heuristic identify the workplace, of the first case in this network. The contact tracing network is shown in Figure 6.9, and the red vertex is the source estimator determined by both algorithms.

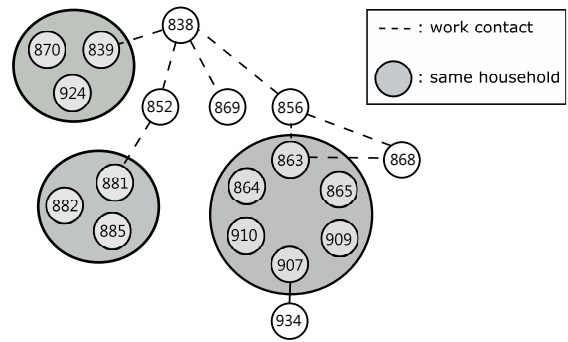

**Figure 6.8:** The infected subgraph of the contact tracing network starts from a single case 838 in Taiwan. The number in each vertex is the case number. Case 838 has the maximum statistical centrality, and it is the first domestic case in this cluster.

**Comparison with Approaches in Literature**    In addition to the rumor centrality approach, we have selected two other approaches, Dynamical Age [47], and Jordan Centrality [170], to compare to Algorithm SCT. We use the average distance-error to measure the performance of each algorithm. The simulation results are shown in Table 6.4. In each simulation, we repeat the following process: generate $G_n$, find estimators, and compute errors five hundred times in each type of network. All datasets in Table 6.4 are available at [90], [97] or can be generated by networkx [64]. Note that the SDC estimator significantly outperforms other approaches in circulant graphs and grid graphs since the SDC estimator is designed to solve the maximum likelihood estimation problem on finite regular graphs with cycles.

## 6.2   Rumor Source Detection in Twitter and Infodemics

The excess of information during a crisis, including the rapid spread of fake messages and unfounded rumors through various channels, makes it difficult for people to trust accurate information. In fact, rumors and epidemics are often interconnected as rumors can spread quickly during epidemics to influence people's behaviors and perceptions about the disease. For example, during the COVID-19 (Coronavirus Disease 2019) pandemic, half-truths and lies related to COVID-19 were circulating and spreading in online social networks, exacerbating the problem of vaccination and hindering efforts to control the pandemic. Indeed, the World Health Organization (WHO) identified the COVID-19 Infodemic, which is the spread of misinformation related to the COVID-19 pandemic, as another unprecedented crisis of global scale. The risks of infodemics include risk-taking behaviors, mistrust in health authorities, and lengthened outbreaks [6], [54], [100].

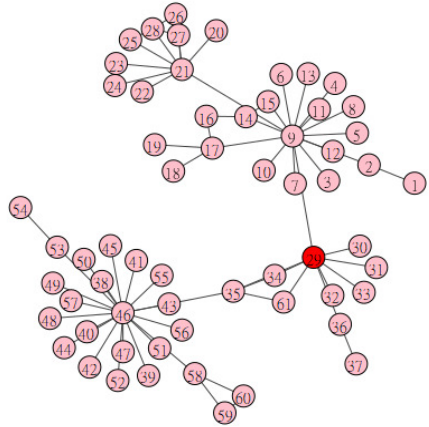

**Figure 6.9:** The contact tracing network of COVID-19, at the beginning of May 2021, in Taiwan. The vertex in red is the workplace of the person with the earliest symptom onset in this cluster.

**Table 6.4:** Average error (in terms of number of hops) comparing Algorithm 5 (SDC) to BFS heuristic version of Rumor Centrality (BFS-RC), Jordan Center (JC) and Dynamical Age (DA) in different network topologies.

| Network | $|G|$ | $|G_n|$ | SDC | BFS-RC | JC | DA |
|---|---|---|---|---|---|---|
| Circulant Graph(6000,6) | 6,000 | 400 | **1.67** | 2.75 | 2.87 | 1.90 |
| $100 \times 100$ Grid Graph | 10,000 | 150 | **1.87** | 3.79 | 2.04 | 2.08 |
| Random 3-regular Graph | 5,000 | 200 | **1.26** | 1.42 | 1.57 | 1.32 |
| Barabási-Albert(5000,3) | 5,000 | 300 | **2.74** | 4.25 | 2.75 | 2.96 |
| Canada Road Network | 1,965,206 | 100 | **3.15** | 3.41 | 3.19 | 4.03 |
| LastFM Asia Social Net. | 7,624 | 100 | **2.47** | 2.59 | 2.61 | 2.75 |
| Western U.S. Power Grid | 4,941 | 200 | **4.26** | 4.87 | 4.46 | 4.82 |

Actively debunking misinformation monitoring may be necessary to promote social resilience to viral misinformation by developing the science of fact-checking and rumor source detection [45], [63], [66], [135], [149]. Quantifiable measures like the infodemic risk index in [54], [66] can be used to track the magnitude of exposure to unreliable COVID-19 related news in Twitter, which is one of the most popular online social networks [11], [15], [34], [45], [62], [101], [150], [166]. Factors that influence the quick diffusion of misinformation in online social networks like Twitter are still not fully understood due to the large number of subscribers, the nature of real-time messaging and the presence of faked accounts and spambots [66]. The framework of network centrality as statistical inference can be useful to analyze infodemic risks by providing quantifiable measures to assess the spread of misinformation. The rumor centrality has been used in [166] to develop an online tool called "Trumor" to identify influential spreaders on Twitter. Each node in a dynamic retweet network, crawled by the Twitter API, is assigned a Trumor Score, which is based on the normalized version of rumor centrality to reveal the influence of a node on specific topics numerically [166]. Let us illustrate in the following the application of the epidemic centrality in [132], [165] to a real-world rumor-spreading dataset on Twitter.

**Scientific Rumor Spreading on Twitter: Higgs Twitter Dataset**   To verify the efficacy of the source estimator in [132], [165], [166], we select a rumor spreading dataset on Twitter provided in [34], [97]. The authors in [34] considered all the tweets related to the discovery of a Higgs boson-like particle. The dataset information include the Twitter users' activities including "retweet", "mention", and "reply" to study the spatio-temporal patterns of information spreading related to the specific Tweet information. The whole dataset contains 456,626 nodes and 14,855,842 edges, where the nodes and edges represent users and users' activities, respectively. In addition to the graph structure, this dataset also provides the type of each directed edge, e.g., "retweet", "mention", "reply", and their corresponding timestamps. The information-spreading process in Higgs Twitter Dataset was divided into four periods during the week between the 1st and the 7th of July 2012.

Since there are multiple information sources in the Higgs Twitter Dataset in each period, we assume that there is an implicit super node which is the source that spreads the rumor initially. Let $v_s$ denote the super node, and for each source in the original graph, we add an edge between the source and $v_s$. The degree of $v_s$ is 219, i.e., there are 219 sources in the original graph. The resultant graph $G_n$ is a connected spread graph with a single source $v_s$. According to the timestamp, during the first period (1st of July 2012) of the rumor spreading, $G_n$ has 2267 nodes, 3143 edges, and 2580 triangles. The diameter of $G_n$ is 10, and the maximum degree is 457. The degree rank plot and the graph topology are shown in Figure 6.10. Since the computation of the estimator in [132] requires computing $n!$ where, $n$ is the number of nodes in the graph, which may cause overflow when $n$ exceeds 150. To construct smaller spreading graphs, we can remove nodes from $G_n$ based on their timestamps.

In the following, we compare three source estimators based on rumor centrality (RC) [131]–[133], stochastic distance centrality (SDC) [165], and graph eccentricity (JC) [170] with one another. Note that when computing JC, the chances are high that there is more than one node with the same eccentricity. To resolve this issue, we select the node with the minimum distance from $v_s$ as the source estimator. Since the algorithm of rumor centrality needs to be re-designed to handle the overflow during the computation, we only conduct simulations on a small subgraph of $G_n$ using the rumor centrality estimator.

From Table 6.5, we can observe that when the size of $G_n$ is small, the performance of RC and SDC are the same; however, when $n$ is large, we need to handle the overflow issue during the computation of RC. On the other hand, to compute SDC, we need additional time to compute the smallest chordless cycle for each node in advance. Lastly, the shortcoming of JC is that usually, there are multiple Jordan centers in a non-tree graph.

## 6.3 Conclusions and Remarks

In this section, we examined how the theories and algorithms in the previous sections can be applied to real-world applications in digital contact tracing and rumor source detection. In particular, we focused on the epidemic centrality, which generalizes the rumor centrality for the general graph case and can be computed using message-passing algorithms. The epidemic centrality can be used to identify the contagion source on regular tree networks with irregular boundary conditions and also graphs with multiple cycles. We then examined how epidemic centrality-based algorithms (e.g., SDC algorithm) can be used for digital contact tracing of superspreaders during the COVID-19 pandemic. The computational complexity of the SDC algorithm is $O(d \cdot n^3)$, which is slower than the message passing algorithm proposed in [132]. This issue can be viewed as a trade-off between accuracy and running time. We also discussed another practical application of the epidemic centrality to rumor source detection in online social networks using a specific real-world instance of rumor spreading on Twitter.

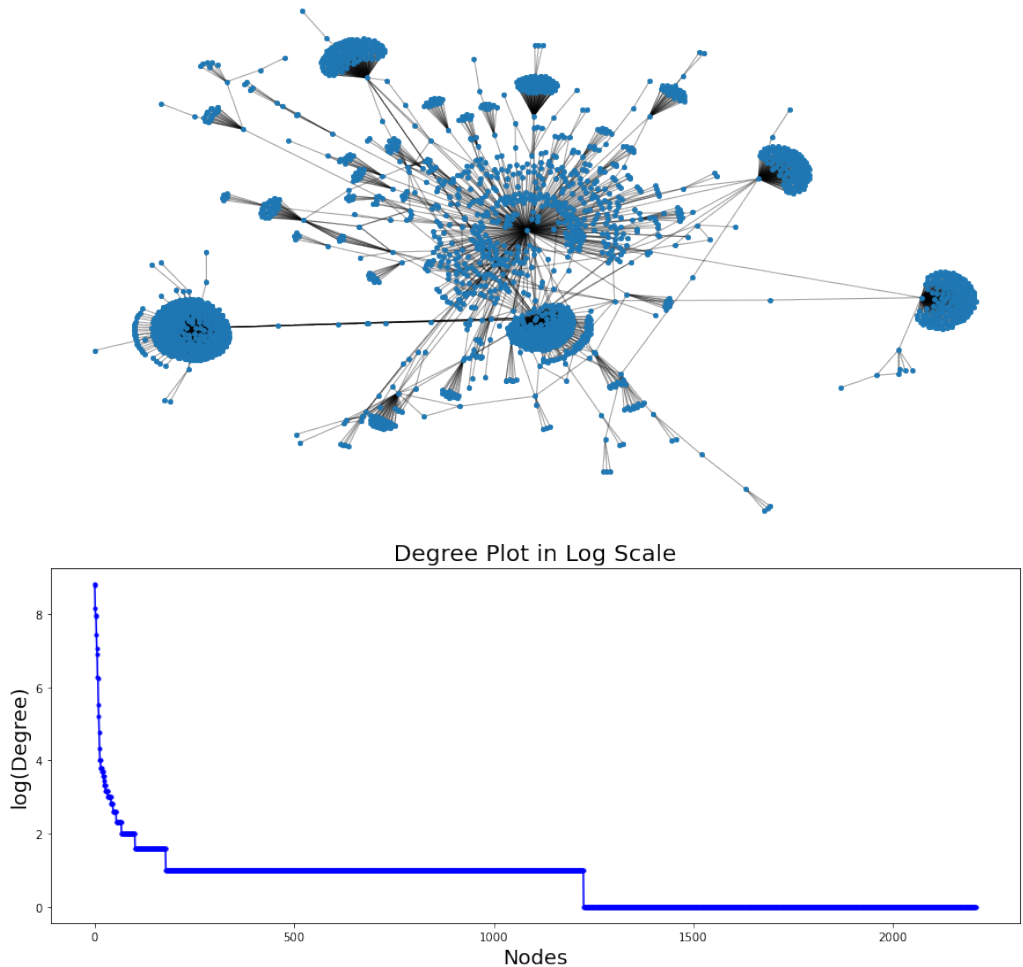

**Figure 6.10:** The above graph is the network topology of the spread graph (infected subgraph) $G_n$ on 1st July 2012. The second graph ranks nodes according to their degree in descending order, where the degrees are in the $\log_2(\cdot)$ scale.

**Table 6.5:** Detection error (in terms of the number of hops) comparing SDC, BFS heuristic RC and JC with different values of $n$ on the first period of the spreading over the Higgs Twitter dataset.

| n | RC | SDC | JC |
|------|------|-----|----|
| 70 | 0 | 0 | 1 |
| 100 | 0 | 0 | 1 |
| 150 | 0 | 0 | 0 |
| 500 | N.A. | 0 | 0 |
| 2267 | N.A. | 0 | 0 |

# 7

## Further Discussions and Open Issues

### 7.1 Related Research Topics and Open Issues

Contagion source detection in large networks structures has become significant in the wake of the COVID-19 pandemic. While research related to source detection began a decade ago, several issues remain unresolved. These issues are primarily due to the assumptions made about the network structures or spreading models being considered. As statistical inference involves collecting data and using statistical assumptions to model the process that generated the data, this approach can lead to strong empirical claims about causality and solution structure. However, the validity of these claims does not come from the presuppositions, which are often difficult to verify. There are open issues about the helpfulness of statistical inference models even when definitive answers cannot be expected. In this section, we first introduce some related research topics on viral spreading in networks, and then identify some of these open issues in the problem of contagion source detection and discuss potential future directions.

### 7.1.1  Related Research Topics

Prior to Shah's seminal work [132] on the source detection problem, earlier studies predominantly focused on exploring how network topology and factors such as infection and recovery rates contribute to the viral spreading [5], [22], [32], [55], [61], [121], [122], [125], [140] and not on the statistical inference aspects. The source detection problem was first introduced as a maximum likelihood estimation problem on an observed infected network in [132], where each node in the network is considered as a statistical estimator. Since then, various other epidemic models have been investigated for contagion source detection, including models that incorporate reinfection, recovery, and different types of contagion spread. The Susceptible-Infected-Recovered (SIR) model was considered in [22], [24], [102], [169], [170]; and the Susceptible-Infected-Susceptible (SIS) model was considered in [107], [155]. A Susceptible-Infected-Recovered-Infected (SIRI) was consider in [67], [112]. Regardless of the specific epidemic model being considered, most of the literature mentioned above assumes homogeneity between nodes in order to simplify the mathematical analysis. For example, all susceptible nodes with infected neighbors are assumed to have the same probability of being infected in the next time slot [107], [155], [170]. The oversimplification might limit the applicability of models to real-world scenarios where individuals can vary widely in terms of their connectivity and susceptibility to infection. The work in [138] considered the quickest detection of the contagion source estimation problem using a dynamic model based on noisy and incomplete measurements.

Aside from the aforementioned epidemic spreading models, opinion dynamics models in social networks have also been considered as a piece of ideal information spreading models [1], [2], [148]. Another commonly considered information-spreading model is the independent cascade (IC) model proposed in [10]. The IC model has been widely used in the literature to study information and contagion spreading in networks. The study in [169] investigates the source detection problem using the IC model and provides a maximum a posterior (MAP) based solution. Moreover, the influence maximization problem is also considered in the IC model [79], [80], [151]. The work in [119] considered an even more fundamental question: given a set of infected nodes and the underlying network, can we distinguish whether this is an epidemic spreading or just a randomly occurring illness among nodes?

In addition to source detection and influence maximization, other issues like privacy protection, competitive information spreading and noisy observation have been explored in the literature. For example, works in [41]–[44] consider the problem of obfuscating the spreading source in online social networks. By contrast, the work in [129] states that when three or more independent snapshots are available, it is possible to correctly detect the source with a constant probability under the adaptive diffusion model. A competitive information spreading problem was studied using game theory in [99], which depicts two information sources spreading conflicting information in opposition to one another under a linear threshold model. Lastly, under the assumption of noisy observation, i.e., some infected nodes are undetectable (or asymptomatic), the source detection problem is considered in [106]. The work in [22] modeled the spreading of COVID-19 with asymptomatic cases, and the works in [116], [117] addressed the problem of detecting cascading phenomena in networked structures.

### 7.1.2 Open issues: Networks with Irregularities

In Section 4, we address some of the open issues associated with the limitations in the models in [131]–[133] that assumes a countably infinite number of susceptible users, i.e., an "infected" user always has a susceptible neighbor. When this no longer holds, the analyses in [24], [35], [53], [85], [131]–[133], [154] are no longer applicable. In practice, the nonlinear features of the contagion source detection problem cannot be ignored as the number of users is always countably finite (e.g., the world's population is about 7 billion), and online social network users who receive but do not spread messages can be effectively modeled as end vertices in the infection graph. Looking at the simplest case, irregularities come in the form of a graph with cycles or a finite underlying degree-regular graph where the end vertices (i.e., the susceptible users with only a single neighbor) introduce nonlinearity to the constraint set of the maximum likelihood estimation. Nonlinear irregularities can cause counter-intuitive results such as nodes near the graph boundary having a higher likelihood than the graph center [161], [162]. Hence, the number of such kinds of vertices and their locations in the graph significantly shape spreading as well as the source detection performance.

In Section 4, we have resolved these special cases and used its solution to design heuristics for graphs with more complex boundary effects. The heuristics may have the advantage of computational efficiency but are still suboptimal with respect to optimally solving the maximum likelihood estimation problem. It will be important to establish some form of optimal guarantees of the detection algorithms for the general graph case. It is possible that some of these nonlinear effects can be considered as outliers or may become negligible when the infection graph grows sufficiently large. In such cases, a robust approach is to consider heuristics based on the rumor centrality or the epidemic centrality as a first approximation to attack the general problem and then to study the gap in the estimation performance. In general, the contagion source detection problem remains a very difficult problem to solve optimally, and new theoretical advances beyond the current state-of-the-art are needed.

### 7.1.3 Open issues: Statistical and Computational Challenges

The possibly unprecedented volume of data in the real-world applications can render the computational heuristics in [24], [35], [53], [85], [131]–[133], [154] to be impractical. Novel analytical methods will be needed to overcome the computational complexity barrier of solving the contagion source detection problem in the regime when the infection graph becomes very large. The network centrality as a statistical inference framework in this monograph only considers a static network whose inherent graph-theoretic features do not change. It will be important to generalize this to time-dependent networks where graph features can change over time, possibly affecting the availability of network data for statistical inference [33]. Finding the appropriate network centrality to explain flow patterns or temporal scales of changes in the network to solve stochastic optimization problems concerning the network will be especially interesting.

There are several computational challenges in this field that need improvement due to statistical uncertainty, such as missing information or a mismatch between algorithmic tuning and data statistics. Firstly, processing large graphs can be challenging as their volume may reach a point where it limits the computation of standard graph algorithms. Secondly, data may have local and global statistical dependence, which affects the problem-solving approach and solution quality. For example, when the parameters of a spreading model depend on the underlying network topology, finding a good solution for contagion source detection requires understanding the data statistics. A viable solution would consider the inherent statistics of the data for algorithmic tuning to optimize the computational performance of large networks. A promising area of research is leveraging machine learning techniques to exploit statistical features without incurring significant information loss or degraded solution quality [31], [76], [87].

## 7.2 Reverse and Forward Engineering

We explore how the application of network centrality as a statistical inference framework can address problems and provide solutions in both *reverse and forward engineering* contexts.

### 7.2.1 Reverse Engineering

When it comes to reverse engineering, the question is: What are the statistical inference problem formulations related to spreading that a well-known network centrality implicitly solves? The distance centrality and branch weight centrality, as demonstrated in this monograph, can solve the contagion source detection problem [132]. Betweenness centrality can tackle the single-vaccine estimation problem in [163]. The appropriate network centrality can concisely capture the impact of adding or removing nodes as well as stochastic processes in the graph. This concept can be linked to perturbation analysis in stochastic programming and can be instrumental in comprehending the effect of community in a network, such as how a specific network motif that has been added or removed can impact spreading.

Network centrality algorithms can compute exact or approximate solutions to statistical inference optimization problems. For example, as illustrated in this monograph, rumor centrality is optimal only when the graph is a degree-regular tree, and otherwise serves as a good heuristic to finding approximately good solution. A network centrality perspective thus provides guiding principles on algorithm design even when the original problem is hard to solve. In addition, the mathematical tools of combinatorics, graph theory, probability theory and computational complexity can quantify the performance of statistical inference in finite and asymptotically large network regimes of massive graphs. The value of reverse engineering thus lies in shedding theoretical insights into the solvability and optimality of the problem concerned.

### 7.2.2 Forward Engineering

In the case of *forward engineering*, we ask: Given a stochastic optimization formulation over a network, how to transform it or decompose it to one whose subproblems are graph-theoretic and can utilize network centrality, then solve or approximate the overall problem? Answering these questions thus entails an algorithmic approach that seeks to simplify the original problem, making the problem-solving methodology scalable to accommodate practical situations and to invent new network analytics [2], [19], [21], [26], [128], [136], [145], [152], [163]. For instance, once it has been established that the rumor center (based on the rumor centrality and optimal only for degree-regular tree graphs) is equivalent to the distance center or the centroid in graph theory, this opens doors to new algorithm design associated with new kinds of network centrality. These algorithms can also serve as computationally efficient heuristics to address general graphs. In essence, forward engineering enables the theoretical axiomatization and reuse of pre-existing graph algorithms [2], [19], [21], [26], [128], [136], [145], [152], [163]. This enables proactive intervention measures like node immunization against cascading outbreaks in [19], [152], [163]. We illustrate this approach with the vaccine centrality in [163].

### 7.2.3 Vaccine Centrality for Proactive Intervention

**Vaccine Centrality and Protection Node Placement**

In this section, we formulate the problem of expected outage minimization from cascading failures as a stochastic optimization problem over a graph. Since the protection nodes are "immune" to the cascading failure, the cascading failure is not able to spread through the protection nodes, i.e., the graph is partitioned into several connected subgraphs after removing all the protection nodes. The optimization problem of interest is to evenly partition the graph into smaller subgraphs by placing the protection nodes. Consider modeling this network as an acyclic connected graph with $N$ nodes denoted by $G_N$. Let $V_P$ be the set of protection nodes in $G_N$ protected by a vaccine. We let **Expect**$(|G_n|)$ be the expectation of the number of failed nodes (i.e., the spread of the cascading failure should it happen). Then, the protection node placement problem can be formulated as follows:

$$\begin{aligned}
\underset{v \in V_P \subseteq G_N}{\text{minimize}} \quad & \textbf{Expect}(|G_n|) \\
\text{subject to} \quad & |V_P| = k,
\end{aligned} \tag{7.1}$$

where $k$ is the cardinality of the number of protection nodes. Now, (7.1) is a stochastic program that is hard to solve in general. We shall show that, when $G_N$ has a tree topology, (7.1) can be simplified as a deterministic problem that we can solve using the network centrality which uses partially ordered sets (poset) in graphs in Section 7.2.3 to identify the protection nodes. We call this the *vaccine centrality* for solving the stochastic program in (7.1).

**Optimality Characterization and Bounds**

Let the $C(\{V_P\}) = (C_1^{\{V_P\}}, C_2^{\{V_P\}}, ..., C_m^{\{V_P\}})$ be the sequence of connected components after removing nodes in $V_P$ from $G_N$ (cf. Table 7.1). Assume that the failure starts from a node $v$ uniformly picked in $G_N$, and that the cascading failure stops spreading once the failure affects all nodes in its connected component. In this case, the number of nodes being affected is the number of nodes in the connected component that contains $v$ when all the nodes in $V_P$ are removed from $G_N$. For example, in Figure 2.3, if $v_1$ is protected, then $\mathbf{Expect}(|G_n|) = \frac{1}{7} \cdot (3 \cdot 3 + 3 \cdot 3 + 1 \cdot 0)$. This multiplicative factor $\frac{1}{7}$ is the probability of each node being picked initially. On the righthand-side, $3 \cdot 3$ is the number of nodes being affected by the cascading failure once it starts from $v_2$, $v_4$ or $v_5$ multiplied by $|\{v_2, v_4, v_5\}|$. Hence, the stochastic optimization problem in (7.1) can be equivalently expressed as the following deterministic problem:

$$\begin{aligned} \underset{V_P \subseteq V(G_N)}{\text{minimize}} \quad & (C_1^{\{V_P\}})^2 + (C_2^{\{V_P\}})^2 + ... + (C_m^{\{V_P\}})^2 \\ \text{subject to} \quad & |V_P| = k, \end{aligned} \tag{7.2}$$

where the variable in this optimization problem is a set of nodes in $G_N$, and $m$ is the number of connected components after removing $V_P$ from $G_N$.

**Table 7.1:** Glossary of key notations used in two main sections, namely Section 7.2.3 and Section 7.2.3, on partially ordered set-based results for causal inference and the related stochastic optimization solution to minimize cascade spread graphs.

| Notation | Remark |
|---|---|
| $\mathbf{Expect}(|G_n|)$ | The expected size of a cascade spread graph $G_n$ that occurs randomly in $G_N$ |
| $V_P$ | The set of nodes in $G_N$ to be protected with the vaccine |
| $C(\{V_P\})$ | Sequence of connected components after removing all nodes in $V_P$ from $G_N$ |
| $T_c$ | Centroid tree obtained recursively from the centroid decomposition of $G_N$. Particularly, $T_c$ is a tree abstract data type and $v^\star(T_c) = v^\star(G_N)$. |
| $|t_v^{v^\star(T_c)}|$ | Vaccine centrality of $v$ defined as the size of the subtree $t_v^{v^\star(T_c)}$ in $T_c$. |

In the following, we show how the centroid to (7.2) can be a feasible solution to (7.2) and also demonstrate when it solves (7.2) optimally. Now, the centroid $v^\star(G_N)$ is defined as:

$$v^\star(G_N) = \arg\underset{v \in G_N}{\text{minimize}} \quad \underset{1 \leq i \leq D}{\max} \{C_i^v\}. \tag{7.3}$$

In (7.3), $C_i^v$ is the $i$-th connected component after removing $v$ from $G_N$ and $D$ is defined by $\underset{v \in G_N}{\max} d_v$. Note that, if there is a node $v$ such that $d_v = j < D$, then we define $C_i^v = 0$ for $j \leq i \leq D$. In particular, after we have added a new auxiliary variable $\lambda \in \mathbf{R}^{D \times 1}$ on (7.3), we obtain

$$
\begin{aligned}
\underset{v \in G_N}{\text{minimize}} \quad & \max_{\lambda \in \mathbf{R}^{D \times 1}} \sum_{i=1}^{D} \lambda_i \cdot C_i^v \\
\text{subject to} \quad & \lambda^T \mathbf{1} = 1, \\
& \lambda_i > 0, \ i = 1, \dots, D.
\end{aligned}
\tag{7.4}
$$

Let $\lambda_i$ be defined as $\frac{C_i^v}{N-1}$, for $i = 1 \dots D$. By the definition of $C_i^v$, we have $\sum_{i=1}^{D} C_i^v = N - 1$. Hence, $\sum_{i=1}^{D} \lambda_i = 1$, which implies $\lambda$ is feasible in (7.4). Then (7.4) becomes an upper bound to the optimal value of the following problem:

$$
\underset{v \in G_N}{\text{minimize}} \quad \frac{1}{N-1} \sum_{i=1}^{D} C_i^v \cdot C_i^v.
\tag{7.5}
$$

Observe that (7.5) is the same as the form in (7.2) when $k = 1$. This means that the centroid of $G_N$ is a feasible (but suboptimal) solution for the problem in (7.2) even if we only pick a single node as the protection node. On the other hand, from the relationship between the $\ell_2$-norm and $\ell_\infty$-norm,

$$
\sqrt{(C_1^v)^2 + (C_2^v)^2 + \dots + (C_D^v)^2} \geq \max_{1 \leq i \leq D} C_i^v.
$$

Hence, we have

$$
\min_{v \in G_N} \sum_{i=1}^{D} (C_i^v)^2 \geq \min_{v \in G_N} \max_{1 \leq i \leq D} (C_i^v)^2,
$$

and we have thus established upper and lower bounds of the optimal value in (7.2) given by

$$
\min_{v \in G_N} \max_{1 \leq i \leq D} (C_i^v)^2 \leq \sum_{i=1}^{D} (C_i^v)^2 \leq (N-1) \min_{v \in G_N} \max_{1 \leq i \leq D} C_i^v,
\tag{7.6}
$$

where under the special case of $k$ being 1, the centroid of $G_N$ is the optimal solution corresponding to the optimization problems in the upper and lower bounds.

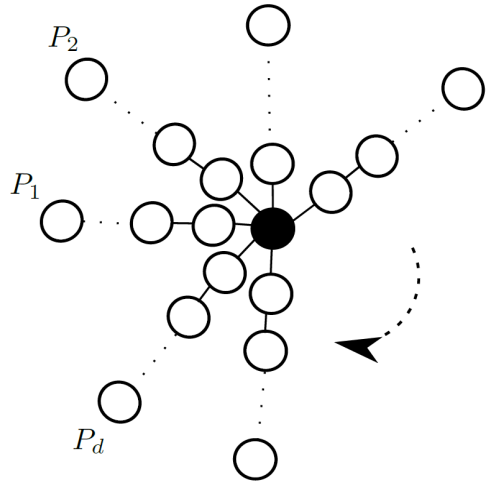

**Figure 7.1:** An example illustrating that the centroid of $G_N$ is the optimal solution for (7.2), the shaded node is the centroid of $G_N$.

**Theorem 7.1.** Let $G_N$ be a graph such that the centroid $v^\star(G_N)$ of $G_N$ is the only node with $d_{v^\star(G_N)} > 2$, i.e., for all $v \in G_N$ and $v \neq v^\star(G_N)$, $d_v \leq 2$, then $v^\star(G_N)$ is the optimal solution for (7.2) when $k = 1$.

Theorem 7.1 can be proved by considering a sufficient condition of the optimality of $v^\star(G_N)$: For any $v \in V(G_n)$, there is an integer $q$ such that, $C_i^v \geq C_i^{v^\star(G_N)}$ for $i = 1, \ldots, q$ and $C_i^v \leq C_i^{v^\star(G_N)}$ for $i = q+1, \ldots, d_{v^\star}$ where $(C_1^{v^\star}, C_2^{v^\star}, \ldots, C_{d_{v^\star(G_N)}}^{v^\star})$ is defined as above but in decreasing order, i.e, $C_i^{v^\star(G_N)} \geq C_j^{v^\star(G_N)}$ whenever $i > j$.

From Theorem 7.1, we deduce that, under some special cases, the centroid is indeed the optimal solution for (7.2). For example, in Figure 7.1, the shaded node is the centroid with degree $d$ and there are $d$ paths connected to the centroid. Note that the length of each of these $d$ paths need not be the same.

**Corollary 1.** If $G_N$ is a tree, then the optimal solution of the optimization problem (7.2) when $|V_P| = 1$ is the node $v_{\mathcal{B}}$ with the maximum betweenness centrality. (See (2.11) for the definition of $v_{\mathcal{B}}$.)

Roughly speaking, the betweenness centrality [51] is proportional to the number of times that a node acts as a "bridge" on the shortest path for any two nodes in the graph. Since its inception in [51] in 1977, the betweenness centrality is often used as a routine in popular algorithms for clustering and community identification and requires a complexity of $O(N + E(G_N))$ space and runs in $O(N \cdot E(G_N))$ time on general graphs [13]. The centroid $v^\star(G_N)$ can possibly be regarded as a heuristic approximation of the betweenness center $v_{\mathcal{B}}$. However, particularly for the special case in Theorem 7.1, we have $d(v^\star(G_N), v_{\mathcal{B}}) = 0$.

## Vaccine Centrality

In this section, we introduce the vaccine centrality of a given node in an induced tree abstract data type (*centroid tree*) based on the well-known graph decomposition method called the *centroid decomposition* [60], [115]. Let $G_N$ be a tree and $T_c$ be the corresponding centroid tree. The definition of the vaccine centrality of a given node $v$ is defined by $|t_v^{v^\star(T_c)}|$ on $T_c$, where $v^\star(T_c)$ is the centroid of $T_c$. Note that the vaccine centrality is only defined on $T_c$ instead of the original graph $G_N$, and each node in $G_N$ has a corresponding node in $T_c$. Moreover, the centroid of $T_c$ is also the centroid of $G_N$ due to the construction rules of $T_c$. Assume that $v$'s parent node is removed from $G_N$, then the vaccine centrality of $v$ measures how large a subtree of $G_N$ can be decomposed when $v$ is removed. Note that $v$ can only be chosen after its parent node in $T_c$ was chosen. For example, in Figure 7.2, assume that node 1 is removed from $G_N$. We have $2_1$, $2_2$ and $2_3$ are children node of 1 in $T_c$, and $|t_{2_1}^{v^\star(T_c)}| = 2$, $|t_{2_2}^{v^\star(T_c)}| = 3$ and $|t_{2_3}^{v^\star(T_c)}| = 6$. Hence, after node 1 is chosen to be protected, the next choice is $2_3$, since it has the maximum vaccine centrality, i.e., the protection of $2_3$ can decompose the 6-nodes subtree into smaller subtrees.

## Approximation Algorithm for $k$ Protection Nodes

So far, the results in Section 4 apply to acyclic graphs, i.e., networks with a tree topology. For the general case of a graph with general topology, e.g., having cycles, we use the Breadth First Search (BFS) heuristic. In the BFS heuristic, we apply Algorithms 1, 6, and 7 on a BFS-induced spanning tree, which is denoted as $T_{\mathsf{BFS}}$, of $G_N$. The intuition is that if the cascading failure starts from a node $v$, then this BFS spanning tree rooted at $v$ would correspond to all the nearest neighbors of $v$ being affected at the earliest time. Our $k$-protection placement algorithm contains three parts, the first part is graph decomposition using centroid decomposition. By leveraging the properties of the centroid of a tree, at each recursion, we can decompose the tree into components that are roughly balanced in size (i.e., each subtree component has a size less than or equal to $N/2$).

The second part is to construct a *centroid tree* $T_c$ from the result of the centroid decomposition. Note that $T_c$ is a tree rooted at the first centroid, i.e., the centroid $v^\star(T_{\mathsf{BFS}})$ of $T_{\mathsf{BFS}}$, and the $\mathsf{height}(T_c) \leq \log_2 N + 1$. Besides, each node in $T_{\mathsf{BFS}}$ has a corresponding node in $T_c$ and $v^\star(T_{\mathsf{BFS}}) = v^\star(T_c)$. In Algorithm 6, we denote the centroid found in the previous level as $v^\star_{previousLV}$.

The third part is selecting $k$ nodes from $T_{\mathsf{BFS}}$ based on their vaccine centrality on $T_c$. We can use Algorithm 1 to compute $|t_v^{v^\star(T_c)}|$ for each $v \in T_c$. For example, in Figure 7.2, $t_{2_2}^1 = 3$ and $t_{2_3}^1 = 6$. After computing $|t_v^{v^\star(T_c)}|$ for all $v$, we sort all the nodes in $T_c$ according to their $|t_v^{v^\star(T_c)}|$ in a decreasing order. Let $\mathsf{Sort}_v$ be the ordered list. Lastly, select the first $k$ nodes in $\mathsf{Sort}_v$ to be the protection nodes set.

In the following, we analyze the computational complexity and the optimality of Algorithm 6 and 7 when $G_N$ is a tree. In Algorithm 6, the computational complexity of line 4 is $O(|T|)$ which is proved in the previous section. The recursion in line 13 executes at most $O(\log_2 N)$ times. Hence, the computational complexity of Algorithm 6 is $O(N \log_2 N)$. In Algorithm 7, line 2 is the message passing algorithm with complexity $O(N)$, and line 3 needs to sort $N$ nodes with complexity $O(N \log_2 N)$. In summary, the total computational complexity is $O(N \log_2 N)$.

**Theorem 7.2.** Let $f(\{V_p\})$ denote the objective function in (7.2) and let $V_p^*$ denote the optimal solution of (7.2). When $G_N$ is a tree, the choice of $V_p$ in Algorithm 7 guarantees that

$$1 \leq \frac{f(\{V_p\})}{f(\{V_p^*\})} \leq \frac{2}{c(1-c)},$$

where $k$ is the size of the protection set $V_p$ and $0 < c < 1$ is a constant such that $k = c \cdot N$.

In Theorem 7.2, it guarantees the performance of the Algorithm 7 in the worst case. For comparison, we use a *degree centrality*-based heuristic

---

**Algorithm 6** Centroid Decomposition and Centroid Tree [163]

---

1: Initially set $currentLV = 0$
2: CENTROID-DECOMPOSITION($T$,$currentLV$,$v^\star_{previousLV}$)
3: $currentLV = currentLV + 1$
4: Compute the centroid $v^\star(T)$ of $T$ (randomly pick one if there are two centroids)
5: $v^\star.lv = currentLV$
6: Decompose $T$ into several subtrees $T'_j s$ by removing $v^\star(T)$ from $T$
7: $V(T_c) = V(T_c) \cup \{v^\star(T)\}$
8: **if** $v^\star(T).lv \neq 1$ **then**
9:     $E(T_c) = E(T_c) \cup \{(v^\star(T), v^\star_{previousLV})\}$
10: **end if**
11: **for** each subtree $T_j$ **do**
12:     **if** $|T_j| > 1$ **then**
13:         CENTROID-DECOMPOSITION($T_j$,$currentLV$,$v^\star(T)$)
14:     **else**
15:         $v.lv = currentLV + 1$, $\forall v \in V(T_j)$
16:         $V(T_c) = V(T_c) \cup \{v\}$
17:         $E(T_c) = E(T_c) \cup \{(v, v^\star_{previousLV})\}$
18:     **end if**
19: **end for**

---

**Algorithm 7** Construct a Set of $k$ Protection Nodes $V_P$ [163]

---

1: Input: $T_c$, $k$, Set $V_P = \{\ \}$
2: Compute $t_v^{v^\star(G_N)}$ for each $v \in T_c$
3: Let $\mathsf{Sort}_v$ be the list of nodes in $T_c$ sorted in a decreasing order according to $t_v^{v^\star(G_N)}$
4: **for** $i = 1 \ldots k$ **do**
5:     $V_P = V_P \cup \mathsf{Sort}_v(i)$
6: **end for**

---

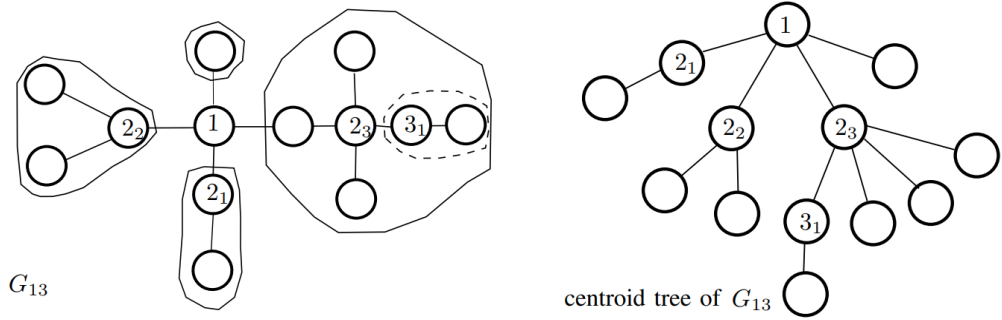

**Figure 7.2:** Example of centroid decomposition of $G_{13}$ and the tree on the right is the centroid tree from the centroid decomposition. After removing 1 from $G_{13}$, we have four connected components. For simplicity, the notation $2_i$ for $i = 1, 2, 3$ are equivalent to the notation $v_{2,i}^*$ used in Section 7.2.3 which are the centroids in the second recursion, and node $3_1$ is the centroid from the third recursion.

that sorts all the nodes according to their degrees. Thus, the *degree centrality* heuristic has $O(N \log N)$ computational complexity to select the protection set. In the following, we give an example illustrating that the performance of the degree-centrality heuristic cannot be bounded above as the size of the protection set increases.

**Example 7.1.** Suppose $G_N$ is composed of two balanced tree graphs (e.g., the graph in Figure 2.3) rooted at $v$ and $u$ respectively and connected by a path $P = (p_1, p_2, \ldots, p_t)$, where $p_i$ is a node on the path for $i = 1, \ldots, t$. Note that $v$ is adjacent to $p_1$ and $u$ is adjacent to $p_t$. Assume the length of $P$ is close to $N$, i.e., $t$ is much larger than the size of the two balanced trees on both sides. In this case, the optimal strategy to place one protection node is to choose the node on the path $P$, and the cost **Expect**$(|G_n|)$ will be $\frac{(N-1)^2}{2}$ or $\frac{N(N-1)}{2}$. The output of Algorithm 7 will be the same as the optimal solution. However, the degree centrality heuristic will output either $v$ or $u$, and **Expect**$(|G_n|)$ is around $N^2$. When the number of protection nodes increases, the output of the degree centrality heuristic will not change too much until it starts to select the node on the path. On the contrary, **Expect**$(|G_n|)$ of Algorithm 7 is bounded above by $\frac{2N^2}{k+1}$ due to the property of the centroid.

In summary, forward engineering is characterized by its use of optimization formulation to model practical defense approaches against contagion. For instance, in [163], minimizing the spread of contagion through the use of vaccine centrality can help avert systemic cascading failures in networked infrastructures, and this approach can be adapted to address other types of contagions as well. By defining an appropriate objective function, it becomes possible to leverage *approximation algorithms* to solve statistical inference problems. The work in [118] considers prioritization policies to optimize the sequence of tracing using a tool from operations research called a "branching bandit". Additionally, careful selection of optimization constraints enables systematic decoupling and decomposition of the original optimization problem into smaller, more manageable sub-problems. By using graph decomposition and search algorithms, such as pruning or clustering, it is possible to further simplify the constraint sets and facilitate the design of algorithmic methods with lower complexity. Ultimately, the value of forward engineering lies in its ability to facilitate algorithm design and computational aspects of optimization problem formulation.

## 7.3 Machine Learning Approach

Machine learning techniques can be used to address computational challenges associated with uncertainties in solving (3.2). For example, a contact tracer may not have the full contagion networked data at the early stage of an epidemic. This means that the contagion graph $G_N$ in (3.2) and the infection spreading dynamics are assumed to be unknown to a contact tracer. This is the chief uncertainty faced by all contact tracers, who therefore have to adopt a strategy to collect this data starting from an index case (i.e., the first documented infected person) and trace his or her close contacts and so on.

Machine learning algorithms have the potential to improve the predictive power of statistical inference in complex network topology by leveraging different network centrality measures to capture statistical measures. Machine learning-based algorithms can reduce generalization errors by using data-driven models that generalize well for new data. Additionally, these models can provide insight into how future epidemics spread and be utilized to design early warning systems. Recent advancements in machine learning, including deep learning and graph neural networks, can be particularly useful for addressing uncertainties in data when solving complex problems like (3.2), which we describe in more detail following the work in [20], [143] below.

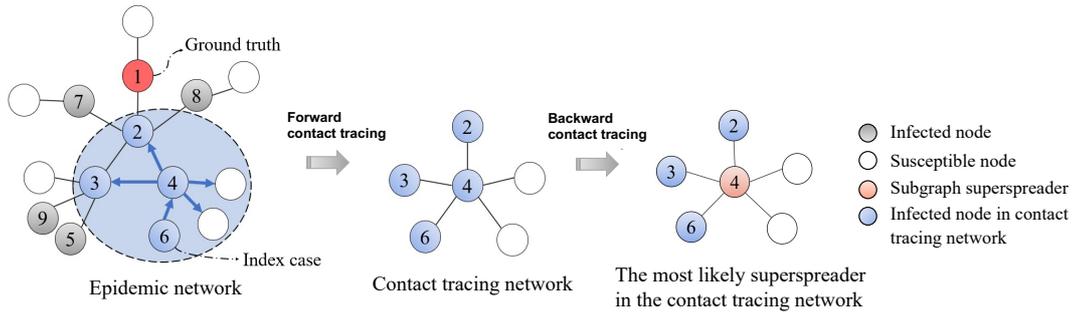

**Figure 7.3:** Illustrating a contagion graph network $G_9$ with nine infections (shaded nodes) whose numbering indicates the infection order starts from the ground truth, i.e., the real superspreader. The contact tracing network $G_4$ (within a dotted circle) starts from the index case node $v_6$ (blue arrows show the tracing directions) by forward contact tracing. The backward contact tracing is to find the node in $G_9$ that is most likely to be the superspreader.

At various stages of contact tracing, we model an instantaneous snapshot of a subgraph of $G_N$ in (3.2) that we call the *contact tracing network* being harvested. This contact tracing network grows by one (infected) node at each stage. If the traced node is infected, we continue to trace the node's neighbors. Otherwise, we stop tracing along this node. Let $G_n$ denote the contact tracing network with $n$ nodes at the $n$-th stage of tracing (the index case is $G_1$). Obviously, $G_n \subseteq G_N \subseteq G$. Contact tracing aims to find the node in $G_N$ that is most likely to be the superspreader, as shown in Figure 7.3. However, the superspreader may not yet be in $G_n$, meaning that the contact tracing effort is still in its early stage or not fast enough (relative to the pandemic spreading speed). In such a case, backward contact tracing should yield an estimate as close as possible to this most-likely superspreader. In other words, given the available data at the $n$-th stage, the contact tracer finds the node in $G_n$ that is the fewest number of hops away from the most-likely superspreader in $G_N$ (i.e., the optimal maximum likelihood estimate had this $G_N$ been given entirely upfront to the contact tracer as has been first studied in [132]).

Given the data $G_n$ harvested by contact tracing at the $n$-th stage, we have the following maximum-likelihood estimation problem:

$$\hat{v} \in \arg \max_{v \in G_n \subseteq G_N \subseteq G} \mathbb{P}(G_n|v), \qquad (7.7)$$

where $\mathbb{P}(G_n|v)$ is the likelihood function and $\hat{v}$ is the most likely superspreader of the outbreak. The key challenge is that $G_N$ is unknown to the contact tracer, who has to consider the following:

*Forward contact tracing*: How can the contact tracing network be constructed efficiently starting from a given index case?

*Backward contact tracing*: How to solve (7.7) to give the best instantaneous estimate of the superspreader given the data?

Answering both the forward and backward contact tracing jointly constitutes an iterative statistical inference process to track the most-likely superspreader in the entire contagion network. Specifically, to answer the forward contact tracing problem, it is a natural idea to grow the contact tracing network with a breadth-first search (BFS) or depth-first search (DFS) graph traversal algorithm from the index case. To answer the backward contact tracing problem, let us suppose that a given node in $G_n$ is assumed to be the superspreader. Then, starting from that node, there are a number of possible ways to infect all the other nodes consistent with $G_n$ harvested by contact tracing at the $n$-th stage. Even though $G_N$ is unknown, the contact tracer starts from the index case $G_1$ and collects more data in a forward manner (i.e., enlarging $G_n$ in (7.7)), the contact tracer also predicts the superspreader for that instant by solving (7.7). Intuitively, as the contact tracing subgraph $G_n$ grows, the contact tracer desires this prediction to be closer (in terms of the number of hops in $G_N$) to the most-likely superspreader in $G_N$. A question arises: how should $G_n$ grow in *forward contact tracing*? The authors in [20], [143] proposed to enlarge $G_n$ using the BFS and DFS graph traversal algorithms. Solving (7.7) thus can be viewed interestingly as solving a maze where $G_N$ is the maze topology. The solver (i.e., contact tracer) has no prior knowledge of this maze whose starting point and exit point correspond to $v_1^*$ (initial index case) and $v_N^*$ (maximum-likelihood estimate had $G_N$ been known a priori) respectively. The $n$th step in this maze corresponds to the centroid of a rooted tree $G_n$. Forward contact tracing and backward contact tracing are thus analogous to the process of maze exploration and maze traversal, respectively [20], [143].

Another strength of machine learning is feature representation that can capture statistical features critical to contagion source detection and lead to algorithms with better computational performance [156]. Automated machine learning (Auto ML) can be used to learn the underlying statistics of the spreading process to improve the algorithmic tuning of machine learning algorithms or message passing algorithms for massive graph dataset (cf. MEGA framework in [66]). It is an open issue on how to establish fundamental performance tradeoff curves to exploit statistical graph features for algorithmic speedup without incurring significant information loss or degraded solution quality. A machine learning approach to the framework of network centrality as statistical inference can connect graph algorithms with the message passing paradigm (e.g., graphical models and causal inference [50], [113]) and with other techniques and algorithms in graph neural networks and graph signal processing [39], [123], which include methods for sampling, filtering or learning over graphs.

## 7.4 Conclusions and Remarks

We have presented the contagion source detection problem that finds applications in digital contact tracing and rumor source detection in epidemics and infodemics, respectively. We introduced the framework of *Network Centrality as Statistical Inference* to provide a theoretically sound and computationally efficient approach to applying inferential statistics to spreading in networks. New network centralities, such as rumor and epidemic centrality, can characterize the global optimal solution of maximum-likelihood estimation with graph constraints associated with spreading in networks.

The thesis is to treat practical algorithms in the field of digital contact tracing or rumor source detection in cyberspace as distributed solutions for finding the contagion source. The approach of reverse engineering a network centrality (e.g., distance centrality, rumor centrality, and epidemic centrality) provides a design strategy based on theory, where a suitable function for estimation or detection is identified first and then a message-passing algorithm is developed as a distributed solution to certain statistical inference optimization problems. Forward engineering a network centrality refers to a top-down approach of using a network centrality to solve an optimization problem with graph constraints. Examples of such approaches are the harmonic influence centrality and the vaccine centrality, which can be employed to design node placement strategies for protecting against viral spreading. Network centrality-based solutions can provide a good approximation to solving these optimization problems and guidelines on scalable algorithm design (e.g., distributed message passing algorithms) for large networks. The benefit of this approach is that the surveillance of spreading in the network is computationally efficient, scalable, and stable.

The framework of "*network centrality as statistical inference*" is a powerful tool for algorithm design, and has implications beyond the analysis of network centrality and has connections to other disciplines, such as machine learning and graph signal processing. By leveraging the confluence of these research directions, we can gain a deeper understanding of the spread of contagions in large networks and develop techniques for monitoring and mitigating their impact, as has been observed in the last 15 years of research in this field.

Many of the mathematical and algorithmic challenges discussed in this monograph are driven by the requirements of epidemic control like digital contact tracing and infodemic risk management. Addressing these challenges would necessitate the development of novel modeling techniques, theoretical advancements, mathematical tools and data-driven methods, which in turn would contribute to the advancement of technology to analyze past contagion behaviors and effectively combat newly emerging contagions.

# Acknowledgements

The authors gratefully acknowledge the collaborations and interactions on this topic with many colleagues, including Hung-Lin Fu, H. Vincent Poor, Mung Chiang, Ron Guanrong Chen, R. Srikant, Cheng-Shang Chang, Wenyi Zhang, Wee Peng Tay, Wenxiang Dong, Felix Ming Fai Wong, Congduan Li, Ching-Nam Hang, Jiasi Chen, Siya Chen, Chung Chan, Michael Fuchs, Weng Kee Wong, Po-Shen Loh, Kwok-Yan Lam, Josip Car, and Show-Li Jan. We thank an anonymous reviewer for the careful reading of this monograph and the many insightful comments and suggestions. This work is supported in part by the Ministry of Science and Technology of Taiwan under Grant 110-2115-M-033-001-MY2 and in part by the Ministry of Education, Singapore, under its Academic Research Fund (No. 022307 and AcRF RG91/22) and a grant from the NTU World Health Organization Collaborating Centre for Digital Health and Health Education.

# References

[1]  D. Acemoglu, G. Como, F. Fagnani, and A. Ozdaglar, "Opinion fluctuations and disagreement in social networks," *Mathematics of Operations Research*, vol. 38, no. 1, 2013, pp. 1–27.

[2]  D. Acemoglu, A. Malekian, and A. Ozdaglar, "Network security and contagion," *Journal of Economic Theory*, vol. 166, 2016, pp. 536–585.

[3]  E. Adar and L. A. Adamic, "Tracking information epidemics in blogspace," in *Proceedings of IEEE/WIC/ACM International Conference on Web Intelligence*, ser. WI '05, pp. 207–214, Washington, DC, USA: IEEE Computer Society, 2005. DOI: 10.1109/WI.2005.151.

[4]  R. Alexandru and P. L. Dragotti, "Rumour source detection in social networks using partial observations," in *Proceedings of IEEE Global Conference on Signal and Information Processing (GlobalSIP)*, pp. 730–734, 2018. DOI: 10.1109/GlobalSIP.2018.8646695.

[5]  N. T. J. Bailey, *The Mathematical Theory of Infectious Diseases and its Applications*, Second. Griffin, 1975.

[6]  J. Bak-Coleman, I. Kennedy, and M. Wack, "Combining interventions to reduce the spread of viral misinformation," *Nature Human Behavior*, vol. 6, 2022, pp. 1372–1380. URL: https://doi.org/10.1038/s41562-022-01388-6.

[7] J. Balogh and G. Pete, "Random disease on the square grid," *Random Structures and Algorithms*, vol. 13, no. 3–4, 1998, pp. 409–422.

[8] Y. Bengio, P. Gupta, T. Maharaj, N. Rahaman, M. Weiss, and et al, "Predicting infectiousness for proactive contact tracing," in *Proc. of Int. Conf. on Learning Representations*, 2021.

[9] F. Bergeron, P. Flajolet, and B. Salvy, "Varieties of increasing trees," in *CAAP '92*, J. .-. Raoult, Ed., pp. 24–48, Berlin, Heidelberg: Springer Berlin Heidelberg, 1992.

[10] S. Bikhchandani, D. Hirshleifer, and I. Welch, "A theory of fads, fashion, custom, and cultural change as informational cascades," *Journal of Political Economy*, vol. 100, no. 5, 1992, pp. 992–1026.

[11] Bloomberg Businessweek, *A Fake AP Tweet Sinks the Dow for an Instant*, 2013. URL: https://www.bloomberg.com/news/articles/2013-04-23/a-fake-ap-tweet-sinks-the-dow-for-an-instant.

[12] W. J. Bradshaw, E. C. Alley, J. H. Huggins, A. L. Lloyd, and K. M. Esvelt, "Bidirectional contact tracing could dramatically improve COVID-19 control," *Nature Communications*, vol. 12, no. 1, 2021, pp. 1–9.

[13] U. Brandes, "A faster algorithm for betweenness centrality," *Journal of Mathematical Sociology*, vol. 25, no. 2, 2001, pp. 163–177.

[14] G. Brightwell and P. Winkler, "Counting linear extensions," *Order*, vol. 8, no. 3, 1991, pp. 225–242. DOI: 10.1007/BF00383444.

[15] P. E. Brown and J. Feng, "Measuring user influence on Twitter using modified $K$-shell decomposition," in *Proceedings of AAAI Conference on Artificial Intelligence*, 2011.

[16] S. Bubeck, L. Devroye, and G. Lugosi, "Finding Adam in random growing trees," *Random Struct. Alg.*, vol. 50, 2017, pp. 158–172.

[17] G. Cencetti, G. Santin, A. Longa, E. Pigani, A. Barrat, C. Cattuto, S. Lehmann, M. Salathe, and B. Lepri, "Digital proximity tracing on empirical contact networks for pandemic control," *Nature Communications*, vol. 12, no. 1, 2021, pp. 1–12.

[18] C. S. Chang, C. J. Chang, W. T. Hsieh, and D. S. Lee, "Relative centrality and local community detection," *Network Science*, vol. 3, no. 4, 2015, pp. 445–479.

[19] C. Chen, H. Tong, B. A. Prakash, C. E. Tsourakakis, T. Eliassi-Rad, C. Faloutsos, and D. H. Chau, "Node immunization on large graphs: Theory and algorithms," *IEEE Transactions on Knowledge and Data Engineering*, vol. 28, no. 1, 2016, pp. 113–126.

[20] S. Chen, P. D. Yu, C. W. Tan, and H. V. Poor, "Identifying the superspreader in proactive backward contact tracing by deep learning," in *Proceedings of the 52nd Annual Conference on Information Sciences and Systems (CISS)*, 2022. DOI: 10.1109/CISS53076.2022.9751196.

[21] W. Chen, S.-H. Teng, and H. Zhang, "A graph-theoretical basis of stochastic-cascading network influence: Characterizations of influence-based centrality," *Theoretical Computer Science*, vol. 824-825, 2020, pp. 92–111.

[22] Y.-C. Chen, P.-E. Lu, C.-S. Chang, and T.-H. Liu, "A time-dependent SIR model for COVID-19 with undetectable infected persons," *IEEE Transactions on Network Science and Engineering*, vol. 7, no. 4, 2020, pp. 3279–3294. DOI: 10.1109/TNSE.2020.3024723.

[23] Y.-D. Chen, H.-C. Chen, and C.-C. King, "Social network analysis for contact tracing," *Infectious Disease Informatics and Biosurveillance*, vol. 27, 2010, pp. 339–358. DOI: 10.1007/978-1-4419-6892-0_15.

[24] Z. Chen, K. Zhu, and L. Ying, "Detecting multiple information sources in networks under the SIR model," *IEEE Transactions on Network Science and Engineering*, vol. 3, no. 1, 2016, pp. 17–31. DOI: 10.1109/TNSE.2016.2523804.

[25] J. Choi, S. Moon, J. Woo, K. Son, J. Shin, and Y. Yi, "Information source finding in networks: Querying with budgets," *IEEE/ACM Transactions on Networking*, vol. 28, no. 5, 2020, pp. 2271–2284.

[26] F. Chung, P. Horn, and A. Tsiatas, "Distributing antidote using pagerank vectors," *Internet Mathematics*, vol. 6, no. 2, 2009, pp. 237–254.

[27]  T. H. Cormen, C. E. Leiserson, R. L. Rivest, and C. Stein, *Introduction to Algorithms 3th ed.* Cambridge, Massachusetts London, England: The MIT Press, 2009.

[28]  D. J. Daley and D. G. Kendall, "Epidemics and rumours," *Nature*, vol. 204, no. 1118, 1964.

[29]  D. J. Daley and D. G. Kendall, "Stochastic rumours," *IMA Journal of Applied Mathematics*, vol. 1, no. 1, 1965, pp. 42–55.

[30]  Q. E. Dawkins, T. Li, and H. Xu, "Diffusion source identification on networks with statistical confidence," in *Proceedings of the 38th International Conference on Machine Learning (ICML)*, vol. 139, pp. 2500–2509, PMLR, 2021.

[31]  J.-C. Delvenne and A.-S. Libert, "Centrality measures and thermodynamic formalism for complex networks," *Physical Review E*, vol. 83, no. 4, 2011, p. 046 117. DOI: 10.1103/PhysRevE.83.046117.

[32]  N. Demiris and P. D. O'Neill, "Bayesian inference for epidemics with two levels of mixing," *Scandinavian Journal of Statistics*, vol. 32, no. 2, 2005, pp. 265–280. URL: http://www.jstor.org/stable/4616877.

[33]  D. Domenico, G. Manlio, C. Granell, M. Porter, and A. Arenas, "The physics of spreading processes in multilayer networks," *Nature Physics*, vol. 12, 2016, pp. 901–906.

[34]  M. D. Domenico, A. Lima, P. Mougel, and M. Musolesi, "The anatomy of a scientific rumor," *Scientific Reports*, vol. 3, no. 2980, 2013. DOI: 10.1038/srep02980.

[35]  W. Dong, W. Zhang, and C. W. Tan, "Rooting out the rumor culprit from suspects," in *Proceedings of IEEE International Symposium on Information Theory*, pp. 2671–2675, 2013.

[36]  M. Draief and L. Massoulié, *Epidemics and Rumors in Complex Networks*, First. Cambridge University Press, 2009.

[37]  K. Durant and S. Wagner, "On the centroid of increasing trees," *Discrete Mathematics and Theoretical Computer Science*, vol. 21, no. 4, 2019.

[38]  R. Durrett, "Some features of the spread of epidemics and information on a random graph," *The Proceedings of the National Academy of Sciences*, vol. 107, no. 10, 2010, pp. 4491–4498.

[39] H. E. Egilmez, E. Pavez, and A. Ortega, "Graph learning from filtered signals: Graph system and diffusion kernel identification," *IEEE Transactions on Signal and Information Processing over Networks*, vol. 5, no. 2, 2019, pp. 360–374. DOI: 10.1109/TSIPN.2018.2872157.

[40] T. Fan and I. Wang, "Rumor source detection: A probabilistic perspective," in *Proceedings of IEEE International Conference on Acoustics, Speech and Signal Processing (ICASSP)*, pp. 4159–4163, 2018.

[41] G. Fanti, P. Kairouz, S. Oh, K. Ramchandran, and P. Viswanath, "Spy vs. spy: Rumor source obfuscation," in *Proceedings of ACM SIGMETRICS International Conference on Measurement and Modeling of Computer Systems*, pp. 271–284, Portland, Oregon, USA: ACM, 2015. DOI: 10.1145/2745844.2745866.

[42] G. Fanti, P. Kairouz, S. Oh, K. Ramchandran, and P. Viswanath, "Metadata-conscious anonymous messaging," *IEEE Transactions on Signal and Information Processing over Networks*, vol. 2, no. 4, 2016, pp. 582–594. DOI: 10.1109/TSIPN.2016.2605761.

[43] G. Fanti, P. Kairouz, S. Oh, K. Ramchandran, and P. Viswanath, "Rumor source obfuscation on irregular trees," in *Proceedings of ACM SIGMETRICS International Conference on Measurement and Modeling of Computer Science*, pp. 153–164, New York, NY, USA: ACM, 2016. DOI: 10.1145/2896377.2901471.

[44] G. Fanti, P. Kairouz, S. Oh, K. Ramchandran, and P. Viswanath, "Hiding the rumor source," *IEEE Transactions on Information Theory*, vol. 63, no. 10, 2017, pp. 6679–6713. DOI: 10.1109/TIT.2017.2696960.

[45] Z. Fei, Y. Ryeznik, A. Sverdlov, C. W. Tan, and W. K. Wong, "An overview of healthcare data analytics with applications to the COVID-19 pandemic," *IEEE Transactions on Big Data*, vol. 8, no. 6, 2022, pp. 1463–1480.

[46] L. Ferretti, C. Wymant, M. Kendall, and et al, "Quantifying SARS-CoV-2 transmission suggests epidemic control with digital contact tracing," *Science*, vol. 368, no. 6491, 2020, pp. 316–329. DOI: 10.1126/science.abb6936.

[47] V. Fioriti and M. Chinnici, "Predicting the sources of an outbreak with a spectral technique," *Applied Mathematical Sciences*, vol. 8, no. 135, 2014.

[48] P. Flajolet and R. Sedgewick, *Analytic Combinatorics*. Cambridge University Press, 2009.

[49] D. A. Freedman, *Statistical Models: Theory and Practice*, 2nd. Cambridge, UK: Cambridge University Press, 2009.

[50] D. A. Freedman, *Statistical Models and Causal Inference: A Dialogue with the Social Sciences*, 1st. Cambridge, UK: Cambridge University Press, 2011.

[51] L. C. Freeman, "Centrality in social networks conceptual clarification," *Social Networks*, vol. 1, 1978-1979, pp. 215–239.

[52] A. Friggeri, L. A. Adamic, D. Eckles, and J. Cheng, "Rumor cascades," in *Proceedings of 8th International AAAI Conference on Weblogs and Social Media*, 2014.

[53] M. Fuchs and P. D. Yu, "Rumor source detection for rumor spreading on random increasing trees," *Electronic Communications in Probability*, vol. 20, no. 2, 2015.

[54] R. Gallotti, F. Valle, N. Castaldo, P. Sacco, and M. D. Domenico, "Assessing the risks of 'infodemics' in response to COVID-19 epidemics," *Nature Human Behaviour*, vol. 4, 2020, pp. 1285–1293.

[55] A. Ganesh, L. Massoulie, and D. Towsley, "The effect of network topology on the spread of epidemics," in *Proceedings of IEEE 24th Annual Joint Conference of the IEEE Computer and Communications Societies.*, vol. 2, 1455–1466 vol. 2, 2005. DOI: 10.1109/INFCOM.2005.1498374.

[56] K. Garimella, G. G. F. Morales, A. Gionis, and M. Mathioudakis, "Political discourse on social media: Echo chambers, gatekeepers, and the price of bipartisanship," in *Proceedings of the World Wide Web Conference*, pp. 913–922, Lyon, France: International World Wide Web Conferences Steering Committee, 2018. DOI: 10.1145/3178876.3186139.

[57] D. F. Gleich, "Pagerank beyond the web," *SIAM Review*, vol. 57, no. 3, 2015, pp. 321–363.

[58]  W. Goffman and V. A. Newill, "Generalization of epidemic theory: An application to the transmission of ideas," *Nature*, vol. 204, 1964, pp. 225–228.

[59]  A. Goldenberg, A. X. Zheng, S. E. Fienberg, and E. M. Airoldi, "A survey of statistical network models," *Foundations and Trends in Machine Learning*, vol. 2, no. 2, 2010, pp. 129–233.

[60]  A. J. Goldman, "Optimal center location in simple networks," *Transportation Science*, vol. 5, no. 2, 1971, pp. 212–221.

[61]  M. Gomez-Rodriguez, J. Leskovec, and A. Krause, "Inferring networks of diffusion and influence," *ACM Trans. Knowl. Discov. Data*, vol. 5, no. 4, 2012, 21:1–21:37. DOI: 10.1145/2086737.2086741.

[62]  N. Grinberg, K. Joseph, L. Friedland, B. Swire-Thompson, and D. Lazer, "Fake news on Twitter during the 2016 US presidential election," *Science*, vol. 363, no. 6425, 2019, pp. 374–378.

[63]  Z. Guo, M. Schlichtkrull, and A. Vlachos, "A Survey on Automated Fact-Checking," *Transactions of the Association for Computational Linguistics*, vol. 10, 2022, pp. 178–206. DOI: 10.1162/tacl_a_00454.

[64]  A. A. Hagberg, D. A. Schult, and P. J. Swart, "Exploring network structure, dynamics, and function using networkx," in *Proc. of the 7th Python in Science Conference*, pp. 11–15, 2008.

[65]  C. N. Hang, Y. Z. Tsai, P. D. Yu, J. Chen, and C. W. Tan, "Privacy-enhancing digital contact tracing with machine learning for pandemic response: A comprehensive review," *Big Data and Cognitive Computing, Special issue on digital health and data analytics in public health, accepted and to appear*, 2023.

[66]  C. N. Hang, P. D. Yu, S. Chen, C. W. Tan, and G. Chen, "Machine learning-enhanced graph analytics for infodemic risk management," *IEEE Journal of Biomedical and Health Informatics*, Preprint, Accepted with minor revision, 2023.

[67]  W. Hu, W. P. Tay, A. Harilal, and G. Xiao, "Network infection source identification under the SIRI model," in *Proceedings of IEEE International Conference on Acoustics, Speech and Signal Processing (ICASSP)*, pp. 1712–1716, 2015.

[68]  F. Ji, W. Tang, and W. P. Tay, "On the properties of gromov matrices and their applications in network inference," *IEEE Transactions on Signal Processing*, vol. 67, no. 10, 2019, pp. 2624–2638.

[69]  F. Ji and W. P. Tay, "An algorithmic framework for estimating rumor sources with different start times," *IEEE Transactions on Signal Processing*, vol. 65, no. 10, 2017, pp. 2517–2530.

[70]  F. Ji, W. P. Tay, and L. R. Varshney, "Estimating the number of infection sources in a tree," in *Proceedings of IEEE Global Conference on Signal and Information Processing (GlobalSIP)*, pp. 380–384, 2016. DOI: 10.1109/GlobalSIP.2016.7905868.

[71]  J. Jiang, S. Wen, S. Yu, Y. Xiang, and W. Zhou, "Identifying propagation sources in networks: State-of-the-art and comparative studies," *IEEE Communications Surveys & Tutorials*, vol. 19, no. 1, 2017, pp. 465–481.

[72]  V. Jog and P. Loh, "Analysis of centrality in sublinear preferential attachment trees via the Crump-Mode-Jagers branching process," *IEEE Transactions on Network Science and Engineering*, vol. 4, no. 1, 2017, pp. 1–12. DOI: 10.1109/TNSE.2016.2622923.

[73]  V. Jog and P. Loh, "Persistence of centrality in random growing trees," *Random Structures & Algorithms*, vol. 52, no. 1, 2018, pp. 136–157.

[74]  N. L. Johnson and S. Kotz, *Urn models and their application: An approach to modern discrete probability theory*. Wiley, 1977.

[75]  A. Kalvit, V. S. Borkar, and N. Karamchandani, "Stochastic approximation algorithms for rumor source inference on graphs," *Performance Evaluation*, vol. 132, 2019, pp. 1–20.

[76]  U. Kang, S. Papadimitriou, J. Sun, and H. Tong, "Centralities in large networks: Algorithms and observations," in *Proceedings of the 2011 SIAM international conference on data mining*, pp. 119–130, 2011.

[77]  N. Karamchandani and M. Franceschetti, "Rumor source detection under probabilistic sampling," in *Proceedings of IEEE International Symposium on Information Theory*, pp. 2184–2188, 2013.

[78]  N. D. Kazarinoff, *Geometric Inequalities*. American Mathematical Society, 1975.

[79]  D. Kempe, J. Kleinberg, and E. Tardos, "Maximizing the spread of influence through a social network," *Proc. ACM SIGKDD International Conference on Knowledge Discovery and Data Mining*, 2003, pp. 137–146.

[80]  D. Kempe, J. Kleinberg, and E. Tardos, "Influential nodes in a diffusion model for social networks," in *Proceedings of the International Colloquium on Automata, Languages, and Programming (ICALP)*, pp. 1127–1138, 2005.

[81]  J. T. Kemper, "On the identification of superspreaders for infectious disease," *Mathematical Biosciences*, vol. 48, no. 1-2, 1980, pp. 111–127.

[82]  L. Kennedy-Shaffer, M. Baym, and W. P. Hanage, "Perfect as the enemy of good: Tracing transmission with low-sensitivity tests to mitigate SARS-CoV-2 outbreaks," *Lancet Microbe*, vol. 2, 2021, pp. 219–224.

[83]  W. O. Kermack and A. G. McKendrick, "A contribution to the mathematical theory of epidemics," *Proc. R. Soc. Lond. A*, vol. 115, no. 772, 1972, pp. 700–721. DOI: 10.1098/rspa.1927.0118.

[84]  H. Kesavareddigari, S. Spencer, A. Eryilmaz, and R. Srikant, "Identification and asymptotic localization of rumor sources using the method of types," *IEEE Transactions on Network Science and Engineering*, vol. 7, no. 3, 2019, pp. 1145–1157. DOI: 10.1109/TNSE.2019.2911275.

[85]  J. Khim and P. Loh, "Confidence sets for the source of a diffusion in regular trees," *IEEE Trans. Network Science and Engineering*, vol. 4, no. 1, 2017, pp. 27–40.

[86]  S. Kojaku, L. Hébert-Dufresne, E. Mones, S. Lehmann, and Y.-Y. Ahn, "The effectiveness of backward contact tracing in networks," *Nature Physics*, vol. 17, no. 5, 2021, pp. 652–658.

[87]  D. Koller and N. Friedman, *Probabilistic Graphical Models: Principles and Techniques*. Cambridge, MA: The MIT Press, 2009.

[88]  F. Krzakala, "Belief propagation for the (physicist) layman," Lecture Notes, 2011.

[89]   A. Kumar, V. S. Borkar, and N. Karamchandani, "Temporally agnostic rumor-source detection," *IEEE Transactions on Signal and Information Processing over Networks*, vol. 3, no. 2, 2017, pp. 316–329.

[90]   J. Kunegis, "KONECT – The Koblenz Network Collection," in *Proc. Int. Conf. on World Wide Web Companion*, pp. 1343–1350, 2013.

[91]   S. Landau, *People Count: Contact-Tracing Apps and Public Health*. MIT Press, 2021.

[92]   D. M. J. Lazer, M. A. Baum, Y. Benkler, A. J. Berinsky, K. M. Greenhill, F. Menczer, M. J. Metzger, B. Nyhan, G. Pennycook, J. L. Zittrain, and et al, "The science of fake news," *Science*, vol. 359, no. 6380, 2018, pp. 1094–1096. DOI: 10.1126/science. aao2998.

[93]   V. Lecomte, G. Ódor, and P. Thiran, "The power of adaptivity in source identification with time queries on the path," *Theoretical Computer Science*, vol. 911, 2022, pp. 92–123. DOI: 10.1016/j.tcs. 2021.09.034.

[94]   D. Leith and S. Farrell, "A measurement-based study of the privacy of Europe's COVID-19 contact tracing apps," in *Proceedings of IEEE International Conference on Computer Communications*, 2021.

[95]   D. J. Leith and S. Farrell, "Coronavirus contact tracing: Evaluating the potential of using bluetooth received signal strength for proximity detection," *ACM SIGCOMM Computer Communication Review*, vol. 50, no. 4, 2020, pp. 66–74.

[96]   K. Lerman and R. Ghosh, "Information contagion: Empirical study of the spread of news on Digg and Twitter social networks," in *Proceedings of 4th International AAAI Conference on Weblogs and Social Media*, 2010.

[97]   J. Leskovec and A. Krevl, *SNAP Datasets: Stanford large network dataset collection*, 2014. URL: http://snap.stanford.edu/data.

[98]   J. Leskovec, M. McGlohon, C. Faloutsos, N. Glance, and M. Hurst, "Patterns of cascading behavior in large blog graphs," in *Proceedings of SIAM International Conference on Data Mining*, pp. 551–556, 2007. DOI: 10.1137/1.9781611972771.60.

[99] Y. Lim, A. Ozdaglar, and A. Teytelboym, "Competitive rumor spread in social networks," *SIGMETRICS Perform. Eval. Rev.*, vol. 44, no. 3, 2017, pp. 7–14. DOI: 10.1145/3040230.3040233.

[100] S. van der Linden, "Misinformation: Susceptibility, spread, and interventions to immunize the public," *Nature Medicine*, vol. 28, 2022, pp. 460–467.

[101] X. Liu, A. Nourbakhsh, Q. Li, R. Fang, and S. Shah, "Real-time rumor debunking on Twitter," in *Proceedings of the 24th ACM International on Conference on Information and Knowledge Management*, ser. CIKM '15, pp. 1867–1870, Melbourne, Australia, 2015. DOI: 10.1145/2806416.2806651.

[102] A. Y. Lokhov, M. Mézard, H. Ohta, and L. Zdeborová, "Inferring the origin of an epidemic with a dynamic message-passing algorithm," *Phys. Rev. E*, vol. 90, 1 2014.

[103] A. Louni and K. P. Subbalakshmi, "Who spread that rumor: Finding the source of information in large online social networks with probabilistically varying internode relationship strengths," *IEEE Transactions on Computational Social Systems*, vol. 5, no. 2, 2018, pp. 335–343.

[104] L. Lovász, "Random walks on graphs: A survey," *Combinatorics*, vol. 2, 1993, pp. 1–46.

[105] W. Luo and W. P. Tay, "Identifying infection sources in large tree networks," in *Proceedings of the 9th Annual IEEE Communications Society Conference on Sensor, Mesh and Ad Hoc Communications and Networks (SECON)*, pp. 281–289, 2012. DOI: 10.1109/SECON.2012.6275788.

[106] W. Luo and W. P. Tay, "Estimating infection sources in a network with incomplete observations," in *Proceedings of IEEE Global Conference on Signal and Information Processing*, pp. 301–304, 2013.

[107] W. Luo and W. P. Tay, "Finding an infection source under the SIS model," in *Proceedings of IEEE International Conference on Acoustics, Speech and Signal Processing*, pp. 2930–2934, 2013.

[108] W. Luo, W. P. Tay, and M. Leng, "Identifying infection sources and regions in large networks," *IEEE Trans. Signal Processing*, vol. 61, no. 11, 2013, pp. 2850–2865.

[109] W. Luo, W. P. Tay, and M. Leng, "How to identify an infection source with limited observations," *IEEE Journal of Selected Topics in Signal Processing*, vol. 8, no. 4, 2014, pp. 586–597. DOI: 10.1109/JSTSP.2014.2315533.

[110] W. Luo, W. P. Tay, and M. Leng, "Rumor spreading maximization and source identification in a social network," in *Proceedings of IEEE/ACM International Conference on Advances in Social Networks Analysis and Mining (ASONAM)*, pp. 186–193, 2015.

[111] W. Luo, W. P. Tay, and M. Leng, "Infection spreading and source identification: A hide and seek game," *IEEE Transactions on Signal Processing*, vol. 64, no. 16, 2016, pp. 4228–4243.

[112] W. Luo, W. P. Tay, M. Leng, and M. K. Guevara, "On the universality of the Jordan center for estimating the rumor source in a social network," in *Proceedings of IEEE International Conference on Digital Signal Processing (DSP)*, pp. 760–764, 2015.

[113] D. J. C. Mackay, *Information Theory, Inference and Learning Algorithms*, First. Cambridge University Press, 2003.

[114] T. Matsuta and T. Uyematsu, "Probability distributions of the distance between the rumor source and its estimation on regular trees," in *Proceedings of the 37th Symposium on Information Theory and its Applications (ISTIA)*, pp. 605–610, 2014.

[115] N. Megiddo, "An $O(nlog_2n)$ algorithm for the $k$th longest path in a tree with applications to location problems," *SIAM Journal on Computing*, vol. 10, no. 2, 1979, pp. 328–337.

[116] E. Meirom, C. Milling, C. Caramanis, S. Mannor, A. Orda, and S. Shakkottai, "Localized epidemic detection in networks with overwhelming noise," *ACM SIGMETRICS Performance Evaluation Review*, vol. 43, 2014. DOI: 10.1145/2796314.2745883.

[117] E. A. Meirom, C. Caramanis, S. Mannor, A. Orda, and S. Shakkottai, "Detecting cascades from weak signatures," *IEEE Transactions on Network Science and Engineering*, vol. 5, no. 4, 2018, pp. 313–325. DOI: 10.1109/TNSE.2017.2764444.

[118] M. Meister and J. Kleinberg, "Optimizing the order of actions in a model of contact tracing," *The Proceedings of the National Academy of Sciences (PNAS) Nexus*, vol. 2, no. 3, 2023.

[119] C. Milling, C. Caramanis, S. Mannor, and S. Shakkottai, "Network forensics: Random infection vs spreading epidemic," *ACM SIGMETRICS Performance Evaluation Review*, vol. 40, no. 1, 2012, pp. 223–234. DOI: 10.1145/2318857.2254784.

[120] S. Negahban, S. Oh, and D. Shah, "Rank centrality: Ranking from pairwise comparisons," *Operations Research*, vol. 65, no. 1, 2016.

[121] M. E. J. Newman, "The spread of epidemic disease on networks," *Physical Review E*, vol. 66, 2002, p. 016 128.

[122] P. D. O'Neill, "A tutorial introduction to Bayesian inference for stochastic epidemic models using Markov chain Monte Carlo methods," *Mathematical Biosciences*, vol. 180, no. 1, 2002, pp. 103–114. DOI: https://doi.org/10.1016/S0025-5564(02)00109-8.

[123] A. Ortega, P. Frossard, J. Kovacevic, J. M. F. Moura, and P. Vandergheynst, "Graph signal processing: Overview, challenges, and applications," *Proceedings of the IEEE*, vol. 106, no. 5, 2018, pp. 808–828. DOI: 10.1109/JPROC.2018.2820126.

[124] L. Page, S. Brin, R. Motwani, and T. Winograd, *The pagerank citation ranking: Bringing order to the web*, 1998.

[125] R. Pastor-Satorras and A. Vespignani, "Immunization of complex networks," *Physical Review E*, vol. 65, 2002.

[126] B. Pittel, "On spreading a rumor," *SIAM J. Appl. Math.*, vol. 47, no. 1, 1987, pp. 213–223. DOI: 10.1137/0147013.

[127] B. A. Prakash, J. Vreeken, and C. Faloutsos, "Efficiently spotting the starting points of an epidemic in a large graph," *Knowledge and Information Systems*, vol. 38, no. 1, 2014, pp. 35–39.

[128] V. M. Preciado, M.Zargham, C. Enyioha, A. Jadbabaie, and G. J. Pappas, "Optimal resource allocation for network protection against spreading processes," *IEEE Trans. Control of Network Systems*, vol. 1, no. 1, 2014, pp. 99–108.

[129] M. Z. Rácz and J. Richey, "Rumor source detection with multiple observations under adaptive diffusions," *IEEE Transactions on Network Science and Engineering*, vol. 8, no. 1, 2021, pp. 2–12. DOI: 10.1109/TNSE.2020.3022621.

[130] R. Richardson and C. Director, "CSI computer crime and security survey," *Computer Security Institute*, vol. 1, 2008, pp. 1–30.

[131] D. Shah and T. Zaman, "Detecting sources of computer viruses in networks: Theory and experiment," *Proc. of ACM SIGMETRICS*, 2010.

[132] D. Shah and T. Zaman, "Rumors in a network: Whos's the culprit?" *IEEE Transactions on Information Theory*, vol. 57, 2011, pp. 5163–5181.

[133] D. Shah and T. Zaman, "Rumor centrality: A universal source detector," *Proc. of ACM SIGMETRICS*, 2012.

[134] Singapore Ministry of Health, "News highlights in March and April 2021," 2021. URL: https://www.moh.gov.sg/news-highlights/details/23-more-cases-discharged-120-new-cases-of-covid-19-infection-confirmed.

[135] S. T. Smith, E. K. Kao, E. D. Mackin, D. C. Shah, O. Simek, and D. B. Rubin, "Automatic detection of influential actors in disinformation networks," *The Proceedings of the National Academy of Sciences*, vol. 118, no. 4, 2021, pp. 1–10.

[136] S. T. Smith, E. K. Kao, K. D. Senne, G. Bernstein, and S. Philips, "Bayesian discovery of threat networks," *IEEE Transactions on Signal Processing*, vol. 62, no. 20, 2014, pp. 5324–5338.

[137] S. Spencer and R. Srikant, "Maximum likelihood rumor source detection in a star network," in *Proceedings of IEEE International Conference on Acoustics, Speech and Signal Processing (ICASSP)*, pp. 2199–2203, 2016. DOI: 10.1109/ICASSP.2016.7472067.

[138] A. Sridhar and H. V. Poor, "Quickest inference of network cascades with noisy information," *IEEE Transactions on Information Theory*, vol. 69, no. 4, 2023, pp. 2494–2522.

[139] R. A. Stein, "Super-spreaders in infectious diseases," *International Journal of Infectious Diseases*, vol. 15, no. 8, 2011, e510–e513.

[140] G. Streftaris and G. J. Gibson, "Statistical inference for stochastic epidemic models," in *Proceedings of International Workshop on Statistical Modelling*, pp. 609–616, 2002.

[141] Taiwan Centers for Disease Control, "Press releases from February 2021 to May 2021," 2021. URL: https://www.cdc.gov.tw/En/Bulletin/Detail/PbMrcxQ2bO7H2cdN2_JUEw?typeid=158.

[142] C. W. Tan, P. Yu, C. Lai, W. Zhang, and H. Fu, "Optimal detection of influential spreaders in online social networks," in *Proceedings of the Conference on Information Science and Systems (CISS)*, pp. 145–150, 2016. DOI: 10.1109/CISS.2016.7460492.

[143] C. W. Tan, P. D. Yu, S. Chen, and H. V. Poor, "DeepTrace: Learning to optimize contact tracing in epidemic networks with graph neural networks," *CoRR*, vol. abs/2211.00880, 2022. URL: https://arxiv.org/abs/2211.00880.

[144] W. Tang, F. Ji, and W. P. Tay, "Estimating infection sources in networks using partial timestamps," *IEEE Transactions on Information Forensics and Security*, vol. 13, no. 12, 2018, pp. 3035–3049. DOI: 10.1109/TIFS.2018.2837655.

[145] S.-H. Teng, "Scalable algorithms for data and network analysis," *Foundations and Trends in Computer Science*, vol. 12, no. 1-2, 2016, pp. 1–274.

[146] T. Uno and H. Satoh, "An efficient algorithm for enumerating chordless cycles and chordless paths," *Proceedings of the International Conference on Discovery Science*, 2014, pp. 313–324.

[147] P. Van Mieghem, J. Omic, and R. Kooij, "Virus spread in networks," *IEEE/ACM Transactions on Networking*, vol. 17, no. 1, 2009, pp. 1–14. DOI: 10.1109/TNET.2008.925623.

[148] L. Vassio, F. Fagnani, P. Frasca, and A. Ozdaglar, "Message passing optimization of harmonic influence centrality," *IEEE Trans. Control of Network Systems*, vol. 1, no. 1, 2014, pp. 109–120.

[149] M. D. Vicario, A. Bessib, F. Zollo, and et al, "The spreading of misinformation online," *The Proceedings of the National Academy of Sciences*, vol. 113, no. 3, 2016, pp. 554–559.

[150] S. Vosoughi, D. Roy, and S. Aral, "The spread of true and false news online," in *Science*, ser. Vol. 359, Issue 6380, pp. 1146–1151, American Association for the Advancement of Science, 2018. DOI: 10.1126/science.aap9559.

[151] C. Wan, W. Chen, and Y. Wang, "Scalable influence maximization for independent cascade model in large-scale social networks," *Data Mining and Knowledge Discovery*, vol. 25, 2012, pp. 545–576.

[152] B. Wang, G. Chen, L. Fu, L. Song, and X. Wang, "DRIMUX: Dynamic rumor influence minimization with user experience in social networks," *IEEE Transactions on Knowledge and Data Engineering*, vol. 29, no. 10, 2017, pp. 2168–2181.

[153] Z. Wang, W. Dong, W. Zhang, and C. W. Tan, "Rumor source detection with multiple observations: Fundamental limits and algorithms," in *Proceedings of ACM International Conference on Measurement and Modeling of Computer Systems*, pp. 1–13, Austin, Texas, USA: ACM, 2014. DOI: 10.1145/2591971.2591993.

[154] Z. Wang, W. Dong, W. Zhang, and C. W. Tan, "Rooting out rumor sources in online social networks: The value of diversity from multiple observations," *IEEE Journal of Selected Topics in Signal Processing*, vol. 9, no. 4, 2015, pp. 663–677.

[155] Z. Wang, W. Zhang, and C. W. Tan, "On inferring rumor source for SIS model under multiple observations," in *Proceedings of the IEEE International Conference on Digital Signal Processing (DSP)*, pp. 755–759, 2015.

[156] J. Waring, C. Lindvall, and R. Umeton, "Automated machine learning: Review of the state-of-the-art and opportunities for healthcare," *Artificial Intelligence in Medicine*, vol. 104:101822, 2020.

[157] K. Wasa, Y. Kaneta, T. Uno, and H. Arimura, "Constant time enumeration of bounded-size subtrees in trees and its application," *Proc. International Computing and Combinatorics Conference*, 2012.

[158] D. B. West, *Introduction to Graph Theory*. Englewood Cliffs, New Jersey: Prentice Hall, 1996.

[159] F. M. F. Wong, C. W. Tan, S. Sen, and M. Chiang, "Quantifying political leaning from tweets, retweets and retweeters," *IEEE Transactions on Knowledge and Data Engineering*, vol. 28, no. 8, 2016, pp. 2158–2172.

[160]   L. Ying and K. Zhu, *Diffusion Source Localization in Large Networks*, ser. Synthesis Lectures on Communication Networks. Morgan & Claypool Publishers, 2018.

[161]   P. D. Yu, C. W. Tan, and H. Fu, "Rumor source detection in finite graphs with boundary effects by message-passing algorithms," in *Proceedings of IEEE/ACM International Conference on Advances in Social Networks Analysis and Mining*, pp. 86–90, Sydney, Australia: ACM, 2017. DOI: 10.1145/3110025.3110028.

[162]   P. D. Yu, C. W. Tan, and H. Fu, "Rumor source detection in unicyclic graphs," in *Proceedings of IEEE Information Theory Workshop (ITW)*, pp. 439–443, 2017. DOI: 10.1109/ITW.2017.8277993.

[163]   P. D. Yu, C. W. Tan, and H. Fu, "Averting cascading failures in networked infrastructures: Poset-constrained graph algorithms," *IEEE Journal of Selected Topics in Signal Processing*, vol. 12, no. 4, 2018, pp. 733–748. DOI: 10.1109/JSTSP.2018.2844813.

[164]   P. D. Yu, C. W. Tan, and H. Fu, "Graph algorithms for preventing cascading failures in networks," in *Proceedings of the 52nd Annual Conference on Information Sciences and Systems (CISS)*, pp. 1–6, 2018. DOI: 10.1109/CISS.2018.8362273.

[165]   P. D. Yu, C. W. Tan, and H. Fu, "Epidemic source detection in contact tracing networks: Epidemic centrality in graphs and message-passing algorithms.," *IEEE Journal of Selected Topics in Signal Processing*, vol. 16, no. 2, 2022, pp. 234–249. DOI: 10.1109/JSTSP.2022.3153168.

[166]   T. Zaman, "Information extraction with network centralities: Finding rumor sources, measuring influence, and learning community structure," *Ph.D. Thesis, Massachusetts Institute of Technology*, 2011.

[167]   Z. Zhang, W. Xu, W. Wu, and D. Du, "A novel approach for detecting multiple rumor sources in networks with partial observations," *Journal of Combinatorial Optimization*, vol. 33, no. 1, 2017, pp. 132–146.

[168]   L. Zheng and C. W. Tan, "A probabilistic characterization of the rumor graph boundary in rumor source detection," in *Proceedings of IEEE International Conference on Digital Signal Processing (DSP)*, pp. 765–769, 2015.

[169]   K. Zhu and L. Ying, "Information source detection in networks: Possibility and impossibility results," in *Proceedings of The 35th Annual IEEE International Conference on Computer Communications (INFOCOM)*, pp. 1–9, 2016.

[170]   K. Zhu and L. Ying, "Information source detection in the SIR model: A sample-path-based approach," *IEEE/ACM Trans. Networking*, vol. 24, no. 1, 2016, pp. 408–421.